\documentclass[lettersize,journal]{IEEEtran}
\usepackage{color}

\usepackage{soul}

\usepackage{amsmath}
\usepackage{amssymb}
\usepackage{graphicx}
\usepackage{colortbl}
\usepackage[table]{xcolor}
\usepackage{utfsym}
\usepackage{enumitem}

\usepackage{multirow, hhline}

\usepackage{lscape}
\usepackage{rotating}
\usepackage{balance}
\usepackage{epstopdf}

\usepackage{array}
\newcolumntype{C}{>{\Centering\arraybackslash}X}

\newcommand{\virgolette}[1]{``#1''}

\usepackage{tabularx,ragged2e,booktabs,caption}
\usepackage{subcaption}
\usepackage{pifont}
\newcommand{\cmark}{\ding{51}}%
\newcommand{\xmark}{\ding{55}}%

\usepackage[graphicx]{realboxes}

\usepackage{pdfrender}

\usepackage{scalerel}
\usepackage{tikz}

\newcommand*\emptycirc[1][1ex]{\tikz\draw (0,0) circle (#1);} 
\newcommand*\halfcirc[1][1ex]{%
  \begin{tikzpicture}
  \draw[fill] (0,0)-- (90:#1) arc (90:270:#1) -- cycle ;
  \draw (0,0) circle (#1);
  \end{tikzpicture}}
\newcommand*\fullcirc[1][1ex]{\tikz\fill (0,0) circle (#1);}

\usetikzlibrary{svg.path}

\definecolor{orcidlogocol}{HTML}{A6CE39}
\tikzset{
  orcidlogo/.pic={
    \fill[orcidlogocol] svg{M256,128c0,70.7-57.3,128-128,128C57.3,256,0,198.7,0,128C0,57.3,57.3,0,128,0C198.7,0,256,57.3,256,128z};
    \fill[white] svg{M86.3,186.2H70.9V79.1h15.4v48.4V186.2z}
                 svg{M108.9,79.1h41.6c39.6,0,57,28.3,57,53.6c0,27.5-21.5,53.6-56.8,53.6h-41.8V79.1z M124.3,172.4h24.5c34.9,0,42.9-26.5,42.9-39.7c0-21.5-13.7-39.7-43.7-39.7h-23.7V172.4z}
                 svg{M88.7,56.8c0,5.5-4.5,10.1-10.1,10.1c-5.6,0-10.1-4.6-10.1-10.1c0-5.6,4.5-10.1,10.1-10.1C84.2,46.7,88.7,51.3,88.7,56.8z};
  }
}

\newcommand\orcidicon[1]{\href{https://orcid.org/#1}{\mbox{\scalerel*{
\begin{tikzpicture}[yscale=-1,transform shape]
\pic{orcidlogo};
\end{tikzpicture}
}{|}}}}

\usepackage[hidelinks]{hyperref}

\usepackage[acronym,shortcuts]{glossaries-extra}

\glssetcategoryattribute{acronym}{nohyper}{true}
\setabbreviationstyle[acronym]{long-short}

\newacronym{can}{CAN}{Controller Area Network}
\newacronym{dos}{DoS}{Denial of Service}
\newacronym{ecu}{ECU}{Electronic Control Unit}
\newacronym{evita}{EVITA}{European E-Safety Vehicle Intrusion Protected Applications}
\newacronym{hmac}{HMAC}{Hash-based Message Authentication Code}
\newacronym{hsm}{HSM}{Hardware Security Module}
\newacronym{its}{ITS}{Intelligent Transportation System}
\newacronym{ids}{IDS}{Intrusion Detection System}
\newacronym{mac}{MAC}{Message Authentication Code}
\newacronym{mmac}{M-MAC}{Mixed Message Authentication Code}
\newacronym{obd}{OBD}{On-Boad Diagnostic}
\newacronym{ts}{TS}{Time Server}
\newacronym{tcp}{TCP}{Transmission Control Protocol}
\newacronym{vanet}{VANET}{Vehicular Ad-hoc Network}

\usepackage[strict]{changepage}
\usepackage{framed}
\definecolor{formalshade}{rgb}{0.85,1,0.85}
\definecolor{darkblue}{rgb}{0.0,0.6,0.30}
\newenvironment{formal}{%
  \MakeFramed{\advance\hsize-\width\FrameRestore}%
  \noindent\hspace{-4.55pt}%
  \begin{adjustwidth}{}{7pt}%
}
{%
  \end{adjustwidth}\endMakeFramed%
}

\begin{document}

\title{A Survey and Comparative Analysis of Security Properties of CAN Authentication Protocols}
\markboth{}%
{Lotto \MakeLowercase{\textit{et al.}}: A Comparative Analysis of Security and Features of CAN Authentication Protocols}

\author{Alessandro Lotto~\orcidicon{0000-0003-3556-4589},~Francesco~Marchiori~\orcidicon{0000-0001-5282-0965},~\IEEEmembership{Student~Member,~IEEE,}\\ Alessandro Brighente~\orcidicon{0000-0001-6138-2995},~\IEEEmembership{Member,~IEEE,} and Mauro Conti~\orcidicon{0000-0002-3612-1934}~\IEEEmembership{Fellow,~IEEE}
 
\thanks{Authors are with the Department of Mathematics, University of Padova, via Trieste 63, Padova, Italy (email: alessandro.lotto@math.unipd.it, francesco.marchiori@math.unipd.it, alessandro.brighente@unipd.it, mauro.conti@unipd.it).

M. Conti is also affiliated with the Faculty of Electrical Engineering, Mathematics and Computer Science, Delft University of Technology, Mekelweg 4, 2628 CD Delft, Netherlands.}
\thanks{This work has been submitted to the IEEE for possible publication. Copyright may be transferred without notice, after which this version may no longer be accessible.}
}

\maketitle

\begin{abstract}
The large number of Electronic Control Units (ECUs) mounted on modern cars and their expansive communication capabilities create a substantial attack surface for potential exploitation.
Despite the evolution of automotive technology, the continued use of the originally insecure Controller Area Network (CAN) bus leaves in-vehicle communications inherently non-secure.
In response to the absence of standardized authentication protocols within the automotive domain, researchers propose diverse solutions, each with unique strengths and vulnerabilities.
However, the continuous influx of new protocols and potential oversights in meeting security requirements and essential operational features further complicate the implementability of these protocols.

This paper comprehensively reviews and compares the 15 most prominent authentication protocols for the CAN bus.
Our analysis emphasizes their strengths and weaknesses, evaluating their alignment with critical security requirements for automotive authentication.
Additionally, we evaluate protocols based on essential operational criteria that contribute to ease of implementation in predefined infrastructures, enhancing overall reliability and reducing the probability of successful attacks.
Our study reveals a prevalent focus on defending against external attackers in existing protocols, exposing vulnerabilities to internal threats.
Notably, authentication protocols employing hash chains, Mixed Message Authentication Codes, and asymmetric encryption techniques emerge as the most effective approaches.
Through our comparative study, we classify the considered protocols based on their security attributes and suitability for implementation, providing valuable insights for future developments in the field.
\end{abstract}


\IEEEpeerreviewmaketitle

\section{Introduction}
\label{sec:introduction}

\IEEEPARstart{T}{he} rising number of vehicles has led to increased traffic congestion and safety concerns for drivers~\cite{HighEfficiency, IoV}.
To address these challenges, industries, and academia are actively developing technologies for \acp{its}.
These systems aim to enhance road coordination through innovative services, including traffic management and vehicle-to-infrastructure communication~\cite{toor2008vehicle, hartenstein2008tutorial, djahel2014communications}.
While \acp{vanet} are seen as a step towards secure \acp{its} and autonomous driving, it is crucial to ensure security at individual network nodes~\cite{checkoway2011comprehensive}.
In this context, securing the in-vehicle communication is paramount, considering the potential impact on network reliability and user safety~\cite{checkoway2011comprehensive}.
Moreover, the automotive market's increasing focus on advanced technology has led to a proliferation of electronic components, such as \acp{ecu}, that facilitate communication between vehicles~\cite{checkoway2011comprehensive}.

The \ac{can}, introduced by Robert Bosch GmbH in 1983, has become a renowned and widely used serial bus, initially designed for automotive applications~\cite{LeiA, CAN, CAN2}.
Still, \ac{can} has found versatile use in fields like medical devices, industrial automation, and robotics, thanks to its reliable, robust, efficient, and flexible communication capabilities~\cite{CAN2, otherCAN1, CANissues}.
Despite its adoption as the \emph{de-facto} standard in modern cars, \ac{can} lacks inherent security features, as its design assumes an isolated and friendly environment ~\cite{WooAuth, CANissues, CANauth, LCAP}.
However, with modern cars featuring various communication channels like Bluetooth, \ac{obd} ports, and cellular technology, interactive systems expose potential vulnerability surfaces for attackers~\cite{kang2018automated, zhang2016controlling, palanca2017stealth}.
The absence of security features in \ac{can}, including confidentiality, encryption, data integrity, and authenticity, renders the system susceptible to exploitation.
This vulnerability was demonstrated by C. Miller and C. Valasek, who remotely hacked a Fiat Chrysler (Jeep) through the internet-connected \virgolette{Uconnect}, exploiting open \ac{tcp} ports on the Sprint mobile network~\cite{bloom2021weepingcan, iehira2018spoofing, bozdal2020evaluation, remoteExploit}.
This breach enabled remote code execution, providing attackers full control over the vehicle and leaving users powerless to regain control.

A crucial yet vulnerable aspect of \ac{can} is its broadcast communication, where each message contains a field specifying data type rather than the sender's identity.
Such a lack of sender identification poses severe authentication issues and security threats, endangering vehicle and passenger safety~\cite{lokman2019intrusion, hossain2020lstm, hanselmann2020canet}.
Researchers have proposed various authentication schemes for securing \ac{can} communications to address this vulnerability.
Despite widespread recognition of the authentication issue in the automotive industry and the proposal of several solutions, a practical lack of adoption persists in modern cars.
This gap exposes vehicles to potential malicious attacks, such as the injection of packets in the \ac{can} bus for vehicle theft~\cite{headlights}.
This further highlights the need for comprehensive analysis and comparison of authentication solutions to understand the reasons behind their limited usage~\cite{VehicleNet}.
Moreover, given the constant introduction of new protocol proposals from researchers, who can often oversee essential security requirements or necessary features for the protocols' implementation, the available body of literature has become particularly cumbersome and convoluted.
This emphasizes the importance of thoroughly assessing the most promising authentication solutions in practical scenarios.

\textbf{Contributions.}
In this paper, we provide a comparative analysis of the 15 most prominent authentication protocols for \ac{can}.
On top of the protocols' descriptions, we offer a comprehensive review and a unified model for comparing these protocols.
Furthermore, we discern the critical security requirements necessary for the automotive framework, many of which have been overlooked or defined with limitations in related works.
We bridge crucial gaps in the existing knowledge by highlighting these limitations and discussing how to address them.
To complement our analysis, we introduce additional comparison criteria that can aid practitioners in selecting the most suitable protocol for real-world implementations.
Lastly, we take a forward-looking approach by proposing enhancements that can fortify the existing protocols, making them more robust and secure.
Additionally, we offer valuable insights for developing future authentication protocols in the context of the \ac{can} bus.
Together, these contributions advance our understanding of \ac{can} authentication protocols and provide practical guidance for their implementation and improvement in automotive systems.
To the best of our knowledge, this survey work marks the first of its kind in the literature, as also demonstrated in Table~\ref{tab:surveycomp}.

\begin{table*}[!htpb]
\caption{Comparison of the surveys' contributions.}
\label{tab:surveycomp}
\centering
\begin{tabular}{lccccccc}
\hline
\multicolumn{1}{c}{\textbf{Survey}} & \textbf{Year} & \textbf{\begin{tabular}[c]{@{}c@{}}Protocols\\Description\end{tabular}} & \textbf{\begin{tabular}[c]{@{}c@{}}Protocols\\Comparison\end{tabular}} & \textbf{\begin{tabular}[c]{@{}c@{}}Security\\Criteria\end{tabular}}  & \textbf{\begin{tabular}[c]{@{}c@{}}Operational\\Criteria\end{tabular}} & \textbf{\begin{tabular}[c]{@{}c@{}}Security\\Analysis\end{tabular}} & \textbf{\begin{tabular}[c]{@{}c@{}}Security\\Classification\end{tabular}} \\ \hline

Wolf et al.~\cite{survey4}          & 2004  & \emptycirc & \emptycirc & \emptycirc & \emptycirc & \emptycirc & \emptycirc    \\ 
Studnia et al.~\cite{survey13}      & 2013  & \emptycirc & \emptycirc & \emptycirc & \emptycirc & \emptycirc & \emptycirc    \\ 
Liu et al.~\cite{survey11}          & 2017  & \emptycirc & \emptycirc & \emptycirc & \emptycirc & \emptycirc & \emptycirc    \\ 
Nowdehi et al.~\cite{survey2}       & 2017  & \halfcirc & \fullcirc & \emptycirc & \fullcirc & \emptycirc & \fullcirc \\ 
Avatefipour et al.~\cite{survey1}   & 2018  & \halfcirc & \emptycirc & \emptycirc & \emptycirc & \emptycirc & \emptycirc    \\ 
Groza et al.~\cite{survey3}         & 2018  & \halfcirc & \emptycirc & \emptycirc & \emptycirc & \emptycirc & \emptycirc    \\ 
Gmiden et al.~\cite{survey7}        & 2019  & \halfcirc & \fullcirc & \halfcirc & \halfcirc & \emptycirc & \emptycirc    \\ 
Bozdal et al.~\cite{survey10}       & 2020  & \emptycirc & \emptycirc & \emptycirc & \emptycirc & \emptycirc & \emptycirc    \\ 
Hartzell et al.~\cite{survey9}      & 2020  & \emptycirc & \emptycirc & \emptycirc & \emptycirc & \emptycirc & \emptycirc    \\ 
Aliwa et al.~\cite{survey6}         & 2021  & \halfcirc & \fullcirc & \halfcirc & \halfcirc & \emptycirc & \emptycirc    \\
Jo et al.~\cite{survey5}            & 2021  & \halfcirc & \emptycirc & \emptycirc & \emptycirc & \emptycirc & \emptycirc    \\
Fakhfakh et al.~\cite{survey8}      & 2022  & \halfcirc & \emptycirc & \emptycirc & \emptycirc & \emptycirc & \emptycirc    \\ 

\textbf{This work} & \textbf{2023} & \fullcirc & \fullcirc & \fullcirc & \fullcirc & \fullcirc & \fullcirc \\ \hline

\end{tabular}
\footnotesize
\vspace{.25em}
\begin{itemize}
    \item[] \qquad \qquad \quad \: \fullcirc\:indicates full consideration of a specific aspect.
    \item[] \qquad \qquad \quad \: \halfcirc\:indicates partial consideration of a specific aspect (i.e., less protocols, security requirements, or complementary features).
    \item[] \qquad \qquad \quad \: \emptycirc\:indicates no consideration of a specific aspect.
\end{itemize}
\end{table*}

The main contributions of our work are summarized as follows:
\begin{itemize}
    \item We present a technical and detailed overview of \textbf{15 cryptographic authentication protocols} for \ac{can}.
    To the best of our knowledge, this is the first contribution to provide a detailed review and unified model for comparison.
    Table~\ref{tab:ProtocolsConsidered} reviews the considered protocols, which are the most surveyed in the literature, with the addition of the two latest protocols.
    
    \item We identify and define \textbf{security criteria} necessary for \ac{can} authentication protocols applied to the automotive framework.
    Some are either not considered in related works or defined with some limitations.
    We highlight such limitations and provide a discussion to address them.
    
    \item We identify and define \textbf{operational criteria} crucial for selecting the protocol for real-world implementation.
    In conjunction with security requirements, these characteristics provide a comprehensive analytical overview of the protocols.
    
    \item We propose \textbf{enhancements} that may be applied to make protocols more robust and secure and provide indications for developing future authentication protocols in the \ac{can} bus.
\end{itemize}

\textbf{Organization.}
The paper is organized as follows. Section~\ref{sec:relatedWorks} presents the related works.
Section~\ref{sec:backgorownd} provides an overview of \ac{can} specifications and the current standard scenario.
Section~\ref{sec:adversarial} presents the possible attacks to the \ac{can} bus and defines the adversarial model.
Section~\ref{sec:hsm} discusses hardware security solutions to introduce cryptographic primitives in \ac{can}.
Section~\ref{sec:protocols} presents all the authentication protocols considered in this work.
Section~\ref{sec:criteria} defines the criteria according to which the protocol comparison will be based.
In Section~\ref{sec:comparison1} and Section~\ref{sec:comparison2} the comparison is presented.
Section~\ref{sec:result} presents an analysis of the results from the comparison, and Section~\ref{sec:Takeaways} provides the takeaway messages.
Finally, Section~\ref{sec:conclusions} concludes the paper.
Furthermore, Appendix~\ref{sec:cryptosec} provides the necessary background knowledge about cryptographic elements that will be introduced in the paper.
\section{Related Works}\label{sec:relatedWorks}
Our research shows that most surveys about \ac{can} security mainly analyze \ac{can} vulnerability surfaces, possible exploits, and corresponding countermeasure approaches.
Furthermore, the large part of existing surveys in the literature that focus on a specific security aspect, thus carrying out a detailed review and comparison of existing security mechanisms, mainly cares about \ac{ids}. 
On the contrary, only a few works specifically survey cryptographic authentication solutions in an extensive and in-depth manner.
The absence of a comprehensive literature review about the authentication framework for the \ac{can} bus and limitations shown by existing works
prevents us from reaching a definite solution for \ac{can} authentication protocols, which highlights the importance of conducting a comprehensive and in-depth comparative analysis between the most discussed and promising authentication protocols for \ac{can}.

Several survey works overview the most common and discussed vulnerability surfaces and corresponding threats and exploits, together with possible countermeasure approaches for security solutions~\cite{survey4, survey13, survey11, survey1, survey10, survey9, survey8}.
From the analysis of these surveys, we can conclude that the current state-of-the-art approaches to secure the \ac{can} bus environment are encryption-based mechanisms (authentication and encryption schemes), \acp{ids}, firewalls, and network segmentation.
However, these papers only conduct a high-level and general discussion about possible countermeasure approaches and mitigation solutions, giving some case studies as examples.
Works~\cite{survey10} and ~\cite{survey8} present a dedicated section about authentication schemes.
However, the protocol analysis and (when present) comparison are carried out from a high-level perspective without providing detailed security analysis and comparison criteria, which is our work's goal.
These works are considerably noteworthy in value and thus will constitute the starting point for our work.
Table~\ref{tab:surveycomp} compares related works to ours, highlighting the contribution of our survey.
Furthermore, Table~\ref{tab:ProtocolsConsidered} compares the protocols considered in this survey to the ones considered in the related works.
We consider the most promising protocols from the literature and related survey works for our analysis.

\noindent\emph{Nowdehi et~al.}~\cite{survey2} conducted an effective comparison analysis between 10 authentication protocols for \ac{can}.
The authors defined five industrial requirements they identified as necessary to satisfy for a security solution to be usable in practice.
Protocol comparison is performed in light of the defined requirements: effectiveness, backward compatibility, support for vehicle repair and maintenance, sufficient implementation details, and acceptable overhead.
The limitations of their survey paper encompass concerns regarding cost-effectiveness in mandating hardware-supported cryptographic primitives for all \acp{ecu} in safety-critical real-time systems, potential backward compatibility issues, and the necessity of acceptable overhead for hardware-based solutions.
We present a more detailed analysis of these limitations in Section~\ref{subsec:limitations}.
It is worth noting that these limitations are rooted in a requirements definition that is implementation-oriented and performance-oriented, as stated by the authors.
On the contrary, our paper prioritizes a security analysis of authentication protocols for \ac{can}, marking it as the first of its kind.

\noindent\emph{Groza et~al.}~\cite{survey3} conducted a survey presenting the evolution of protocol proposals for \ac{can} security in chronological order from 2007 to 2016.
Besides, the authors discussed some cryptographic tools and solutions for the key-sharing problem when using encryption-based mechanisms in \ac{can}.
Some authentication protocol proposals were briefly mentioned as examples of possible authentication approaches, but no detailed description nor protocol analysis were given.
Furthermore, no comparison between the protocols was presented.

\noindent\emph{Gmiden et~al.}~\cite{survey7} work surveys \ac{ids} and cryptographic security solutions designed for the \ac{can} bus.
Regarding the cryptographic solutions, the authors evaluated and compared the considered protocols according to the following criteria: authentication, integrity, confidentiality, backward compatibility, replay attack resistance, and real-time performance.
The survey's limitations include a narrow focus on only six protocols, which does not comprehensively represent the entirety of the literature.
Moreover, the lack of precise definitions for evaluation criteria, particularly regarding backward compatibility, raises ambiguity about protocol compliance.
The generic nature of the protocol evaluation in the survey results in a limited and high-level analysis, potentially overlooking variations in authentication effectiveness in different scenarios and against different types of attackers.

\noindent\emph{Aliwa et~al.}~\cite{survey6} presented the \ac{can}-bus protocol and its limitations with related vulnerabilities and corresponding exploits an attacker can carry out.
Then, they discussed possible approaches to enforce security in the \ac{can} framework - such as authentication, encryption, and \ac{ids} deployment - and presented some case studies.
The authors outlined several protocols for the authentication framework according to their main features and working principles.
Furthermore, they provided a summary table that compares the protocol based on several aspects comprising security and operational features.
We highlight, however, that the considered comparison criteria are less than those we will consider for our analysis.
On the other hand, this work lacks a precise definition of evaluation criteria and a proper security analysis and discussion.

\noindent\emph{Jo et~al.}~\cite{survey5} presented a classification of possible \ac{can} attack surfaces (physical access-based, wireless access-based requiring initial physical access, wireless access-based without physical access) and a categorization of defense approaches (preventative protection, intrusion detection, authentication, and post-protection).
For each defense category identified, the authors gave an overview of several implementation approaches, analyzing their pros and cons by presenting some implementation proposals in the literature.
Regarding the authentication part, the authors focused on solutions for the authentication key sharing and transmission of the authentication tag.
They revised possible approaches for each authentication aspect and pointed out some related works employing them.
However, this survey did not define any protocol evaluation criteria, nor was any protocol comparison performed.
Rather, a conceptual analysis of the authentication methodology, advantages, and drawbacks of the considered protocols was presented.

\noindent Finally, \emph{Pesè et~al.}~\cite{S2-CAN} proposed the S2-CAN authentication protocol.
In their work, the authors compared their proposal with other related works.
However, the compared protocols were generally presented in their main authentication strategy, and the comparison was carried out according to technical details of the induced latency and authentication method, i.e., employed cryptographic algorithms, \ac{mac} length, use of hardware modules.
However, no comprehensive and in-depth security analysis and comparison was present.

\def\anlge{68}

\begin{table*}[!htpb]
\caption{Comparison of the surveys' considered protocols for CAN authentication.}
\label{tab:ProtocolsConsidered}
\centering
\begin{tabular}{lccccccccccccccc}
\hline
\textbf{Survey} & \rotatebox{\anlge}{\textbf{CANAuth}} & \rotatebox{\anlge}{\textbf{Car2X}} & \rotatebox{\anlge}{\textbf{LinAuth}} & \rotatebox{\anlge}{\textbf{MaCAN}} & \rotatebox{\anlge}{\textbf{LCAP}} & \rotatebox{\anlge}{\textbf{CaCAN}} & \rotatebox{\anlge}{\textbf{VeCure}} & \rotatebox{\anlge}{\textbf{Woo-Auth}} & \rotatebox{\anlge}{\textbf{LeiA}} & \rotatebox{\anlge}{\textbf{vatiCAN}} & \rotatebox{\anlge}{\textbf{LiBrA-CAN}} & \rotatebox{\anlge}{\textbf{VulCAN}} & \rotatebox{\anlge}{\textbf{TOUCAN}} & \rotatebox{\anlge}{\textbf{AuthentiCAN}} & \rotatebox{\anlge}{\textbf{S2-CAN}} \\ \hline

Wolf et al.~\cite{survey4}          & \emptycirc & \emptycirc & \emptycirc & \emptycirc & \emptycirc & \emptycirc & \emptycirc & \emptycirc & \emptycirc & \emptycirc & \emptycirc & \emptycirc & \emptycirc & \emptycirc & \emptycirc \\ 
Studnia et al.~\cite{survey13}      & \emptycirc & \fullcirc & \emptycirc & \fullcirc & \emptycirc & \emptycirc & \emptycirc & \emptycirc & \emptycirc & \emptycirc & \fullcirc & \emptycirc & \emptycirc & \emptycirc & \emptycirc \\ 
Liu et al.~\cite{survey11}          & \emptycirc & \emptycirc & \emptycirc & \emptycirc & \emptycirc & \emptycirc & \emptycirc & \fullcirc & \emptycirc & \emptycirc & \emptycirc & \emptycirc & \emptycirc & \emptycirc & \emptycirc \\ 
Nowdehi et al.~\cite{survey2}       & \fullcirc & \fullcirc & \fullcirc & \fullcirc & \emptycirc & \fullcirc & \fullcirc & \fullcirc & \emptycirc & \fullcirc & \fullcirc & \emptycirc & \emptycirc & \emptycirc & \emptycirc \\ 
Avatefipour et al.~\cite{survey1}   & \emptycirc & \emptycirc & \fullcirc & \emptycirc & \emptycirc & \emptycirc & \fullcirc & \emptycirc & \emptycirc & \emptycirc & \emptycirc & \emptycirc & \emptycirc & \emptycirc & \emptycirc \\ 
Groza et al.~\cite{survey3}         & \fullcirc & \emptycirc & \emptycirc & \fullcirc & \emptycirc & \fullcirc & \emptycirc & \fullcirc & \emptycirc & \emptycirc & \fullcirc & \emptycirc & \emptycirc & \emptycirc & \emptycirc \\ 
Gmiden et al.~\cite{survey7}        & \emptycirc & \emptycirc & \emptycirc & \emptycirc & \emptycirc & \fullcirc & \fullcirc & \fullcirc & \emptycirc & \emptycirc & \fullcirc & \fullcirc & \emptycirc & \emptycirc & \emptycirc \\ 
Bozdal et al.~\cite{survey10}       & \emptycirc & \emptycirc & \emptycirc & \emptycirc & \emptycirc & \emptycirc & \fullcirc & \emptycirc & \emptycirc & \emptycirc & \fullcirc & \emptycirc & \emptycirc & \emptycirc & \emptycirc \\  
Hartzell et al.~\cite{survey9}      & \emptycirc & \emptycirc & \fullcirc & \emptycirc & \emptycirc & \emptycirc & \emptycirc & \emptycirc & \emptycirc & \emptycirc & \emptycirc & \emptycirc & \emptycirc & \emptycirc & \emptycirc \\ 
Aliwa et al.~\cite{survey6}         & \fullcirc & \emptycirc & \fullcirc & \emptycirc & \fullcirc & \fullcirc & \fullcirc & \emptycirc & \fullcirc & \emptycirc & \fullcirc & \emptycirc & \fullcirc & \emptycirc & \emptycirc \\ 
Jo et al.~\cite{survey5}            & \emptycirc & \emptycirc & \emptycirc & \emptycirc & \emptycirc & \fullcirc & \fullcirc & \fullcirc & \fullcirc & \fullcirc & \emptycirc & \fullcirc & \fullcirc & \emptycirc & \emptycirc \\ 
Fakhfakh et al.~\cite{survey8}      & \emptycirc & \emptycirc & \emptycirc & \emptycirc & \emptycirc & \emptycirc & \fullcirc & \fullcirc & \emptycirc & \emptycirc & \emptycirc & \emptycirc & \emptycirc & \emptycirc & \emptycirc \\
\textbf{This work} & \fullcirc & \fullcirc & \fullcirc & \fullcirc & \fullcirc & \fullcirc & \fullcirc & \fullcirc & \fullcirc & \fullcirc & \fullcirc & \fullcirc & \fullcirc & \fullcirc & \fullcirc \\ \hline

\end{tabular}
\end{table*}
\section{Background}
\label{sec:backgorownd}

In this section, we provide a background overview of the \ac{can} standard protocol~\cite{CAN, CAN2} in its main and most significant aspects necessary for a complete comprehension of the following discussions (Section~\ref{subsec:canstandard}).
Moreover, we also present two other protocols, CAN+~\cite{CAN+} (Section~\ref{subsec:canplus}) and CAN-FD~\cite{CAN-FD} (Section~\ref{subsec:canfd}), which are two more recent versions of the standard \ac{can} protocol.

\subsection{CAN}
\label{subsec:canstandard}
In this section, we provide a presentation of the four core aspects of the \ac{can} standard: \textit{(i)} the physical layer and \ac{ecu} structure, \textit{(ii)} the structure of a \ac{can} frame, \textit{(iii)} the procedure to access to the bus, \textit{(iv)} the arbitration algorithm.

\subsubsection{\underline{Physical Layer}}
\ac{can} is a multi-master differential two-bus protocol. Every \ac{ecu} of the system is connected to all the others through two wires named CAN\_H and CAN\_L~\cite{CAN, CAN2}.
Being a multi-master protocol, every message is broadcast to all the other \acp{ecu}.
The differential property is related to the physical encoding of the bits into voltage values, which occurs as follows.
\begin{equation*}
    \begin{cases}
    0 \rightarrow CAN\_H - CAN\_L \approx 2V,\\
    1 \rightarrow CAN\_H - CAN\_L \approx 0V.\\
    \end{cases}
\end{equation*}
In the equation, bit 0 is referred to as \emph{dominant} and bit 1 as \emph{recessive}.
This labeling suggests the bus behavior when a 0 and 1 are written simultaneously: as a 2V difference dominates over a 0V difference, the bit 0 will always overwrite the bit 1.
Moreover, thanks to this differential encoding, the electrical current flows in opposite directions for the two wires, resulting in a field-canceling effect.
This makes \ac{can} immune to noise and fault tolerant.

In practice, the signal is imperfect and needs some time to stabilize.
Therefore, when writing and reading the bus, delays must be considered.
The bit transmission time is then divided into 8 to 25 time intervals, and after about $2/3$ of the bit time, the signal sampling is performed.
A synchronization process at the beginning of the transmission is required as well~\cite{CAN+}.
\acp{ecu} synchronize by themselves the position of the sample point to a specific phase in relation to the edges of the monitored bit stream.
Receivers shift the phases of the samplings relatively to the phase of the transmitter's ones.
Thus, a node's specific phase shift depends on the signal delay time from the transmitter to that node~\cite{CAN-FD}.
The typical speed of a \ac{can} bus is 500 Kbit/s with a limit of 1 Mbit/s.

\subsubsection{\underline{ECU Structure}}
A typical \ac{ecu} usually comprises three parts: a \emph{micro-processor}, a \emph{\ac{can} controller}, and a \emph{\ac{can} transceiver}.
Even though messages are always broadcasted, the \ac{can} controller filters out unwanted messages by looking at the identification field of the message, thus understanding if the content is of interest.
\ac{can} transceivers connect the physical transmission medium to the controller and physically operate the two bus lines to generate the logical bits.

\subsubsection{\underline{CAN Frame}}
\begin{figure}[!htpb]
     \centering
     \begin{subfigure}[b]{\linewidth}
         \centering
         \includegraphics[width=.75\textwidth]{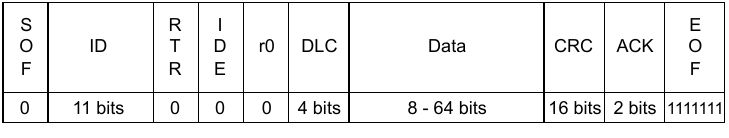}
         \caption{Standard \ac{can} 11-bit identifier.}
         \label{subfig:can11bit}
     \end{subfigure}
     \begin{subfigure}[b]{\linewidth}
         \centering
         \includegraphics[width=\textwidth]{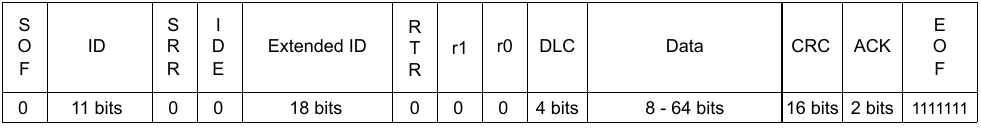}
         \caption{Extended \ac{can} 29-bit identifier.}
         \label{subfig:can29bit}
     \end{subfigure}
     \caption{\ac{can} frame structure.}
     \label{fig:CANframe}
\end{figure}

A \ac{can} data frame is composed of several fields, and generally, there are two formats of the protocol: the standard and extended, with an 11-bit and 29-bit identifier, respectively~\cite{CAN2}.
An overview of these two formats is shown in Figure~\ref{fig:CANframe}.
Each field is described as follows.
\begin{itemize}
    \item \textbf{Start of Frame (SOF)}: single dominant bit marking the beginning of the message.
    \item \textbf{Identifier (ID)}: 11-bit field that establishes the message's priority and discriminates the type of payload, i.e., the type of data sent.
    The extended version also uses the \textit{Extended ID (IDE)} field.
    \item \textbf{Remote Transmission Request (RTR)}: dominant single-bit field when information is required from another node.
    In this case, the ID identifies the type of information requested.
    \item \textbf{Substitute Remote Request (SRR)}: substitutes the RTR bit as a placeholder in the extended version.
    \item \textbf{Identifier Extension (IDE)}: single bit field, dominant if Extended ID is used.
    \item \textbf{r0, r1}: unused, reserved for future standard definitions.
    \item \textbf{Data Length Code (DLC)}: 4-bit field indicating the number of bytes of the payload.
    Values are taken from $0000$ to $1000$.
    \item \textbf{Data}: message payload ranging from 1 to 8 bytes. 
    \item \textbf{Cyclic Redundancy Check (CRC)}: contains the checksum of the data field for error detection.
    \item \textbf{Acknowledge (ACK)}: 2-bit field.
    The first bit is originally set to recessive.
    Every node receiving the message correctly overwrites it with a dominant bit.
    If the bit is left recessive, no \ac{ecu} correctly receives the message, which is discarded and sent again after re-arbitration.
    The second bit is used as a delimiter.
    \item \textbf{End-of-frame (EOF)}: 7 recessive bits marking the end of the frame.
\end{itemize}

\subsubsection{\underline{Arbitration Algorithm}}
\ac{can} is a Carrier-Sense Multiple-Access/Collision Detection and Arbitration Message Priority (CSMA/CD+AMP) communication protocol~\cite{wey2013enhancement}.
CSMA means that each node must wait for a determined period of inactivity before attempting to access the bus and send a message. CD+AMP means collisions with multiple messages sent simultaneously are solved through a bit-wise arbitration algorithm based on the message priority encoded in the identifier field.
When two or more \acp{ecu} try to access the bus simultaneously, only the one with higher priority succeeds, and no other \ac{ecu} attempts to access the bus until complete message transmission~\cite{murvay2017attacks}.
Besides, the arbitration algorithm also assures that no message is lost due to an interruption caused by a bus access attempt from an \ac{ecu} with higher priority than the one currently transmitting.
Notice that, thanks to the differential encoding, a higher priority corresponds to a lower ID, as bit 0 dominates over bit 1.
As a drawback, the arbitration mechanism strongly limits \ac{can} bus bandwidth. Indeed, protocol specifications require that the signal propagation between any couple of nodes is less than half of one-bit time~\cite{CAN2}.
This clearly defines an upper bound for the bit rate and (physical) bus length.

\subsection{CAN+}
\label{subsec:canplus}
\begin{figure}[!htpb]
    \centering
    \includegraphics[width=.825\linewidth]{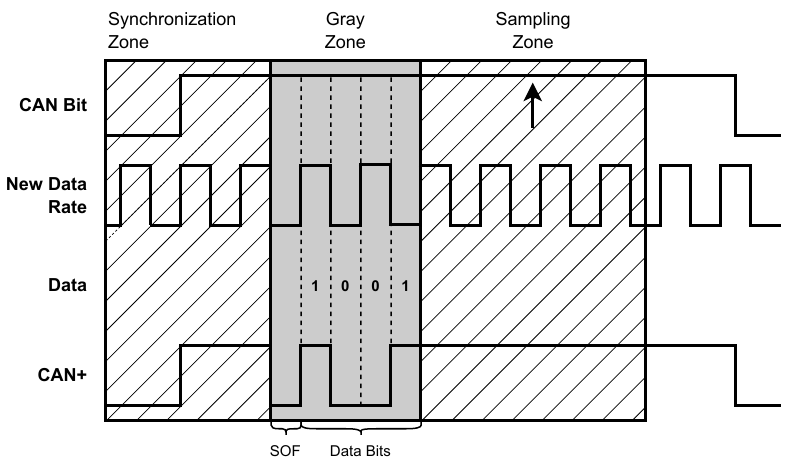}
    \caption{Transmission of a standard \ac{can} bit and of overclocking bits of CAN+.}
    \label{fig:can+}
\end{figure}

In recent years, the number of \acp{ecu} used is rising, and the workload on the bus is getting higher, causing a longer response time and degrading the real-time capability of the system.
The CAN+ protocol was introduced in 2009 as a solution for higher throughput, enabling a data rate up to 16 times higher compared to the standard \ac{can}~\cite{CAN+}.
By analyzing the transmission time of a single bit in classic \ac{can} protocol, researchers found a \emph{gray zone} period during which the bus could take any value without disturbing the normal \ac{can} communication.
This gray zone is delimited by the \emph{synchronization} and \emph{sampling zones}, as shown in Figure~\ref{fig:can+}, and it is the one used by CAN+ to transmit the additional information at a higher data rate.
These additional bits are called \emph{overclocked bits}.
Since overclocking is possible only when we can assure the presence of a single transmitter, we can exchange the overclocked bits during data transmission only.
By defining $f$ as the \emph{overclocking factor}, and recalling that a \ac{can} standard data field is at most 8 bytes, the maximum amount of transmitted bytes per message is: $$N_{total} = 64 + f\times 64 = (1+f) \times 64.$$
A detailed analysis shows that the overclocking factor can take a maximum value of 16, over which no feasible overclocking is possible~\cite{CAN+}.
It is also worth mentioning that to implement the CAN+ protocol properly, transceivers need to support data rates up to 60 Mbit/s~\cite{CAN+}.

\subsection{CAN-FD}
\label{subsec:canfd}
\begin{figure}[!htpb]
     \centering
     \begin{subfigure}[b]{\linewidth}
         \centering
         \includegraphics[width=.75\textwidth]{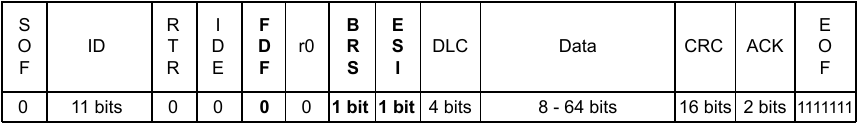}
         \caption{Standard \ac{can}-FD 11-bit identifier.}
         \label{subfig:canfd11bit}
     \end{subfigure}
     \begin{subfigure}[b]{\linewidth}
         \centering
         \includegraphics[width=\textwidth]{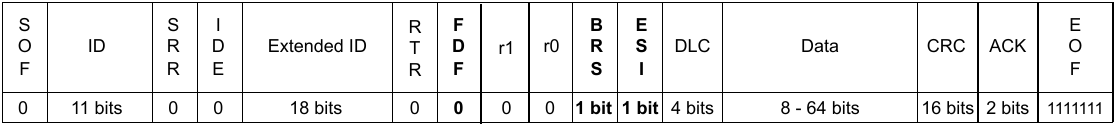}
         \caption{Extended \ac{can}-FD 29-bit identifier.}
         \label{subfig:can2fd9bit}
     \end{subfigure}
     \caption{\ac{can} frame structure.}
     \label{fig:canfd}
\end{figure}

\ac{can} with Flexible Data rate (CAN-FD) was designed in 2011 to \textit{(i)} improve the header-to-payload ratio, and \textit{(ii)} speeding up the frame transmission~\cite{CAN-FD}.
The protocol offers major improvements to the standard \ac{can}~\cite{comparingCAN-FD}.
\begin{itemize}
    \item Extension of Data field up to 64 bytes.
    \item Average transmission rate increased up to 6 Mbit/s.
    \item Higher performance in the CRC algorithm, lowering the risk of undetected errors.
\end{itemize}
As CAN+, CAN-FD exploits the fact that only a single node can access the bus after the arbitration.
Thus, switching to a faster and predefined transmission rate is possible.
It is worth noting that all nodes must switch to the new rate synchronously and return to the standard rate at the end of data transmission.
The speed-up factor is thus determined by \textit{(i)} the speed of the transceivers (if the bit time is too short, it will not be possible to decode it) and \textit{(ii)} the time resolution of the synchronization mechanism.
Other secondary precautions must be taken for complete backward compatibility and functioning of the protocol~\cite{CAN-FD}.
Besides data field extension, the protocol modifies the frame by defining new fields, as shown in Figure~\ref{fig:canfd}. 
\begin{itemize}
    \item \textbf{Flexible Data Format (FDF)}: single dominant bit providing an edge for resynchronization before a possible bit rate switch.
    \item \textbf{Bit Rate Switch (BRS)}: single bit that defines whether to switch into CAN-FD decoding mode.
    If recessive, \acp{ecu} must switch to the faster bit rate.
    \item \textbf{Error State Indicator (ESI)}: single bit set to dominant for error active, recessive for error passive.
\end{itemize}
Furthermore, the length of the payload is still defined by the DLC field but with an additional encoding: values from $1001$ to $1111$ define the data length up to 64 bytes.
Thus, each increase corresponds to a 4-byte increase in the data length (e.g., $1001$ corresponds to 12 bytes, $1010$ to 16 bytes).

\section{System Model}
\label{sec:hsm}
As \ac{can} protocol design was not meant for security, some complications arise when trying to build an authentication protocol on top of it.
These obstacles are mainly related to the introduction of cryptographic primitives and the management of related security parameters.
Besides, because \ac{can} IDs only identify the message's content type when introducing cryptographic primitives, there may be a need for transmitter identification.

In this section, we address the following two main aspects: \textit{(i)} handling cryptographic primitives (Section~\ref{subsec:hardwaresecuritymodules}), and \textit{(ii)} transmitter-receiver identification (Section~\ref{subsec:identificationissue}).

\subsection{Hardware Security Modules}
\label{subsec:hardwaresecuritymodules}
Generally, using cryptographic primitives requires a secure storage of keys and related parameters to offer accountability and confidentiality~\cite{mavrovouniotis2013hardware}.
Besides, when moving into the vehicular framework, we must consider other requirements and limitations, such as hard real-time constraints, power consumption, limited computing and storage resources, and implementation and maintenance costs.
To meet these constraints and limitations, the \ac{evita} project aimed to design building blocks for secure automotive on-board networks for securing security-relevant components~\cite{Evita}.
As such, they designed \acp{hsm} as \emph{root of trust} modules to be easily integrated as an on-chip extension to the \acp{ecu}~\cite{Evita1}. A \ac{hsm} provides secure hardware and cryptographic functions for secure communication between nodes without occurring in high costs~\cite{Evita2}.
Also, engineers developed low-level drivers based on AUTOSAR to interact with \ac{evita} \acp{hsm}~\cite{Evita3}.
AUTOSAR is a set of specifications and standards for Secure On-board Communication Modules to reduce costs and improve \acp{ecu} scalability~\cite{autosar}.

Components of an \ac{hsm} are divided into mandatory and optional, depending on the security requirements that must be fulfilled.
\ac{evita} also specifies three \ac{hsm} variants to meet different security levels and cost-effectiveness.
\begin{itemize}
    \item \textbf{Full \ac{hsm}} -- 
    It focuses on protecting the in-vehicle domain from security vulnerabilities from vehicle-to-everything communications.
    It provides the maximum functionality level, security, and performance among the three \ac{hsm} variants. It is designed to provide a security lifetime of over 20 years.
    It comprises blocks to perform symmetric and asymmetric cryptographic operations, such as hash computations and random number generations~\cite{Evita3}.    
    The major cryptographic building blocks are composed as follows.
    \begin{itemize}
        \item \emph{ECC-256-GF(p)}: high-performance asymmetric cryptographic engine based on a high-speed 256-bit elliptic curve arithmetic using NIST-approved prime field parameters~\cite{mcivor2006hardware}.
        \item \emph{WHIRLPOOL}: AES-based hash function, as proposed by NIST~\cite{stallings2006whirlpool}.
        \item \emph{AES-128}: symmetric block encryption/decryption engine using the official NIST advanced encryption standard~\cite{dworkin2001advanced}.
        \item \emph{AES-PRNG}: pseudo-random number generator, seeded with a truly random seed from a true internal physical random source.
        \item \emph{Counter}: a 64-bit monotonic counter function block that serves as a simple, secure clock alternative.
    \end{itemize}
    
    \item \textbf{Medium \ac{hsm}} --
    It aims to secure communications only within the vehicle. Compared to the full version, it is not provided with a hardware Error Correction Code (ECC) engine and a hardware hash engine.
    This module can quickly execute symmetric and non-time-critical asymmetric operations.
    Security credentials are permanently protected since they are kept from the application CPU~\cite{Evita3}.
    
    \item \textbf{Light \ac{hsm}} --
    It focuses on protecting \acp{ecu}, sensors, and actuators.
    The security scheme is limited to a single very-specialized symmetric AES hardware accelerator, while the application CPU handles all security credentials.
    This fulfills the strict cost and efficiency requirements typical of sensors and actuators~\cite{Evita3}.
\end{itemize}

\subsection{Transmitter-Receiver Identification}
\label{subsec:identificationissue}
The other issue that naturally occurs when introducing cryptography on the \acp{ecu} into the \ac{can} framework is identification.
Indeed, several protocols we consider expect some direct communication between nodes (i.e., \virgolette{node A sends to node B}).
However, direct communication is not supported in a broadcast environment without sender/receiver identification.
Hence, some enhancements to the standard are required.
Since the ID field does not identify the sender or receiver but the payload data type, direct communications should require the sender and receiver identifiers to be attached to the frame.
Without such information, the receiver could not identify the transmitter and the correct cryptographic key for decryption.
As a result, we need \ac{ecu}-specific internal identifiers to provide such functionalities.
Since not all the considered protocols mention this fact explicitly, we assume \acp{ecu} to be provided with internal specific identifiers when not directly reported.

One next step would be to consider embedding these identifiers into \ac{can} frames.
We identify using the extended \ac{can} ID field as the optimal solution since it is very unlikely that all the $2^{29}$ possible values are used for the standard \ac{can} IDs.
Therefore, available bits may be used for this purpose.
This prevents adding further bits in frames, which would lead to an increase in the bus load.
On the other hand, manufacturers can arbitrarily decide their specifications since there is no global standard to this scope yet.
We report the existence of the \emph{SAE J1939} protocol, widely adopted in industrial applications as a high-level protocol built on top of \ac{can}, which also gives specifications on how to structure the extended ID field to include \ac{ecu}-specific identifiers~\cite{SAEprotocol, burakova2016truck, murvay2018security}.
\section{Threat Model}
\label{sec:adversarial}
This section defines the adversarial model we consider for our analysis and investigates the possible attacks on an adversary may pursue on the \ac{can} bus.
Considering what has already been discussed in the related works, known limitations and vulnerabilities of the \ac{can} protocol, we classify attacks in two classes: \textit{injection} and \textit{\ac{dos}}~\cite{WooAuth, survey1, S2-CAN, survey5}.

The first class of attacks envisions the injection of malicious packets into the bus to modify the behavior or take control of the system.
Among injection attacks, we distinguish between \emph{replay attacks} and \emph{masquerade attacks}. A replay attack consists of injecting previously recorded legitimate messages without modification.
In a masquerade (or fabrication) attack, the attacker properly builds the content of the malicious packet to be transmitted, choosing the header and payload of the message.
This attack involves sending messages with forged IDs, impersonating any other legitimate \ac{ecu} authorized to send those specific IDs.
It is worth noting that the replay attack is a special case of the masquerade attack.
Still, we prefer to distinguish between the two since, in replay attacks, the attacker takes no action on the message apart from recording and re-transmitting it afterward.

\ac{dos} attacks consist of preventing legitimate \acp{ecu} (or a specific target \ac{ecu}) from accessing the bus and transmitting messages.
This happens because a malicious entity may flood the bus with high-priority messages, not allowing the transmission of frames with lower ID values.
Consequently, a \ac{dos} attack prevents the correct functioning of the system.
However, \ac{dos} attacks are out of the scope of this work as we cannot prevent them with cryptographic authentication mechanisms alone.
Rather, we would need a higher perspective system, such as an \ac{ids}, capable of monitoring and analyzing the ongoing traffic in the bus and recognizing whether a \ac{dos} attack is currently running.

After defining the possible attacks on the \ac{can} bus, we need to define how the attacker can perform an injection attack.
To get access to the \ac{can} bus, an attacker may \textit{(i)} make use of an external device or \textit{(ii)} compromise a legitimate \ac{ecu}.
In the first case, we define an external device as any instrument not belonging to the original system.
This can be either an external non-legitimate \ac{ecu} physically plugged into the network or a diagnostic instrument attached to the \ac{obd} port through which it is possible to inject packets.
In the second case, the adversary manages to compromise a legitimate \ac{ecu} by physically tampering with the device or remotely hacking it.
The two most common attacker behaviors are described as follows.

\begin{itemize}
    \item \emph{Target Injection} --
    Using the compromised \ac{ecu}, the attacker sends packets with ID values the \ac{ecu} is allowed to send, but faking the data field's values.
    We refer to this scenario as \emph{target injection} because we do not classify it as a masquerade attack but as false data transmission.
    In this case, the attacker must tamper with exactly an \ac{ecu} using the target IDs.
    We do not consider this scenario in our discussion since detecting and avoiding such attacks simply with an authentication scheme is impossible.
    It would be necessary, in fact, to have a specifically designed IDS.
    \item \emph{Masquerade Injection} --
    In this scenario, the attacker exploits the tampered \ac{ecu} to send messages with IDs it was not supposed to use, thus performing a masquerade or replay attack.
\end{itemize}
We assume the attacker can remotely compromise an \ac{ecu} and perform a masquerade injection in all scenarios.
We can justify our claim thanks to the fact that, in general, modern vehicles are equipped with an \ac{ecu} that keeps remote access ports open for external networks such as WiFi or Bluetooth~\cite{survey5}.
Besides, it is much harder to implement a target injection remotely than physically compromising the target \ac{ecu}.
However, this does not mean it is not possible.

Finally, we assume the following conditions to hold~\cite{survey6}.
\begin{itemize}
    \item \textbf{The attacker fully knows the underlying system}.
    The attacker knows if security mechanisms are in place and, in case, the specifics of the mechanism in use.
    
    \item \textbf{The attacker has full access to the bus}.
    The attacker can read and record exchanged messages.
    This assumption is reasonable given that potential attackers may have physical access to the \ac{can} bus through multiple points within the vehicle, both internal and external, including locations like headlights~\cite{headlights}.
    
    \item \textbf{It is not possible for the attacker to access data stored in protected memory from the outside.}
    The attacker cannot access the protected data of an \ac{ecu} from another \ac{ecu} or by physically tampering with the target device.
    Rather, they can access the protected data of the compromised \ac{ecu}.
    This assumption is justified by the discussion in the next section introducing tamper-proof memories~\cite{wolf2012design}.
\end{itemize}
\section{Authentication Protocols}
\label{sec:protocols}

In this section, we present the protocols considered for our comparison.
We provide a detailed description of their key working points, whereas information related to design decisions and performance results can be found, if not reported, in the referenced papers.
Furthermore, in this section, we only focus on the protocols' implementation explicitly stated in the papers.
We instead present additional considerations in Section~\ref{sec:comparison1} and Section~\ref{sec:comparison2}.
Table~\ref{tab:ProtocolSummary} summarizes the protocols' authentication approaches and provides the organization of this section.

\def\yearwidth{0.55cm}
\def\canwidth{1.1cm}
\def\approachwidth{2.1cm}
\def\header{15pt}
\def\oneline{12.5pt}
\def\twoline{20pt}
\def\threeline{27.5pt}
\def\fourline{37.5pt}
\def\fiveline{45pt}

\begin{table}[!htpb]
    \centering
    \caption{List of considered \ac{can} authentication protocols and their descriptions.}
    \label{tab:ProtocolSummary}
    \begin{tabular}{l|l|p{\yearwidth}|p{\canwidth}|p{\approachwidth}} \hline
         \parbox[][\header][c]{0.5cm}{\textbf{Sec.}} & \parbox[][\header][c]{1cm}{\textbf{Protocol}} & \parbox[][\header][c]{\yearwidth}{\textbf{Year}} & \multicolumn{1}{l|}{\textbf{Standard}}  & \multicolumn{1}{l}{\textbf{Authentication}} \\\hline
         \rowcolor{gray!15}
         \ref{subsec:canauth} &
         CANAuth~\cite{CANauth} &
         2011 &
         \parbox[][\oneline][c]{2cm}{CAN+}
         &
         \parbox{\approachwidth}{HMAC}
         \\\hline
         \ref{subsec:car2x} &
         Car2X~\cite{car2x} &
         2011 &
         \parbox[][\oneline][c]{2cm}{CAN}
         &
         \parbox{\approachwidth}{MAC}
         \\\hline
         \rowcolor{gray!15}
         \ref{subsec:lcap} &
         LCAP~\cite{LCAP}
         &
         2012 &
         \parbox[][\oneline][c]{2cm}{CAN}
         &
         \parbox{\approachwidth}{Hash chain}
         \\ \hline 
         \ref{subsec:linauth} &
         LinAuth~\cite{LinAuth}
         &
         2012 &
         \parbox[][\oneline][c]{2cm}{CAN}
         &
         \parbox{\approachwidth}{MAC}
         \\ \hline
          \rowcolor{gray!15}
         \ref{subsec:macan} &
         MaCAN~\cite{MaCAN}
         &
         2012 &
         \parbox[][\oneline][c]{2cm}{CAN}
         &
         \parbox{\approachwidth}{MAC}
         \\ \hline
         \ref{subsec:cacan} &
         CaCAN~\cite{CaCAN}
         &
         2014 &
         \parbox[][\oneline][c]{2cm}{CAN}
         &
         \parbox{\approachwidth}{Centralized HMAC}
         \\ \hline
        \rowcolor{gray!15}
         \ref{subsec:vecure} &
         VeCure~\cite{VeCure}
         &
         2014 &
         \parbox[][\oneline][c]{2cm}{CAN}
         &
         \parbox{\approachwidth}{Trust-group MAC}
         \\ \hline
         \ref{subsec:woo} &
         Woo-Auth~\cite{WooAuth}
         &
         2015 &
         \parbox[][\oneline][c]{2cm}{CAN}
         &
         \parbox{\approachwidth}{HMAC}
         \\ \hline
         \rowcolor{gray!15}
         \ref{subsec:leia} &
         LeiA~\cite{LeiA}
         &
         2016 &
         \parbox[][\twoline][c]{2cm}{CAN}
         &
         \parbox{\approachwidth}{Lightweight cryptography}
         \\ \hline
         \ref{subsec:vatican} &
         vatiCAN~\cite{vatiCAN}
         &
         2016 &
         \parbox[][\oneline][c]{2cm}{CAN}
         &
         \parbox{\approachwidth}{HMAC}
         \\ \hline
         \rowcolor{gray!15}
         \ref{subsec:libra} &
         LiBrA-CAN~\cite{Libracan}
         &
         2017 &
         \parbox[][\threeline][c]{2cm}{CAN\\CAN+\\CAN-FD}
         &
         \parbox{\approachwidth}{M-MAC}
         \\ \hline
         \ref{subsec:vulcan} &
         VulCAN~\cite{VulCAN}
         &
         2017 &
         \parbox[][\oneline][c]{2cm}{CAN}
         &
         \parbox{\approachwidth}{MAC}
         \\ \hline
         \rowcolor{gray!15}
         \ref{subsec:toucan} &
         TOUCAN~\cite{Toucan}
         &
         2019 &
         \parbox[][\oneline][c]{2cm}{CAN}
         &
         \parbox{\approachwidth}{Chaskey algorithm}
         \\ \hline
         \ref{subsec:authentican} &
         AuthentiCAN~\cite{authentiCAN}
         &
         2020 &
         \parbox[][\oneline][c]{2cm}{CAN-FD}
         &
         \parbox{\approachwidth}{Asymmetric}
         \\ \hline
         \rowcolor{gray!15}
         \ref{subsec:s2} &
         S2-CAN~\cite{S2-CAN}
         &
         2021 &
         \parbox[][\twoline][c]{2cm}{CAN}
         &
         \parbox{\approachwidth}{Payload cycling shifting encoding}
         \\ \hline
    \end{tabular}
\end{table}

\subsection{CANAuth}
\label{subsec:canauth}

CANAuth (2011) bases its security upon \acp{hmac} and Session Keys~\cite{CANauth}.
Because of the limited payload size of standard frames and the hard real-time requirement for an in-vehicle communication protocol, the authors decided not to put inside (or attach to) the data frame of the \ac{hmac} nor to send a long data packet over multiple messages.
Therefore, they decided to build the authentication mechanism on top of CAN+.

\subsubsection{\underline{Session Key Generation}}
The key establishment procedure assumes that each node $i$ possesses one or more 128-bit pre-shared master keys $K_i$.
Also, the protocol introduces the concept of \emph{Group messages} $\mathcal{G}_{i}$, where $\mathcal{G}$ is the group and $i$ is the group index.
This allows the authentication of a group of related messages using the same key, called \emph{Group key}.
Keys are stored in a tamper-proof memory.
The session key $K_{S,i}$ for the group $\mathcal{G}_{i}$ is generated by the first node that attempts to send message $M_{i} \in \mathcal{G}_{i}$.
In case multiple nodes transmit $M_{i}$, the key established is the one generated by the node with the lowest ID.
The session key $K_{s,i}$ is derived as:
$$K_{S, i} = HMAC(K_i, ctrA_{i} \parallel r_{i})\;\mod\,2^{128},$$
where $ctrA_{i}$ is a counter for message $M_i$ stored in a non-volatile memory, and $r_{i}$ is a random number.
By applying the $mod\,2^{128}$ operation, only the 128 least significant bits of the \ac{hmac} are considered.

\begin{figure}[!htpb]
    \centering
    
    \begin{subfigure}[b]{\linewidth}
        \centering
        \includegraphics[width=.75\linewidth]{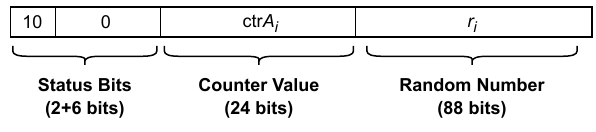}
        \caption{Frame in the first part of session key establishment.}
        \label{fig:CANauth1}
    \end{subfigure}

    \begin{subfigure}[b]{\linewidth}
        \centering
        \includegraphics[width=.75\linewidth]{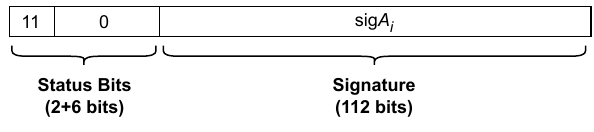}
        \caption{Frame in the second part of session key establishment.}
        \label{fig:CANauth2}
    \end{subfigure}
    
    \caption{CANAuth session key generation exchanged messages.}
    \label{fig:CANauth_key}
\end{figure}

By knowing the counter and the random value, every node with the master key $K_i$ can generate $K_{S,i}$.
Since it is the first sender to define $ctrA_{i}$ and $r_{i}$, it must also deliver them to all the other nodes.
Hence, the session key establishment process develops into two phases, as shown in Figure~{\ref{fig:CANauth_key}}.

\begin{itemize}
    \item \emph{Transmission of parameters.} The transmitter broadcasts a \ac{can} message with attached a CAN+ message structured as in Figure~\ref{fig:CANauth1}. The first bit of the status frame is recessive to signal that the key establishment phase is occurring. The second bit is set dominant to signal that this is the first of two messages. The remaining six bits are dominant and are unused.
    \item \emph{Authentication of the key establishment.} The transmitter broadcasts a second message structured as shown in Figure \ref{fig:CANauth2} to prove the effective knowledge of the session key. In this case, both the first two bits are recessive. The payload of the message consists of a 112-bit signature $$sigA_{i} = HMAC(K_{S, i}, ctrA_{i} \parallel r_{i})\;mod\,2^{112}.$$
\end{itemize}

If any receiving \acp{ecu} raises an error in response to any of the two messages, the transmitting node will restart the transmission from the first message but change the counter (the random number could be the same as before).

\subsubsection{\underline{Message Authentication}}
\begin{figure}[!htpb]
    \centering
    \includegraphics[width=.75\linewidth]{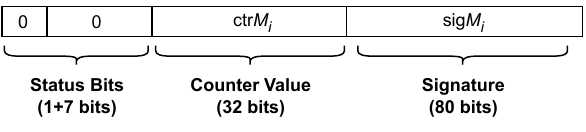}
    \caption{CANAuth data frame during message authentication.}
    \label{fig:CANauth3}
\end{figure}

Once the session key establishment process has been completed, messages can be authenticated.
The authentication of the \ac{can} frame $M_i$ is performed by transmitting a CAN+ message whose payload comprises a counter $ctrM_i$, used for replay attack resistance, and the signature $sigM_i$.
An overview of the CANAuth frame used for authentication is shown in Figure~{\ref{fig:CANauth3}}.

Notice that the counter $ctrM_i$ is independent of $ctrA_i$ (which is used for session key establishment), and it must be increased at least by one for every authenticated message.
Receiving nodes accept the message if the received $ctrM_i$ value exceeds the stored one.
If so, they check for the correctness of the signature as follows:
$$sigM_{i} = HMAC(K_{S, i}, ctrM_{i} \parallel M_{i})\;mod\,2^{80}.$$

\subsection{Car2X}
\label{subsec:car2x}

Car2X (2011) is a cost-effective approach to ensuring trust in Car2X applications using in-vehicle symmetric cryptography.
The protocol uses \ac{mac}-based authentication and allows \ac{mac} truncation.
The security solution establishes trust between \acp{ecu} through an open and symmetric cryptography approach.
It integrates into vehicles' existing infrastructure, requiring only the addition of security hardware modules essential for cryptographic acceleration.

\subsubsection{Architecture}

The security framework employs \acp{hsm} to store symmetric keys, ensuring secure and controlled key access.
Metadata, including expiration dates and use-flags, allows for differentiated functionalities.
The system uses dynamic key exchanges through a Key Master (KM), distributing keys for authentication and transport encryption.
Multiple KMs support domain-specific key sharing.
Integrated into a broader security framework, the system provides API abstractions for entity authentication, secure communication, and access control.
The communication module facilitates flexible and secure communication channels, supporting various attributes and security levels through key exchanges and secure channel establishment managed by the access control framework and key distribution protocol.

\subsubsection{Protocol}

The key distribution and entity authentication process involves secure communication between an \ac{ecu} (e.g., $e_1$) and a group of \acp{ecu} (e.g., $g_x$).
A session key pair ($k_{s,g}$, $k_{s,v}$) is locally generated at $e_1$, with $k_{s,v}$ being exportable.
This key is exported as an encrypted key blob ($b_1$) to the Key Master (KM) using the transport key ($k_{1,t}$) and authenticated with $k_{1,a}$.
These two keys are pre-shared with the KM.
The KM authorizes communication and distributes the session key to group members, ensuring cryptographic operations occur within \acp{hsm}.
The protocol incorporates security features, preventing impersonation and imposing a limited key lifetime of 48 hours.
Indeed, thanks to \ac{hsm}'s use-flags, an attacker cannot impersonate a group's sender despite using symmetric keys.
Due to constraints like limited message payload in CAN buses, a data segmentation approach is employed for transmitting cryptographic keys and signatures.
This segmentation introduces delays and relies on packet sequencing to address transmission challenges.
The protocol is augmented with a mandatory security header indicating cryptographic payload details to enhance security.
Privacy and security requirements vary, prompting the option for MAC truncation for efficiency, considering factors like bus speed and load.
A minimum 32-bit MAC length is deemed reasonable within specified environmental limitations, balancing security and resource constraints.

\subsection{LCAP}
\label{subsec:lcap}

Lightweight \ac{can} Authentication Protocol (LCAP) (2012) provides source authentication by using a hash chain mechanism~\cite{LCAP}. 
For each pair of Sender-Receiver \ac{ecu} $(S,R)$, there exists a pre-shared 128-bit secret key $K_{SR}$ assumed to be stored in a protected memory.
The authors then define \emph{data messages}, the class corresponding to the standard \ac{can} messages, and \emph{handshake messages}, the class of messages that identify a specific action.
Within the handshake messages and for each sender-receiver pair, we distinguish between 5 actions, each with the same 11-bit ID field but different IDE value: \emph{Channel Setup Request}, \emph{First Response Message}, \emph{Consecutive Response Message}, \emph{Soft Synchronization Request}, \emph{Hard Synchronization Request}.
This implies that in addition to the standard \ac{can} identifiers, $5\cdot n\cdot m$ identifiers must be added.
The last two bytes of each message contain the checksum of the first six.
The protocol can work in two possible operating modes: \emph{extended mode}, where the hash value is sent using the extended identifier field, and \emph{standard mode}, where the hash value is sent within the payload.

\subsubsection{\underline{Initialization}}
Each sender creates the following elements.

\begin{itemize}
    \item $H_C$ -- Handshaking hash chain.
    \item $C_C$ -- Channel hash chain.
    \item $K_S$ -- 80-bit session key.
    \item $K_H$ -- 16-bit \ac{hmac} key.
    \item $M_C^i$ -- Hash chain for message type $i$.
    One chain is created for each message type $i$ the \ac{ecu} can send.
\end{itemize}
We use the subscript $C$ to define the whole chain and the subscript $j$ to refer to the $j^{th}$ element of the chain.

\subsubsection{\underline{Channel Setup}}
\begin{figure}[!htpb]
    \centering
    \includegraphics[width=.5\linewidth]{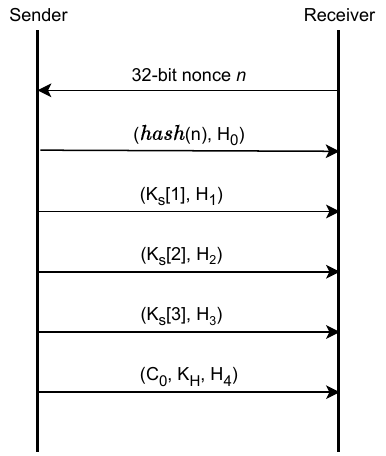}
    \caption{LCAP Channel Setup performed the first time.}
    \label{fig:LCAP_ChannelSetup}
\end{figure}

The sender delivers $K_S$, $C_0$ and $K_H$ to each receiver separately with the following steps:

\begin{itemize}
    \item The receiver generates a \emph{Channel Setup Request} to the sender with a message containing a 32-bit nonce $n$.
    \item The sender replies with a \emph{First Response} message that contains the hash value of the nonce truncated to 16 bits, together with the correct element $H_j \in H_C$.
    \item At this point, the session key needs to be exchanged.
    It is worth noting that since it is 80 bits long, a single \ac{can} frame is not enough.
    Hence, it is split into three parts.
    The sender forwards three \emph{Consecutive Response} messages, each containing a part of $K_S$ and $H_{j+\{1, 2, 3\}}$.
    The receiver can verify the authenticity of each message by checking if $hash(H_i) = H_{i-1}$.
    \item The sender transmits $C_0$ and $K_H$.
\end{itemize}
All the messages are encrypted with $K_{SR}$, except for the last one that uses $K_S$.
An overview of the process is shown in Figure~{\ref{fig:LCAP_ChannelSetup}}.

\subsubsection{\underline{Message Setup}}
For every message type the \ac{ecu} can send, it broadcasts a series of \emph{data messages} containing $M^i_0 \in M_C^i$.
Each message $i$ is authenticated using the corresponding element $C_j$: $C_1$ can be verified since $C_0$ was previously shared.  

\subsubsection{\underline{Data Exchange}}
True data are exchanged and authenticated by appending the element $M_j^i \in M_C^i$ and encrypted with the session key $K_S$.
Each receiver decrypts the message and verifies its authenticity using $K_H$: if $hash(M^i_j) = M^i_{j-1}$, the message is accepted.

\subsubsection{\underline{Chain Refresh}}
Since it is periodically necessary to refresh the hash chain $M^i_C$ before it is completely consumed, the sender must generate a new one in advance.
A possible solution may be to compute a new value each time one chain element is used.
Then, the \ac{ecu} provides the new chain to its receivers and authenticates the message with the last hash value of the current chain.

\subsubsection{\underline{Soft Synchronization}}
In case of message losses, the sender and the receiver will be out of synchronization with respect to the hash values of the chain.
Hence, the receiver asks for re-synchronization, and the sender replies by delivering the current hash values for each message it sends to that receiver.
These are encrypted with $K_S$ and authenticated as in the \emph{Channel Setup} phase.
An overview is shown in Figure~{\ref{fig:LCAP_Synchronization}}.

\subsubsection{\underline{Hard Synchronization}}
\begin{figure}[!htpb]
    \centering
    \includegraphics[width=.75\linewidth]{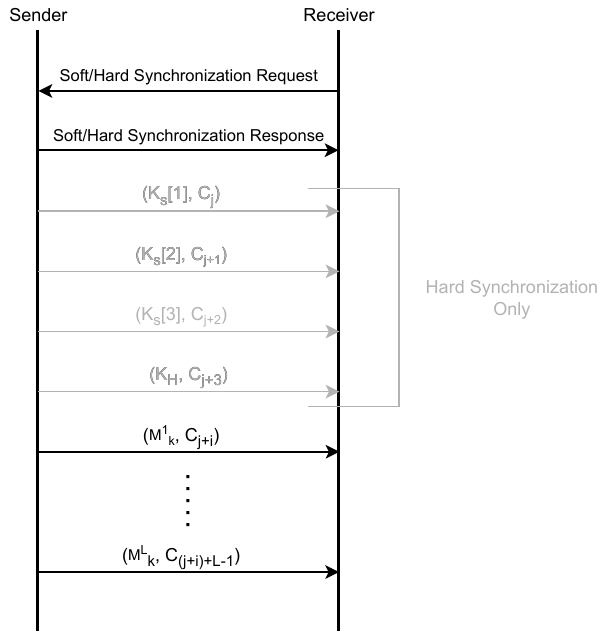}
    \caption{LCAP Soft/Hard Synchronization. Element $k$ refers to the current value of the chain.}
    \label{fig:LCAP_Synchronization}
\end{figure}

This phase allows the receiver to ask for the session and \ac{hmac} keys in case of message loss during the handshake.
This phase develops as in \emph{Soft Synchronization}, with the only difference that before the $M^i_j$ value, the $K_S$ and $K_H$ are exchanged.
Authentication is performed as in \emph{Soft Synchronization} as well.
An overview is shown in Figure~{\ref{fig:LCAP_Synchronization}}.

\subsection{LinAuth}
\label{subsec:linauth}

LinAuth (2012) is a pair-wise \ac{mac}-based authentication protocol~\cite{LinAuth}.
This implies that for each pair of \acp{ecu}, $\left(ECU_i, ECU_j\right)$, there exists a shared secret key $K_{i,j}$.
Besides, given a message $m$ with ID $k$ ($m_k$), a transmitter \ac{ecu} computes the \ac{mac} values for authentication only for those receiver nodes that will be interested in receiving frames with message ID $k$.
It follows that \textit{(i)} to let the receiver understand which is the correct key to use for verification, the authenticated message must also carry the \ac{ecu}-specific identifier $ECU_{ID}$ of the transmitting node; \textit{(ii)} to know which are the nodes for which the transmitter has to compute the \ac{mac} value, there is the need to maintain an ID-table that associates the message IDs to the corresponding interested \acp{ecu}~\cite{LinAuth}.

When an \ac{ecu} has to transmit a message $m_k$, it computes the corresponding \ac{mac} value for each receiver associated with $ID_k$ listed in the ID-table.
Then, it attaches the counter $C_k$ and the collection $\left\{MAC_{k,j}\right\}$ for all the receivers $j$, of the \acp{mac} computed.
It is worth noticing that, to let the receivers understand which \ac{mac} value they need to retrieve for verification, the order in which the \acp{mac} are attached must be pre-defined.

To face possible message loss leading to counter out of synchronization counter and to reduce the bus load, authors decide to split the counter into the most and least significant bits: $C_k = C^M_k || C^L_k$.
Only $C^L_k$ is transmitted, and the following rules apply for message verification (transmitter $ECU_i$, receiver $ECU_j$)~\cite{LinAuth}:
\begin{itemize}
    \item If $C^L_{k,i} > C^L_{k,j}$, the receiver verifies the \ac{mac} value using as counter $C_k = C^M_{k, j} || C^L_{k,i}$.
    If the verification holds, the receiver accepts the message and updates its counter: $C^L_{k,j} = C^L_{k,i}$.
    \item If $C^L_{k,i} \leq C^L_{k,j}$, the receivers verifies the \ac{mac} value using as counter $C_k = C^M_{k, j}+1 || C^L_{k,i}$.
    If the verification holds, the receiver accepts the message and updates its counter: $C_{k,j} = C^M_{k, j}+1 || C^L_{k,i}$.
    If, instead, the verification fails, the message is deemed as replayed and thus discarded.
\end{itemize}

Lastly, to reduce communication overhead, the authors also designed LinAuth with the possibility of using a group keys approach, thus assigning the \acp{ecu} to several~\cite{LinAuth}.

\subsection{MaCAN}
\label{subsec:macan}

MaCAN (2012) relies on \acp{mac} for establishing trust in the origin of signals on the \ac{can} bus~\cite{MaCAN}.
The primary reason for choosing \acp{mac} over asymmetric cryptography is the inherent constraint of the \ac{can} frame's payload size, which makes the latter option impractical.
The core components of the MaCAN security framework encompass three key elements: \textit{(i)} the degree of \ac{mac} truncation, \textit{(ii)} the signing protocol, and \textit{(iii)} the selection of the freshness element.

\subsubsection{\underline{MAC Truncation}}
To ensure the system's real-time capabilities, MaCAN opts for \ac{mac} truncation, limiting the signature length to 32 bits.
This truncated signature is included in a single \ac{can} frame alongside the signal. Several factors influence this design choice.

\begin{itemize}
    \item \textit{\ac{can} Bus Speed} -- The relatively low speed of the \ac{can} bus restricts the number of authentication attempts an attacker can make within a given time frame.
    \item \textit{Session Keys} -- Short-lived session keys are employed for \ac{ecu} communication, limiting the exposure of keys over time.
\end{itemize}
The typical \ac{mac} authentication element length matches the underlying block cipher's block length (e.g., 128 bits with AES).
MaCAN, however, considers using lightweight ciphers like PRESENT in real-world implementations~\cite{bogdanov2007present}.
The number of attacker guesses possible determines the truncation level of 32 bits during a session key's lifespan.
NIST guidelines usually recommend a minimum 64-bit \ac{mac} element, but in MaCAN, due to the slow bus speed and short-lived session keys, 32 bits suffice.
This leads to a 1 in 500 risk of guessing a single signature during the key's lifetime.
To mitigate frequent attempts, MaCAN suggests monitoring actual vs. expected message frequencies and limiting responses to false signatures.

\subsubsection{\underline{Signing Protocol}}
MaCAN offers two message authentication formats, accommodating multiple signals within most Original Equipment Manufacturer (OEM) defined \ac{can} frames.
\begin{itemize}
    \item \textit{Dedicated \ac{can} Frame} -- This format involves a dedicated \ac{can} frame that carries only one signal and its signature as payload.
    While this format increases the number of frames needed to transmit the same number of messages, it can be chosen for a limited number of critical signals.
    However, the limited number of available \ac{can}-IDs and message overhead must be considered.
    \item \textit{Newly Introduced \ac{can}-ID} -- In this format, a new \ac{can}-ID is introduced, and it is included alongside an unauthenticated standard \ac{can} frame.
    This combined frame simplifies the mapping between signals and signatures, making it suitable for on-demand authentication of typically unauthenticated signals.
\end{itemize}

\subsubsection{\underline{Message Freshness}}

To safeguard against replay attacks in a unidirectional environment with event-driven signals and potential \ac{ecu} unavailability, MaCAN employs timestamps for message freshness.
A \ac{ts} is introduced to synchronize time signals across nodes, transmitting timestamps at regular intervals and providing on-demand authenticated timestamps.
MaCAN's protocols allow flexibility in defining which signals are signed and the signature frequency relative to message frequency, adapting the security system to the signal's characteristics and resource availability in various scenarios.

\subsubsection{\underline{Protocols}}

MaCAN defines three crucial protocols to enhance \ac{can} bus security: signal authentication, key distribution, and authenticated time distribution.
\begin{itemize}
    \item \textit{Message Format} -- MaCAN introduces a new 6-bit \ac{ecu} identification token (\ac{ecu}-ID) for secure communication.
    A \virgolette{crypt frame} partitioning scheme is used within the \ac{can} payload to accommodate this.
    This format allows for direct \ac{ecu} addressing, consisting of a destination field and crypt message flags.
    Despite the addition of these security features, the crypt frame format maintains compatibility with standard \ac{can} frames, making it compatible with hardware \ac{can}-ID filters.
    \item \textit{Signal Authentication} -- This protocol governs signal authentication on the \ac{can} bus.
    The process involves defining the desired signal and its signature frequency in relation to the message frequency. The protocol employs two message types: request messages (i.e., initial message specifying the requested signal and signature frequency) and authenticated signal messages (i.e., response to the control message).
    \item \textit{Key Distribution Protocol} -- Manages session keys for pairs or groups of \acp{ecu}.
    It introduces a key server (KS) that shares a symmetric long-term key (LTK) with security-enabled nodes on the bus.
    The protocol's non-time-critical nature allows challenges to ensure message freshness.
    \item \textit{Authenticated Time Distribution Protocol} -- Tackles issues related to non-bidirectional freshness elements like counters or timestamps.
    A Time Server (TS) provides a reliable, monotonically increasing time signal to all nodes on the bus while safeguarding it cryptographically.
    Nodes can request authenticated time signals to synchronize their counters or detect tampering.
\end{itemize}

\subsection{CaCAN}
\label{subsec:cacan}

Centralized authentication in \ac{can} (CaCAN) (2014) follows a centralized authentication approach with the introduction of a central Monitor Node (MN) having the purpose of authenticating all the \acp{ecu} of the network and verifying the authenticity of the exchanged messages~\cite{CaCAN}.
Despite the term centralized may also suggest a topology modification of the network, moving from broadcasting to a star topology, this is not the case: the MN is central just from an operational point of view.
Besides, the MN is equipped with a special \ac{hmac}-\ac{can} controller that can detect and destroy unauthorized frames at runtime by overwriting them with an error frame.
To verify messages' authenticity, the protocol requires the MN and \acp{ecu} to share a 512-bit cryptographic key, thanks to which the \acp{mac} are computed.
This key is computed from a pre-shared secret $S$ assumed to be stored in a Read-Only-Memory (ROM) together with the unique identifier of the \ac{ecu}.

\subsubsection{\underline{Authentication and Key Distribution}}
The protocol expects a two-way authentication in a challenge-response fashion.
Let us consider $ECU_j$ that needs to authenticate to MN.
Assuming that the \acp{ecu} can identify each other:

\begin{itemize}
    \item The MN sends a random nonce $n$ to $ECU_j$.
    \item As the nonce is received, both the MN and the \ac{ecu} start to compute the Authentication Code (AC) as
    $$AC = \text{SHA-256}\,(S\parallel n),$$
    where SHA-256 is the hash function the authors decided to employ.
    It returns a 256-bit value.
    \item After waiting a worst-case computation time for AC, the monitor node sends a data frame to $ECU_j$, whose payload comprises a few bits of the AC.
    \item $ECU_j$ checks for correctness and sends a response with a few continuation bits of the received ones.
    \item MN checks for correctness.
\end{itemize}
The AC computed is also used as the cryptographic key for \ac{mac} computation.
It is worth noting that some key bits are leaked and thus discarded because of the authentication procedure.

\subsubsection{\underline{Monitoring of Data Frames}}
When messages are exchanged, the MN checks if they carry the \ac{mac} by checking the message's ID.
If so, the monitor node immediately starts computing the \ac{hmac} and validating the \ac{mac} of the message.
If an error is detected, it overwrites the message with an error frame by the end of the transmission.
The \ac{mac} attached to the message is 1 byte long and derived from keeping just 8 bits of an \ac{hmac} algorithm.
$$MAC_i = HMAC\,(ECU_{ID}, msg_i, FC_i, K_j),$$
where $ECU_{ID}$ is the unique identifier of the \ac{ecu}, $msg_i$ is the payload of the message (without any counter attached), $FC_i$ is the full counter for message $i$, $K_j$ is the cryptographic key for sender $j$.
The authors decided to attach a counter to the message payload to prevent reply attacks.
However, due to its 32-bit length (full counter $FC = UC \parallel LC$), it cannot be fully transmitted within the payload.
Thus, only the lowest few bits ($LC$) are attached.
The MN stores the full counters for each sender and, when receiving a message, checks for counter freshness according to the following policy:

\begin{itemize}
    \item If $LC_i = OLC_i$, the message is discarded as it is classified as a reply attack (where $OLC_i$ are the holding lowest bits of the monitor node).
    \item If $LC_i > OLC_i$, the message is accepted and $OC_i$ is updated.
    \item If $LC_i < OLC_i$, the message is accepted, $OLC_i$ is updated ($OLC_i = LC_i$), and the holding upper value $OUC_i$ of the counter is increased by 1.
\end{itemize}

\subsection{VeCure}
\label{subsec:vecure}
VeCure (2014) is a \ac{mac}-based authentication protocol in which \acp{ecu} are divided into different \emph{trust groups} according to their trust level~\cite{VeCure}.
The trust level of an \ac{ecu} identifies how easily that \ac{ecu} could be compromised.
There is no specification on the number of groups that one can define.
Trust groups are hierarchically structured: nodes in the lowest-trust group do not have any secret key, while nodes belonging to the group \emph{i} are provided with all the (symmetric) secret keys of groups $j < i$, and the secret key $K_i$ of the group itself.
By doing so, nodes in group $i$ can verify the authenticity of messages transmitted by nodes belonging to groups $j \leq i$.
Without loss of generality, for VeCure design, authors define two trust groups: high-trust and low-trust~\cite{VeCure}.

\subsubsection{\underline{Initialization}}
VeCure provides \ac{ecu}-authentication.
Thus, at the initialization phase, each node in the network receives a unique identifier $ECU_{ID}$ of 1 byte.
Besides, \acp{ecu} belonging to the high-trust group are provided with: (i) a 128-bit secret key $K$, a 2-byte session number $c_s$, and message counter $c_m$.
The session number is a driving session parameter.
Thus, \acp{ecu} will update it each time the vehicle is started, while the message counter is updated at each message authentication.
As $c_m$ only allows authenticating 65536 messages per session before overflow, \acp{ecu} are also required to store an overflow counter $c_o$ that is incremented each time $c_m$ overflows (per session).
The triple $\left(c_s, c_m, c_o\right)$ uniquely identifies any message that was sent by a specific \ac{ecu} since the vehicle was initialized~\cite{VeCure}.

Since each message is bound to the $ECU_{ID}$, every node also needs to keep track of the counters of all the other nodes in the same group to verify the authenticity of the received messages.

\subsubsection{\underline{Authentication Scheme}}
Message authentication is achieved via \ac{mac} computation over the transmitted data.
When a node in the high-trust group has to send some data, it transmits two consecutive messages: the first one is the \ac{can} standard frame and contains the actual data to transmit.
In contrast, the second message is structured as $\left(ECU_{ID}, c_m, MAC, 0xFF\right)$, where $0xFF$ is a marker used to help the receiver verify that is the follow-up message for authentication, and $MAC$ is 4-byte \ac{mac} value.

Due to the strict requirements for induced delay for authentication, authors design the \ac{mac} computation in two phases: a light-weight online calculation and a heavy-weight offline calculation.
The \ac{mac} value is thus computed as follows: $$hash = H\left(ECU_{ID}, c_s, c_o, c_m, K\right),$$ $$MAC = BME\left(hash, m\right),$$
where $H\left(\cdot\right)$ is a one-way cryptographic hash function, $m$ is the data transmitted, and $BME$ (Binding, Mapping, Extraction) is a function performing which ensures that there is a one-to-one mapping between $m$ and the $MAC$ value~\cite{VeCure}.
It is worth noticing that the computation of the $hash$ parameter is independent of the data message $m$; thus, it can be computed offline beforehand for both the sender and receivers. 
This reduces the induced delay in the communication for \ac{mac} computation and verification.

\subsection{Woo-Auth}
\label{subsec:woo}

Woo-Auth (2014) is an \ac{hmac} and session keys-based authentication protocol~\cite{WooAuth}.
The gateway \ac{ecu} (GW) is assumed to have a higher computing power than standard \acp{ecu}.
Each ECU $i$ and the GW are provided with their own unique identifier $ECU_{ID_i}$ and $GW_{ID}$.

\subsubsection{\underline{Loading Long-Term Symmetric Keys}}
Each $ECU_i$ loads in a secure storage and through a secure channel, its long-term symmetric key $K_i$ and $K_{GW}$, while the GW loads all the keys $\{K_i\}$ for each node $i$ and its own key $K_{GW}$.
Key $K_i$ is used between the GW and $ECU_i$, $K_{GW}$ instead is shared among GW and all the \acp{ecu}.

\subsubsection{\underline{Distribution of Initial Session Keys}}
When starting the vehicle, each \ac{ecu} begins a session key derivation process with the GW in a predefined order according to the following steps.

\begin{itemize}
    \item $ECU_i$ selects a random value $R$ and sends it to GW.
    \item GW selects a random number $S$ and generates the \ac{mac}:
    $$MAC_1 = H1_{K_i}(ECU_{ID_i}, GW_{ID}, R, S),$$
    where $H1$ is the keyed hash function using $K_i$ as key and giving a 64-bit output.
    Then, it sends $MAC_1$ to $ECU_i$ together with $S$.
    \item $ECU_i$ verifies $MAC_1$ and computes the initial session keys.
    $$K_E \parallel K_A \parallel K_{EK} \parallel K_G \parallel K_U = KDF(K_{GW}, S),$$
    where $KDF(K_{GW}, S)$ is the keyed one-way hash function using key $K_{GW}$ and $S$ as parameters; $K_E$ is the encryption session key for \ac{can} data frames; $K_A$ is the authentication session key for \ac{can} data frames and session key update; $K_{EK}$ is the encryption key for the session key update phase; $K_G$ is the key used for the following session key derivation process; $K_U$ is a key used for external devices communication.
    \item $ECU_i$ generates the following:
    $$MAC_2 = H2_{K_i}(ECU_{ID_i}, S),$$ $$MAC_3 = H1_{K_A}(ECU_{ID_i} \parallel K_{GW}),$$
    where $H2$ is a keyed hash function using key $K_i$ and returning a 32-bit output.
    Then, it sends $MAC_2, MAC_3$ to GW.
    \item GW verifies $MAC_2$ to check if $S$ was correctly received.
    Afterward, it starts the session keys derivation as done by $ECU_i$.
    \item With the session keys, GW can now verify $MAC_3$ to be sure they share the same keys.
\end{itemize}

\subsubsection{\underline{Authentication of CAN Frames}}
Due to the restricted payload size of \ac{can} frames, authors propose two possible solutions for sending messages: a \emph{basic method} in which the output of the encryption is truncated to fit the payload size, and an \emph{enhanced method} in which the encryption result is split into two parts and sent through two subsequent messages.
For both approaches, the encryption and decryption procedures are the same.
The authors propose to use AES-128 for encryption and Keyed-Hash \ac{mac} for authentication.

Let us assume to be at the $k^{th}$ session.
Sender $ECU_i$ generates the ciphertext $C$ and the authentication tag $MAC_i$ as follows:
$$C=\text{AES-128}_{K_{E,k}}(CTR_{ECU_i}) \oplus M,$$
$$MAC_i = H2_{K_{A,k}}(ECU_{ID_i}\parallel C\parallel CTR_{ECU_i}),$$
where $M$ is the message payload, and $CTR_{ECU_i}$ is the counter of $ECU_i$ used for freshness.
The \ac{mac} is split and inserted into the first 16 bits of the IDE field (the two remaining unused bits are set to zero) and the remaining part in the CRC field.
Notice that, as the CRC field contains part of the \ac{mac}, which aims to provide integrity and authentication, the CRC field still can be considered as proof that the message has not been altered.
The receiving \ac{ecu} first verifies the \ac{mac} received and then decrypts the message as follows.
$$M = \text{AES-128}_{K_{E,k}}(CTR_{ECU_i}) \oplus C.$$
Lastly, it increments the counter associated with the sender \ac{ecu}.

\subsubsection{\underline{Session Key Update}}
Encryption and authentication keys are periodically updated according to the following mechanism.

\begin{enumerate}
    \item The GW selects a new random value $S_{k+1}$ and broadcasts a \emph{Key Request Message} whose payload is given by $(C\parallel MAC)$, where
    $$C = \text{AES-128}_{K_{EK,k}}(CTR_{GW})\oplus S_{k+1},$$
    $$MAC = H_{K_{A,k}}(GW_{ID}\parallel C\parallel CTR_{GW}).$$
    \item Every \ac{ecu}, receiving the \emph{Key Request Message}, verifies the message and derives the session keys to be used in the $(k+1)^{th}$ session $K_{E,k+1}\parallel K_{A,k+1}\parallel K_{EK,k+1}\parallel K_{G,k+1} = KDF_{K_{G, k}}(S_{k+1}).$
    Also, frame counters are initialized to zero.
    It is worth noting that to derive the new session keys for $k \geq 2$, we use the key $K_{G, k-1}$.
    \item Each \ac{ecu} generates a \emph{Key Response Message} with payload
    $$M = H1_{K_{A,k+1}}(ECU_{ID_i}\parallel S_{k+1}),$$
    and transmits it to GW to confirm the reception of the previous request.
    \item After receiving the confirmation response, the gateway sets the counters for the corresponding \acp{ecu} to zero.
    Then, when all response messages are received, it can set its counter to zero.
\end{enumerate}

\subsection{LeiA}
\label{subsec:leia}

Lightweight Authentication Protocol for \ac{can} (LeiA) (2016) makes use of session keys and lightweight cryptographic primitives to assure a good level of security~\cite{LeiA}.
Moreover, LeiA is the first protocol compliant with AUTOSAR specifications (release 4.2)~\cite{autosar}.
Each \ac{ecu} needs to store a tuple $(ID_i, K_i, e_i, K_{S,i}, c_i)$ for each message type $i$, and the security parameter $\eta,$ where:

\begin{itemize}
	\item $ID_i$ is the \ac{can} ID for data type $i$;
	\item $K_i$ is a 128-bit long-term symmetric key used to generate the session key;
	\item $e_i$ is a 56-bit epoch value incremented at each vehicle start-up or when the counter $c_i$ overflows;
	\item $K_{S,i}$ is a 128-bit session key used to generate the \acp{mac};
	\item $c_i$ is a 16-bit counter included in the \ac{mac} computation. 
\end{itemize} 

\subsubsection{\underline{Setup}}
\acp{ecu} are initialized by generating a tuple $(s, n_s)$ from the security parameter $\eta$.
The parameter
$$s = \;<K_0, ..., K_{n-1}>$$
is a collection of master keys, one for each ID, while parameter
$$n_{s} = \;<(c_0,e_0), ..., (c_{n-1},e_{n-1})>$$
is public and is a collection of epoch and counter values couples.
Both counters and epochs are initialized to zero.

\subsubsection{\underline{Session Key Generation}}
For each $ID_i$ the \emph{Key Generation Algorithm} (KGA) is used to derive the corresponding session keys according to the following procedure:

\begin{itemize}
    \item $e_i = e_i + 1$;
    \item $K_{S,i} = KGA(K_i,e_i)$;
    \item $c_i = 0$.
\end{itemize}

\subsubsection{\underline{Sending Authenticated Messages}}
After the generation of the session keys, the \ac{ecu} can authenticate messages.
It must first update counter $c_i$ and generate the following authentication tag.
$$MAC_i = AGA(K_{S,i}, c_i, msg),$$
where $AGA$ is the \emph{Authentication Generation Algorithm} and $msg$ is the payload of the message.
If $c_i$ overflows, $e_i$ is incremented to compute a new session key.
Afterward, the sender will transmit $c_i$, $msg$ $MAC_i$.
On the other hand, after reading and checking for counter freshness, the receiver updates it and verifies the \ac{mac}.
For any message, the counter is placed in the extended identifier field preceded by a 2-bit command code that specifies the content of the payload according to the encoding: 

\begin{itemize}
    \item $00$ for normal data;
    \item $01$ for the \ac{mac} of the data;
    \item $10$ for the epoch value;
    \item $11$ for the \ac{mac} of the epoch.
\end{itemize}
From this encoding, we can understand that the \ac{mac} is sent in a subsequent message with respect to the one containing $msg$.
Moreover, \acp{mac} are transmitted with a different ID with respect to the message they are authenticating.
The authors propose to set $ID_{MAC}=ID_{data}+1$.

\subsubsection{\underline{Resynchronization}}
The resynchronization procedure is used when a \ac{mac} cannot be verified, and the receiver sends an error signal.
In this case, the sender broadcasts a message containing its current $e_i$, $c_i$, and the authentication tag $MAC(e_i)$ for the epoch value.
This allows receivers to resynchronize their epoch and counter.
It is worth noting that receivers will update the two parameters only if the received values are higher than the stored ones.
Indeed, if the counter's value were lower than the receiver's, it would be a sign of a replay attack.
In this case, the \ac{ecu} must not update the stored counter.

\subsection{vatiCAN}
\label{subsec:vatican}

Vetted Authenticated \ac{can} (vatiCAN) (2016) is based on \ac{hmac} authentication, and to reduce bandwidth overhead, only safety critical messages are authenticated~\cite{vatiCAN}.
The choice of which IDs to protect is manually performed during development.
Each \ac{ecu} is provided with a table with IDs expected to be authenticated and 128-bit cryptographic keys for each message type it can transmit among them.
Besides, the protocol also allows grouping related IDs together to use a unique key.
Of course, the key for each \ac{ecu} or group needs to be provisioned to any other \ac{ecu} that will receive the corresponding messages.
These keys are assumed to be stored in \acp{ecu} at manufacturing time.
A 64-bit \ac{mac} can authenticate exchanged messages returned from an \ac{hmac} function (authors suggest using the SHA-3 function).
The \ac{mac} for $ID_i$, or group $\mathcal{G}_j$, is computed as follows.
$$MAC_i = HMAC(ID_i \parallel msg \parallel c_i),$$
where $ID_i$ is the \ac{can} identifier of the message type $i$, $msg$ is the payload and $c_i$ is an ID-specific counter.
The ID of the message is considered for \ac{mac} computation to prevent the same payloads with different identifiers from having the same \ac{mac} if they share the same key.
The \ac{mac} is sent in a subsequent frame, which is done with a different ID concerning one of the authenticated messages.
To maintain the priority relationship between messages, the choice is to pose $ID_{MAC} = ID_{message} + 1$.
Because of possible losses, an \ac{ecu} may remain out of synchronization, i.e., not aligned with the counter $c_i$.
This situation will lead the considered receiver to reject the messages related to $ID_i$ or $\mathcal{G}_j$ since the \acp{mac} it computes will not correspond to those received.
For this reason, authors introduced a \emph{Nonce Generator} (NG) node.
The NG periodically broadcasts a random global nonce $g$ to be used as the new starting value for all counters.
In other words, the NG node resets the counters, giving a new starting point.
The authors suggest a generation frequency of the new counter of $50$ ms.

\subsection{LiBrA-CAN}
\label{subsec:libra}

Lightweight Broadcast Authentication for \ac{can} (LiBrA-CAN) (2017) is designed to protect against internal attackers, and it is based on two paradigms: \emph{key splitting} and \emph{\ac{mac} mixing}~\cite{Libracan}.
The protocol is designed to work upon both CAN-FD and CAN+ and in doing so, it benefits from their features, achieving better performance compared with the standard \ac{can}.
The following assumptions hold.

\begin{itemize}
    \item The adversary can be an external device plugged into the network or a legitimate \ac{ecu} that has been compromised.
    In both cases, the corrupted nodes are assumed to be in the minority.
    \item Cryptographic keys are assumed to be stored in secure memory and distributed to the nodes through a secure key distribution mechanism.
    Keys are renewed at each protocol initialization and refreshed at periodic intervals.
    There is no discussion on the specific key generation and sharing algorithms.
    \item Since the protocol uses counters, a resynchronization mechanism to align nodes to the correct counter value is assumed to be in place.
    \item \acp{ecu} $i$ are provided with unique identifiers $N_i$.
\end{itemize}
The authors also propose two other approaches based on master and distributed-oriented authentication.
However, for our work, we present only the main proposed scheme.

\subsubsection{\underline{Mixed Message Authentication Codes}}
The \acp{mmac} technique is at the basis of the protocol and consists of aggregating multiple \acp{mac} into a single one.
The \ac{mmac} uses an array of keys to build a tag that is verifiable by any of these keys.
The \ac{mmac} mechanism is required to satisfy two fundamental security properties: \emph{unforgeability}, which is standard for MACs and ensures that an adversary is not able to forge an authentication tag correctly, and \emph{strong non-malleability}, which allows a verifier to detect whenever an adversary had tampered with any part of the \ac{mmac}.
A mixed message authentication code is a tuple $(Gen, Tag, Ver)$ of probabilistic polynomial-time algorithms such that:

\begin{itemize}
    \item $\Bar{K} \leftarrow Gen(1^l, s)$ is the key generation algorithm.
    It takes in input a security parameter $l$ and outputs a key set $\Bar{K} = \{K_1, ..., K_s\}$ of $s$ keys.
    
    \item $\tau \leftarrow Tag(\Bar{K}, \Bar{M})$ is the \ac{mac} generation algorithm.
    It takes as input the key set $\Bar{K}$ and the message tuple $\Bar{M}= (m_1, ..., m_s)$, and outputs a tag $\tau$.
    
    \item $v \leftarrow Ver(k_i, m_i, \tau)$ is the verification algorithm.
    It takes in input the key $k_i \in K$, the message $m_i$, the tag $\tau$ and outputs a single bit $v$, equal to 1 if and only if the tag is valid with respect to the used key and input message.  
\end{itemize}
The easiest way to build the \ac{mmac} is by concatenating multiple tags.

\subsubsection{\underline{LiBrA-CAN Main Scheme}}
Given $n$ nodes placed in groups of size $g$, the protocol develops into the following steps.

\begin{itemize}
    \item \textbf{Setup} -- This is the key setup phase that generates $t$ random $l$-bit keys.
    The parameter $t$ is given as $t= \binom{n}{g}$, representing the number of subsets of size $g$ out of $n$ nodes.
    Each node is then provided with the keys for its group.
    Let us define $\Bar{K^i} = \{K_1^i, ..., K_{t'}^i\}$ the set of keys for node $N_i$, where $t'= \binom{n-1}{g-1}$, and $\Bar{K^{i,j}} = \{K_1^i, ..., K_{t''}^i\}$ be the set of share keys of node $N_i$ with $N_j$, where $t''= \binom{n-2}{g-2}$.
    
    \item \textbf{Send authenticated message} -- When node $N_i$ wants to broadcast message $m$, it increments its local counter and computes the corresponding \ac{mmac} using its key set $\Bar{K^i}$.
    Then, the message is sent together with its \ac{mmac}.
    It is worth noting that in this case, the \ac{mmac} construction receives as input $\Bar{M} = (m, m, ..., m)$ composed by $s$ repetition of the same message $m$.
    
    \item \textbf{Verification} -- Node $N_j$ receives an authenticated message sent by node $N_i$.
    Then, it first checks for freshness, and then if $v = 1$ for all the keys belonging to $\Bar{K^{i,j}}$, the message is then assumed to be authentic, and the counter is updated.
\end{itemize}

\subsection{VulCAN}
\label{subsec:vulcan}

VulCAN (2017) employs \acp{mac} to provide strong message authentication for secure communication.
The paper articulates an attacker model where malicious parties have remote access to the car's internal network and can thus perform arbitrary message manipulation and code execution.
Based on the problem statement, authors identify four requirements closely following previous research: \textit{(i)} message authentication, \textit{(ii)} lightweight cryptography, \textit{(iii)} replay attack resistance, and \textit{(iv)} backward compatibility.
On top of these requirements, they provide five guarantees that authors claim were not achieved by state-of-the-art authentication schemes: \textit{(i)} real-time compliance, \textit{(ii)} component isolation, \textit{(iii)} component attestation, \textit{(iv)} dynamic key update, and \textit{(v)} secure legacy \ac{ecu} integration.

\subsubsection{\underline{MAC Generation}}
VulCAN associates a symmetric 128-bit cryptographic key with each authenticated \ac{can} identifier for message authentication.
This key generates a 64-bit \ac{mac} over the message identifier (ID) and payload, including a monotonically increasing counter to protect against replay attacks.
The \ac{mac} generation process is as follows.
\begin{itemize}
    \item For a message with identifier $i$, payload $p$, and counter $c_i$, the \ac{mac} is computed as
    $$m = MAC\left(key_i, \left(i | p | c_i\right)\right).$$
    \item This \ac{mac} is transmitted as the payload of a separate \ac{can} message with the ID $i + 1$.
\end{itemize}
This approach avoids priority inversion issues in the \ac{can} bus, as \ac{can} identifiers also serve as priority indicators during bus arbitration.
In cases where the identifier $i + 1$ is already in use by a legacy application, a different application-specific authentication ID is selected to maintain compatibility with legacy \acp{ecu}.

\subsubsection{\underline{Nonce Initialization and Resynchronization}}
VulCAN includes nonce values in the \ac{mac} computation to address replay attacks and packet loss.
Nonces ensure the same counter values are not reused under the same cryptographic key and associated data.
Nonce initialization is typically handled with short-term session keys generated at system boot time or when the session counter overflows.
Nonce resynchronization mechanisms are implemented to deal with packet loss scenarios, ensuring that sender and receiver nonces remain synchronized.
Depending on the application, unused parts of the message payload may be used to encode nonce values.

\subsection{TOUCAN}
\label{subsec:toucan}

TOUCAN (2019) makes use of the Chaskey \ac{mac} algorithm and AES-128 encryption to authenticate and secure \ac{can} communications, and it is designed to be AUTOSAR compliant (release 4.3.1)~\cite{Toucan}.
Authors assume that all the needed cryptographic material is already available to each \ac{ecu}, which then shares 128-bit cryptographic keys for authentication and encryption.
The authentication tag is computed using the Chaskey \ac{mac} algorithm, an efficient permutation-based \ac{mac} algorithm~\cite{chaskey}.
Chaskey receives in input a 128-bit key and a message split into 128-bit blocks, to which it applies a permutation $\pi$.
If the last block is incomplete, padding values are inserted to reach the desired length.
The permutation $\pi$ is applied eight times for each block $m_i$ performing an Addition-Rotation-XOR (ARX) permutation.
This means that a sequence of three operations is applied for each block: addition $\mod 2^{32}$, bit rotation, and XOR operation.
The output of the Chaskey algorithm is a tag $\tau$ of $t < n$ bits, where $n$ is the length of the key.
The authors propose to dedicate 40 bits out of the 64 available for the payload and the remaining 24 bits for the authentication tag.

\subsection{AuthentiCAN}
\label{subsec:authentican}

AuthentiCAN (2020) is a protocol built on CAN-FD and implements encryption and authentication by using asymmetric cryptographic primitives~\cite{authentiCAN}.
This choice is motivated by the authors wanting to avoid using a single shared key since this could be a serious issue in case of a compromised \ac{ecu}.
\acp{ecu} are assumed to be provided with hardcoded key pairs and their unique identifiers $ECU_{ID}$ in a secure memory at the manufacturing phase.
The protocol develops through four different \emph{Transmission States} discriminated by the first two Most-Significant-Bits (MSB) in the frame's payload.

\subsubsection{\underline{Broadcast Public Key (00)}}
The payload is the transmitter's public key sent in clear so any node can read it.
This phase is performed in a predefined order.
The choice of broadcasting the public key rather than hardcoding the other nodes' public keys on the \ac{ecu} is because faulty nodes can be detected and replaced, or other new nodes can be added without the need to hardcode the new keys.

\subsubsection{\underline{Send Nonce List (01)}}
A node entering this state wants to send a nonce list after exchanging the public keys or because a previous list is exhausted.
In the first case, the \ac{ecu} generates a list of all the other nodes and encrypts it with the respective public key.
In the second case, only the new list for a transmission with a specific node is generated and delivered.
It is worth noting that node $A$, to communicate with a node $B$, uses the nonces in the list sent from $B$, and vice-versa.
Of course, they can send only as many messages as many nonces in their lists, and when the list is exhausted, $B$ needs to request $A$ for a new list.
The authors decided to set the size of the nonces to $8$ bits and to fill the list with as many nonces as possible.
Moreover, they also decided to design the protocol to avoid making the nodes send the nonce list at the startup but rather just when actually needed for communication.

\subsubsection{\underline{Send Message (10)}}
A node enters this state when it wants to communicate with another node by sending actual data.
In this case, the payload contains the encrypted message consisting of the concatenation between the plaintext and the correct nonce.
After message delivery, the sender erases the nonce used.
Thanks to encryption with the receiver's public key, only the legitimate receiver can decrypt it and check if the nonce matches what is expected.

\subsubsection{\underline{Synchronize Nonce List (11)}}
A node enters this state when it has no more nonces available for communication with another node or after a fixed number of rejected messages due to a possible message loss, which makes the two \acp{ecu} misaligned with respect to the counter list.
The payload contains the receiver's $ECU_{ID}$ in plaintext.
The receiver is then triggered to enter in the \emph{Send Nonce List} state.
As it may also happen that a node sends a resynchronization request before the actual nonce list is exhausted, the receiver locally deletes its internal list and generates a new one.

\subsection{S2-CAN}
\label{subsec:s2}

Unlike all the others presented so far, S2-CAN (2021) is an authentication protocol that aims to offer authenticity, confidentiality, and freshness without using cryptography~\cite{S2-CAN}.
Indeed, the security of the protocol depends on \textit{(i)} randomly generated internal IDs and counters to offer authenticity and freshness, \textit{(ii)} frames payload cyclic shifting of a random integer to offer secrecy and confidentiality.
Despite the protocol not using cryptography for message authentication, it is a session-based protocol.
Hence, some cryptographic primitives for securing the session parameters are still needed.
To this purpose, \acp{ecu} are equipped with pre-installed symmetric keys.
Furthermore, the protocol also requires a logical ordering among \acp{ecu}.
The session parameters consist of the following.

\begin{itemize}
    \item Global 3-byte encoding parameter $f$.
    \item \ac{ecu}-specific integrity parameter $int\_ID_j$.
    It is a 1-byte randomly generated internal identifier of the j-th \ac{ecu} $\left(ECU_j\right)$.
    \item \ac{ecu}-specific integrity parameter $pos_{int,j}$ for $ECU_j$.
    It specifies the random position within the \ac{can} payload where the internal identifier will be located.
    This parameter is needed since the \ac{ecu} internal ID is embedded inside the free space of the \ac{can} payload, which can change each time.
    Authors studied in detail how much free space is available on average~\cite{S2-CAN}.
    \item \ac{ecu}-specific 2-byte counter value $cnt_j$ for $ECU_j$.
    It is randomly generated and represents the starting value for the counters used to avoid replay attacks.
\end{itemize}

\subsubsection{\underline{Handshake}}
\begin{figure}[!htpb]
    \centering
    \includegraphics[width=.75\linewidth]{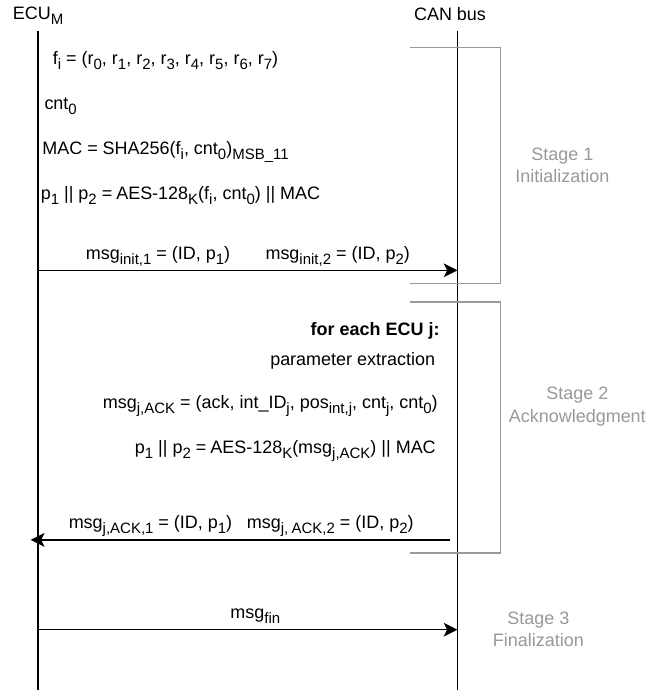}
    \caption{S2-CAN handshake process.}
    \label{fig:S2CAN_Handshake}
\end{figure}

The handshake phase, depicted in Figure~{\ref{fig:S2CAN_Handshake}}, develops through three stages for each new session $S_i$, where $i$ is the number of the session, and repeats with a fixed periodic interval of period $T$.
The gateway \ac{ecu}, also named Master \ac{ecu} ($ECU_M$), handles the handshake phase and establishes new sessions.

\begin{itemize}
    \item \textbf{Stage 1 - Initialization}.
    $ECU_M$ sends an initialization message $msg_{init}$, identified by a specific standard \ac{can} identifier, to indicate the beginning of the handshaking phase for establishing a new session $S_i$.
    The payload of this message includes a randomly generated encoding parameter $f_i = (r_0, r_1, ..., r_8)$, where $r_l \in \left[0,7\right]$ represents the bit rotation number for the $l^{th}$ byte in the $8$-byte \ac{can} payload.
    Each $r_l$ can be represented with $3$ bits.
    Moreover, a $2$-byte counter $cnt_0$ (not to be confused with the \ac{ecu}-specific session parameter $cnt_j$ previously defined) is also attached for replay attack defense.
    The message is encrypted with the pre-shared symmetric key $K$ and authenticated thanks to a 32-byte SHA256-\ac{hmac} of the previous 5 bytes.
    It is worth noting that since only 3 bytes are left, the \ac{mac} should be truncated.
    However, as authors claim that a 3-byte \ac{mac} is too short, they decided to split $msg_{init}$ into two consecutive messages $msg_{init,1}$ and $msg_{init,2}$.
    These messages contain payload $p_1$ and $p_2$, respectively.
    Payload $p_2$ is composed then by the following $8$ bytes of the \ac{mac}.
    This results in a total of $11$ bytes \ac{mac}. Finally, AES-128 is used for encryption.
    
    \item \textbf{Stage 2 - Acknowledgment}.
    \acp{ecu} decrypt $p_1$ and $p_2$ and extract the encoding parameter $f_i$.
    Afterward, following the established ordering, each $ECU_j$ sends back an acknowledgment message $msg_{j, ACK}$ containing a $1$-byte positive acknowledgment code $ack$ and the three \ac{ecu}-specific parameters ($int\_ID_j$, $pos_{int,j}$, $cnt_j$).
    Besides, to provide integrity and freshness protection, the acknowledgment message must include a $2$-byte handshake counter $cnt_i$ and the truncated \ac{hmac} of the message.
    As in \emph{Stage 1}, two messages $msg_{j, ACK, 0}$, $msg_{j, ACK, 1}$ are needed.
    It is worth noting that, to avoid internal identifier collisions, an \ac{ecu} must discard those IDs that were generated by the previous \acp{ecu}.
    
    \item \textbf{Stage 3 - Finalization}.
    $ECU_M$ completes the handshake phase by sending a final message $msg_{\rm fin}$ to signal it has successfully received all the acknowledgment messages from the \acp{ecu}.
    A specific \ac{can} ID identifies this message, and its payload comprises a random non-zero payload (again split into two messages).
\end{itemize}
If any acknowledgment message delay exceeds a certain threshold, the handshake times out, and $ECU_M$ will restart it from \textit{Stage 1}.
If the handshake is still unsuccessful after several attempts, all the \acp{ecu} can revert to the standard \ac{can} communication until the next start of the vehicle.

\subsubsection{\underline{Operation}}
This consists of the regular exchange of data.
To save space in the payload, a 2-byte parameter $q_j$ is computed from the previously stored parameters.
$$q_j = LEFT\_ZERO\_PAD(int\_ID_j, 8) \oplus cnt_j.$$
The message's payload is first logically XORed with $q_j$, and then a Circular Shift (CS) operation is applied according to the previously stored encoding parameter $f_i$.
This CS operation is a byte-wise bit rotation to the $l^{th}$ byte according to the value of the corresponding $l^{th}$ element of $f_i$.
The encoded message is then broadcast in the bus with the \ac{ecu}-specific counter $cnt_j$, which will be incremented for the next message.
The receiving \acp{ecu} will reverse the encoding process and check for the message's authenticity, integrity, and freshness.
Possible losses may lead to an out-of-synchronization scenario.
For this reason, receivers can accept packets with a counter value greater than expected within a certain threshold, depending on the packet loss rate.
\section{Comparison Criteria}
\label{sec:criteria}

This section outlines the comparison criteria we identify to highlight the strengths and weaknesses of the considered protocols.
We distinguish between two classes of criteria: \textit{security criteria} and \textit{operational criteria}.
The security criteria identify those properties necessary for a secure and effective authentication protocol for \ac{can}.
The operational criteria are features, apart from the security capabilities, that one may consider when deciding which authentication protocol to adopt.
These operational features do not change the ability of the system to be resistant or not to some vulnerability.
Rather, they can help enhance the security and reliability of the system.
For example, using session keys does not make a system resistant to replay attacks, but it can harden for the attacker to succeed in the attack~\cite{groza2019tricks}.
This way, the operational criteria integrate into the security requirements, offering a complete view of the protocols.

Section~\ref{subsec:limitations} outlines the identified limitations of the defined requirements in the related works. Section~{\ref{subsec:secreq}} and Section~{\ref{subsec:compfeat}} present the identified security requirements and operational criteria, respectively, upon which we carry out our comparative analysis of the protocols.
A summary of the comparison criteria is presented in Table~\ref{tab:criteria}.

\def\typewidth{2cm}
\def\criteriawidth{3.5cm}
\def\descriptionwidth{11cm}
\def\secheight{80pt}
\def\compheight{75pt}

\begin{table*}[!htpb]
\caption{Security and operational criteria definition for the protocols' comparative analysis.}
\label{tab:criteria}
\centering
\begin{tabular}{p{\typewidth}|p{\criteriawidth}|p{\descriptionwidth}}
\hline

\parbox[][\header][c]{\typewidth}{\textbf{Type}} & \multicolumn{1}{c|}{\textbf{Property}} & \multicolumn{1}{c}{\textbf{Description}} \\ \hline

\multirow{5}{*}
{\parbox[][\secheight][c]{\typewidth}{\textbf{Security\\Criteria}}}
& \cellcolor{gray!15} \parbox[][\oneline][c]{\criteriawidth}{Atomic authentication} & \cellcolor{gray!15} \parbox[][\oneline][c]{\descriptionwidth}{Authentication data is transmitted together with authenticated message} \\
\hhline{~--}
& \parbox[][\twoline][c]{\criteriawidth}{Replay attack resistance} & \parbox[][\twoline][c]{\descriptionwidth}{Nodes can discriminate the freshness of the received messages and discard already transmitted ones} \\
\hhline{~--}
& \cellcolor{gray!15} \parbox[][\oneline][c]{\criteriawidth}{Masquerade attack resistance} & \cellcolor{gray!15} \parbox[][\oneline][c]{\descriptionwidth}{The receiver can discriminate whether a legitimate or rogue ECU sends the message} \\
\hhline{~--}
& \parbox[][\oneline][c]{\criteriawidth}{Hard real-time compliance} & \parbox[][\oneline][c]{\descriptionwidth}{Employed cryptographic algorithms must be considered as fast algorithms} \\ 
\hhline{~--}
& \cellcolor{gray!15} \parbox[][\twoline][c]{\criteriawidth}{Resynchronization capability} & \cellcolor{gray!15} \parbox[][\twoline][c]{\descriptionwidth}{Ability of the system to restore the alignment with respect to the parameters between devices in case of message loss}\\ \hline
\multirow{4}{*}
{\parbox[][\compheight][c]{\typewidth}{\textbf{Operational\\Criteria}}}
&
\parbox[][\oneline][c]{\criteriawidth}{Backward compatibility} & \parbox[][\oneline][c]{\descriptionwidth}{No hardware or topology modifications applied to the already existing one} \\ 
\hhline{~--}
& \cellcolor{gray!15} \parbox[][\threeline][c]{\criteriawidth}{Group keys} & \cellcolor{gray!15} \parbox[][\threeline][c]{\descriptionwidth}{A same key is used to authenticate correlated message IDs. In particular: \\ $\bullet$ Use of one key for each message ID, or \\ $\bullet$ ECUs that can send any $ID \in G_i$ can also send all the other IDs belonging to that group} \\
\hhline{~--}
& \parbox[][\oneline][c]{\criteriawidth}{Session keys} & \parbox[][\oneline][c]{\descriptionwidth}{Periodic refresh of the authentication keys} \\ 
\hhline{~--}
& \cellcolor{gray!15} \parbox[][\twoline][c]{\criteriawidth}{Additional storage memory} & \cellcolor{gray!15} \parbox[][\twoline][c]{\descriptionwidth}{Additional amount of memory needed for the handling of the authentication parameters and keys} \\ \hline
\end{tabular}
\end{table*}

\subsection{Related Works Limitations}
\label{subsec:limitations}

\emph{Nowdehi et~al.}~\cite{survey2} define five industrial requirements that compare the considered protocols: effectiveness, backward compatibility, support for vehicle repair and maintenance, sufficient implementation details, and acceptable overhead.
However, we point out the following limitations regarding the definition of these requirements.
\begin{itemize}
     \item \emph{Cost effectiveness} -- 
     According to \emph{Nowedhi et~al.}~\cite{survey2} definition: \virgolette{[...] requiring all \acp{ecu} to have hardware-supported cryptographic primitives to achieve necessary performances for safety-critical real-time systems is not cost-effective}.
     In the definition, the authors are mostly referring to \acp{hsm}, since from the analysis of the protocols given in Section~\ref{sec:protocols}, these are the hardware modules that are used to store the cryptographic primitives.
     While on the one hand, we acknowledge that their requirements increase the cost of production, on the other, it is important to point out that without those modules, any authentication protocol based on secret keys would be completely insecure.
     Indeed, probing attacks can be carried out in these unprotected scenarios, and the loss of the key confidentiality would make any authentication attempt useless~\cite{wang2017probing, pinto2019demystifying}.
     Although the cost of an \ac{ecu} equipped with an \ac{hsm} might vary depending on its computational capabilities~\cite{AG}, some works in the literature are focused on the minimization of the number of those modules in automotive applications~\cite{xie2017hardware, zou2018hardware, xie2019price, xie2021security}.
     Furthermore, as detailed in Section~\ref{sec:hsm}, different types of \acp{hsm} exist based on their security level, granting additional flexibility during the design process of the vehicle network.

     \item \emph{Backward compatibility} -- According to \emph{Nowedhi et~al.}~\cite{survey2} definition: \virgolette{[...] backward compatibility means that a new solution must be able to co-exist with existing technologies and implementations.
     [...] Consequently, only a message authentication solution that manages the authentication in an extra frame can be considered backward compatible}.
     Limitations of such a definition of backward compatibility are \textit{(i)} the fact that nodes that are not required to provide authentication can be firmware updated in such a way they ignore the authentication part, and \textit{(ii)} providing authentication in a subsequent frame may raise delay issues, as we will point out better later in our analysis.
     Besides, the authors did not consider the trend of shifting to the CAN-FD protocol (which is considered backward compatible) as the new standard, which would allow for a much larger payload transmission~\cite{CAN-FD, NadhirMansour}.
     Hence, the authentication bits can be inserted into the additional space in the payload, and those \acp{ecu} that do not deal with authentication can ignore those bits. 
\end{itemize}

Lastly, we highlight that the requirements defined by \emph{Nowdehi et~al.}~\cite{survey2} are implementation and performance-oriented.
At the same time, they can not provide an understanding of the security level and possible vulnerabilities of the considered protocols.
Therefore, no security analysis is carried out, meaning the security criteria and possible vulnerabilities of the proposed authentication mechanisms are not evaluated.
On the contrary, providing a security analysis of authentication protocols for \ac{can} is the major purpose of this work.
To the best of our knowledge, ours is the first work of this kind.

\subsection{Security Criteria}
\label{subsec:secreq}

In the following, we define the security requirements an authentication scheme for \ac{can} shall comply with for in-vehicle communications.

\begin{itemize}
    \item \textbf{Atomic Authentication} --
    The authentication process must be an atomic operation~\cite{davis2011controller}.
    This means that any extra data needed for the authentication process cannot be separated from the message it belongs to.
    Non-compliance to this requirement would \textit{(i)} increase the bus load and \textit{(ii)} delay the overall process, possibly causing serious issues as the receiving node could not process the received data~\cite{tindell1994guaranteeing, tindell1995calculating, tindell1994analysing}.
    
    \item \textbf{Replay Attack Resistance} --
    This property ensures that the system can discriminate the freshness of a received message, thus correctly deciding whether to accept or discard it.
    The verifying nodes should discard messages intended to replay previous authenticated messages.
    The replay attack is one of the most common, effective, and threatening attacks that can be easily performed on the standard \ac{can} bus~\cite{thirumavalavasethurayar2021implementation}.
    It can be performed even in a trustworthy environment and may affect the vehicle's security in critical scenarios~\cite{jo2021survey}.
    
    \item \textbf{Masquerade Attack Resistance} --
    It is an attack in which a malicious entity pretends to communicate with the victim by impersonating another legitimate entity~\cite{lin2012cyber}.
    Specifically for the \ac{can} framework, a malicious \ac{ecu} impersonates another one by transmitting frames with IDs it is not expected to transmit.
    The rogue \ac{ecu} can either be an external one (external masquerade attack)~\cite{sagong2018cloaking} or a compromised legitimate one (internal masquerade attack)~\cite{choi2018identifying, salmani2005contribution}, as discussed in Section~\ref{sec:adversarial}.
    Since \ac{can} IDs do not identify the transmitter but the message's priority, an authentication protocol shall allow the receiver to discriminate whether it is a legitimate or rogue \ac{ecu} sending the message.
    Within the internal masquerade attacks, we distinguish between two types of attacks:
    \begin{itemize}
        \item \underline{\textit{General-internal}}: this is the general case of an internal masquerade attack in which a rogue \ac{ecu} attempts to transmit and authenticate messages with IDs that the \ac{ecu} was not supposed to transmit.
        Particular attention should be given when using the group keys approach.
        Indeed, it may be the case in which an \ac{ecu} belongs to a certain group $\mathcal{G}$ but for which it can transmit only a subset of the IDs of that group.
        Suppose the \ac{ecu} is compromised.
        In that case, the membership to the group $\mathcal{G}$ allows the attacker to know the cryptographic key also for the IDs the \ac{ecu} was not supposed to use, but which it can correctly authenticate anyway as it has the correct secret key and related parameters.
        \item \underline{\textit{Selective-internal}}: this is a special case in which symmetric keys are employed, and an \ac{ecu} is interested in receiving message(s) with certain ID(s), but that it can not transmit messages with those ID(s).
        Although the \ac{ecu} does not transmit those types of messages, to verify their authenticity at reception, the \ac{ecu} must be provided with the corresponding secret keys anyway, thus opening possible breaches for an internal masquerade injection.
        Indeed, by having the key, the \ac{ecu} may be able to authenticate forged messages correctly.
    \end{itemize}

    During our analysis, we will refer to \emph{Internal Masquerade Attack} when no distinction between \textit{General-internal} and \textit{Selective-internal} is made, thus considering both scenarios.
    
    \item \textbf{Hard Real-Time Compliance} --
    As on-board networks handle safety-critical messages whose arrival time is a critical aspect, the hard real-time constraint is a must~\cite{zhang2018new}.
    To meet this requirement, the cryptographic mechanisms used for the authentication shall be fast and do not add any large overhead~\cite{groza2013efficient}.
    We make the following assumptions.
    \begin{enumerate}
        \item Comparison, counter increment, and message build operations are fast with negligible execution time.
        Therefore, we only focus on cryptography-related operations.
        \item Protocol setup or session parameter exchange phases are amortized in time as they are periodically performed and not at every message authentication.
        Thus, we focus only on the authentication transmission.
        \item \ac{hmac} algorithms are fast.
        This assumption is sustained by the results presented in \emph{Groza et al.}~\cite{survey3}, in which the authors report that encryption algorithms like AES-128 or SHA-256 are suitable for practical applications.
    \end{enumerate}
    
    \item \textbf{Resynchronization Capability} --
    The system can restore the alignment with respect to the parameters between devices in case of message loss~\cite{groza2012broadcast}.
    Indeed, losses may lead \acp{ecu} to be out-of-synchronization, causing the impossibility of successfully authenticating any message, which in turn makes \acp{ecu} continuously discard them.
    Hence, a proper resynchronization mechanism is necessary for system reliability.
\end{itemize}

\subsection{Operational Criteria}
\label{subsec:compfeat}

In the following, we define operational features we consider for a complete evaluation of a protocol for \ac{can} authentication.

\begin{itemize}
    \item \textbf{Backward Compatibility} --
    This is one of the biggest problems encountered while developing an authentication protocol for \ac{can}~\cite{cena2015improving}.
    From our perspective, one of the most important aspects is that to be fully backward compatible with the \ac{can} standard protocol, no hardware or topology modifications should be applied.
    We exclude from our definition of hardware modification the inclusion of possible \acp{hsm}, as we believe these modules are essential to provide authentication in in-vehicle communications, and future \acp{ecu} shall be provided with \acp{hsm} by default.
    While other surveys discussed in Section~\ref{sec:relatedWorks} also consider software updates on \acp{ecu} to invalidate this compatibility, we argue that the ability of a system to introduce authentication just by updating its firmware makes it backward compatible since it does not require physical modification~\cite{backcomp}.
    Furthermore, some surveys also consider the addition of specific frames in the payload or other fields of the \ac{can} message to invalidate the backward compatibility since other \acp{ecu} might not be able to interpret them correctly~\cite{survey2}.
    However, since some fields of the \ac{can} messages are left unused for this exact purpose, \ac{ecu} that are not involved in the authentication process ignore them by default.
    Moreover, given that \acp{ecu} use specific encodings to exchange messages, the payload often has free bits that are not utilized, providing free space for the exchange of the \ac{mac} values~\cite{opendbc}.
    
    \item \textbf{Group Keys} --
    It is reasonable to think that if the same key is used to authenticate correlated message IDs, then the size of key storage will reduce.
    In general, ensuring this requirement is not very demanding~\cite{CANauth}.
    However, this solution must be carefully designed and weighted against the advantages it could bring.
    As anticipated, we stress that using the same key for authenticating more than one ID provides opportunities for masquerade attacks: a compromised \ac{ecu} could send a message with an ID that was not expected to but that belongs to the same group of another ID that is allowed to send~\cite{musuroi2021fast, jain2016physical}.
    Hence, that \ac{ecu} can authenticate the message correctly as it possesses the correct key. This would make the attack impossible to detect unless with any additional sophisticated control mechanism~\cite{wang2018delay}.
    Nonetheless, additional \acp{ids} might also introduce new attack vectors to exploit~\cite{sagong2019mitigating}.
    The worst case occurs when using just one single key.
    Because of these motivations, we decide to redefine the group keys feature as either \textit{(i)} the use of one key for each message ID, \textit{(ii)} or as the case in which the \acp{ecu} that can send any $ID \in G_i$ can also send all the other IDs belonging to that group.
    
    \item \textbf{Session Keys} --
    To defend against replay attacks, the most common solution is introducing a counter value somewhere during the authentication or encryption process to ensure the freshness of the incoming message.
    To make this solution effective, the counter shall not be reused.
    Note that the counter may overflow its maximum value if there are many authenticated messages.
    A common solution is to adopt a session key approach that allows the reuse of the same counter value when the encryption key changes.
    Also, this limits the damages due to key disclosures for the key's lifetime, after which the new key must be learned.
    
    \item \textbf{Storage Overhead} --
    The storage memory available is an important aspect to consider when designing an in-vehicle system~\cite{hoppe2009applying}.
    The available storage for an \ac{ecu} impacts the authentication protocol's security level since it influences several security parameters, such as the length of the nonces and encryption keys.
    Besides, the amount of available storage per \ac{ecu} also influences the overall system's implementation cost, which is another important aspect to consider.
    In our analysis, we consider only the additional storage, with respect to the standard \ac{can}, required by the \acp{ecu} to implement the authentication protocol.
    We do not consider temporary values like intermediate parameters or authentication tags/signatures, as they are of interest and stored just for the time needed, after which the space can be easily freed.    
\end{itemize}
\section{Evaluation of Security Criteria}
\label{sec:comparison1}
In this section, we provide a comparison of the security criteria offered by the presented protocols in light of the security requirements defined in the previous section, which are atomic authentication (Section~\ref{subsec:atomicauth}), replay attack resistance (Section~\ref{subsec:replayattack}), masquerade attack resistance (Section~\ref{subsec:masquerateattack}), hard real-time compliance (Section~\ref{subsec:hard-real}), and resynchronization capability (Section~\ref{subsec:resync}).
Table~\ref{tab:comparisonProtocol1} summarizes results from our analysis.

\def\yearwidth{0.6cm}
\def\canwidth{1.1cm}
\def\approachwidth{2.4cm}
\def\header{15pt}
\def\oneline{20pt}
\def\twoline{27.5pt}
\def\threeline{35pt}
\def\fourline{42.5pt}
\def\fiveline{50pt}

\def\headerheight{17.5pt}

\begin{table*}[!htpb]
    \renewcommand{\arraystretch}{1.25}
    \caption{Security criteria comparison of the different protocols.}
    \label{tab:comparisonProtocol1}
    \centering
    \begin{tabular}{l|c|c|c|c|c|c|c}
        \hline
        \multirow{2}{1cm}{\textbf{Protocol}} & \multicolumn{1}{c|}{\multirow{2}{1.75cm}{\begin{tabular}[c]{@{}c@{}} \textbf{Atomic}\\\textbf{Authentication} \end{tabular}}} & \multirow{2}{1.75cm}{\begin{tabular}[c]{@{}c@{}} \textbf{Replay Attack}\\\textbf{Resistance} \end{tabular}} & \multicolumn{3}{c|}{\textbf{Masquerade Attack Resistance}} & \multirow{2}{2cm}{\begin{tabular}[c]{@{}c@{}} \textbf{Hard Real-Time}\\\textbf{Compliance} \end{tabular}} & \multirow{2}{2.25cm}{\begin{tabular}[c]{@{}c@{}} \textbf{Resynchronization}\\\textbf{Capability} \end{tabular}} \\ \cline{4-6}
        &  &  & \multicolumn{1}{c|}{Gen. Int.} & \multicolumn{1}{c|}{Sel. Int.} & \multicolumn{1}{c|}{Ext.} &  &  \\
        \hline
        \rowcolor{gray!15}
        CANAuth~\cite{CANauth} & \multicolumn{1}{c|}{\cmark} & \multicolumn{1}{c|}{\cmark} & \multicolumn{1}{c|}{\cmark**} & \multicolumn{1}{c|}{\xmark} & \multicolumn{1}{c|}{\cmark} & \multicolumn{1}{c|}{\cmark} & \multicolumn{1}{c}{\cmark}\\
        \hline
        Car2X~\cite{car2x} & \multicolumn{1}{c|}{\xmark} & \multicolumn{1}{c|}{\xmark} & \multicolumn{1}{c|}{\cmark} & \multicolumn{1}{c|}{\cmark} & \multicolumn{1}{c|}{\cmark} & \multicolumn{1}{c|}{\xmark***}& \multicolumn{1}{c}{Not needed****}\\
        \hline
        \rowcolor{gray!15}
        LCAP~\cite{LCAP} & \multicolumn{1}{c|}{\cmark} & \multicolumn{1}{c|}{\cmark} & \multicolumn{2}{c|}{$(2^{16}-n)^{-1}$-secure} & \multicolumn{1}{c|}{\cmark} & \multicolumn{1}{c|}{\cmark}& \multicolumn{1}{c}{\cmark}\\
        \hline
        LinAuth~\cite{LinAuth} & \multicolumn{1}{c|}{\cmark} & \multicolumn{1}{c|}{\cmark} & \multicolumn{1}{c|}{\cmark} & \multicolumn{1}{c|}{\cmark} & \multicolumn{1}{c|}{\cmark} & \begin{tabular}[c]{@{}c@{}} With group keys: \cmark \\ Without group keys: \xmark \end{tabular} & \multicolumn{1}{c}{\cmark}\\
        \hline
        \rowcolor{gray!15}
        MaCAN~\cite{MaCAN} & \multicolumn{1}{c|}{\cmark} & \multicolumn{1}{c|}{\begin{tabular}[c]{@{}c@{}}If not using\\ group keys\end{tabular}} & \multicolumn{1}{c|}{$500^{-1}$-secure} & \multicolumn{1}{c|}{\xmark} & \multicolumn{1}{c|}{\cmark} & \multicolumn{1}{c|}{\cmark}& \multicolumn{1}{c}{\cmark}\\
        \hline
        CaCAN~\cite{CaCAN} & \multicolumn{1}{c|}{\cmark} & \multicolumn{1}{c|}{\cmark*} & \multicolumn{1}{c|}{$2^{-8}$-secure} & \multicolumn{1}{c|}{\cmark} & \multicolumn{1}{c|}{\cmark} & \multicolumn{1}{c|}{\cmark} & \multicolumn{1}{c}{\xmark}\\
        \hline
        \rowcolor{gray!15}
        VeCure~\cite{VeCure} & \multicolumn{1}{c|}{\cmark} & \multicolumn{1}{c|}{\cmark} & \multicolumn{1}{c|}{\xmark} & \multicolumn{1}{c|}{\xmark} & \multicolumn{1}{c|}{\cmark} & \multicolumn{1}{c|}{\cmark} & \multicolumn{1}{c}{\xmark}\\
        \hline
        Woo-Auth~\cite{WooAuth} & \multicolumn{1}{c|}{\begin{tabular}[c]{@{}c@{}}Basic: \cmark\\ Enhanced: \xmark\end{tabular}} & \multicolumn{1}{c|}{\cmark} & \multicolumn{1}{c|}{\xmark} & \multicolumn{1}{c|}{\xmark} & \multicolumn{1}{c|}{\cmark} & \multicolumn{1}{c|}{\cmark} & \multicolumn{1}{c}{\xmark}\\
        \hline
        \rowcolor{gray!15}
        LeiA~\cite{LeiA} & \multicolumn{1}{c|}{\xmark} & \multicolumn{1}{c|}{\cmark} & \multicolumn{1}{c|}{\xmark} & \multicolumn{1}{c|}{\xmark} & \multicolumn{1}{c|}{\cmark} & \multicolumn{1}{c|}{\xmark***} & \multicolumn{1}{c}{\cmark}\\
        \hline
        vatiCAN~\cite{vatiCAN} & \multicolumn{1}{c|}{\xmark} & \multicolumn{1}{c|}{\xmark} & \multicolumn{1}{c|}{\cmark**} & \multicolumn{1}{c|}{\xmark} & \multicolumn{1}{c|}{\cmark} & \multicolumn{1}{c|}{\xmark***} & \multicolumn{1}{c}{\cmark}\\
        \hline
        \rowcolor{gray!15}
        LiBrA-CAN~\cite{Libracan} & \multicolumn{1}{c|}{\cmark} & \multicolumn{1}{c|}{\cmark} & \multicolumn{1}{c|}{\cmark} & \multicolumn{1}{c|}{\cmark} & \multicolumn{1}{c|}{\cmark} & \multicolumn{1}{c|}{\cmark} & \multicolumn{1}{c}{\cmark}\\
        \hline
        VulCAN~\cite{VulCAN} & \multicolumn{1}{c|}{\xmark} & \multicolumn{1}{c|}{\cmark} & \multicolumn{1}{c|}{$2^{-6}$-secure} & \multicolumn{1}{c|}{\xmark} & \multicolumn{1}{c|}{\cmark} & \multicolumn{1}{c|}{\xmark***} & \multicolumn{1}{c}{\cmark}\\
        \hline
        \rowcolor{gray!15}
        TOUCAN~\cite{Toucan} & \multicolumn{1}{c|}{\cmark} & \multicolumn{1}{c|}{\cmark} & \multicolumn{1}{c|}{$2^{-24}$-secure} & \multicolumn{1}{c|}{\xmark} & \multicolumn{1}{c|}{\cmark} & \multicolumn{1}{c|}{\cmark} & \multicolumn{1}{c}{Not needed****}\\
        \hline
        AuthentiCAN~\cite{authentiCAN} & \multicolumn{1}{c|}{\cmark} & \multicolumn{1}{c|}{\cmark} & \multicolumn{1}{c|}{\cmark} & \multicolumn{1}{c|}{\cmark} & \multicolumn{1}{c|}{\cmark} & \multicolumn{1}{c|}{\cmark} & \multicolumn{1}{c}{\cmark}\\
        \hline
        \rowcolor{gray!15}
        S2-CAN~\cite{S2-CAN} & \multicolumn{1}{c|}{\cmark} & \multicolumn{1}{c|}{\cmark} & \multicolumn{1}{c|}{\xmark} & \multicolumn{1}{c|}{\xmark} & \multicolumn{1}{c|}{\cmark} & \multicolumn{1}{c|}{\cmark} & \multicolumn{1}{c}{\cmark}\\
        \hline
    \end{tabular}
    \footnotesize
    
\vspace{.25em}
    \begin{itemize}
        \item[] \quad \textit{Gen. Int. = General Internal; Sel. Int. = Selective Internal; Ext. = External.}
        \item[] \quad * Lower security strength compared to the other protocols.
        \item[] \quad ** If proper group design.
        \item[] \quad *** Possible delays due to non-atomic authentication.
        \item[] \quad **** No use of nonces/counters.
    \end{itemize}
\end{table*}

\subsection{Atomic Authentication}
\label{subsec:atomicauth}
\textbf{CANAuth} does not properly send the authentication tag with the authenticated message.
Still, since it is sent through the overclocked bits of CAN+, they can be considered unique messages.
Indeed, no other \ac{ecu} can interpose between the transmission of the two messages as the CAN+ message is sent during the standard \ac{can} transmission.

\noindent\textbf{Car2X} employs data segmentation to transmit cryptographic keys and data packets.
Thus, for $n$ segments, there is at least $n$ times the packet transmission and handling time with respect to a single packet.
Furthermore, other \acp{ecu} might interpose in between during packet transmission due to arbitration, delaying packet reception.
Although authors state that for some applications requiring a less strict security level, \ac{mac} truncation can be adopted because an even level of security must be guaranteed across the whole \ac{can} bus, atomic authentication is not guaranteed. For these reasons, Car2X does not comply with this requirement.

\noindent\textbf{LinAuth} attaches the employed counter and all the \acp{mac} for authentication to the authenticated frame.
However, since the protocol expects that a transmitting \ac{ecu} computes a \ac{mac} value for each of the interested \acp{ecu}, the number of \acp{mac} computed may exceed the available space in a single \ac{can} frame payload.
Since there is no mention of how to handle such a scenario in the protocol specification, we assume the easiest and best scenario possible.
Hence, we assume that in case the \acp{mac} do not fit into the authenticated message, the \ac{ecu} immediately transmits a second frame in which the payload contains the remaining \acp{mac}.
To prevent other \acp{ecu} accessing the bus in between the transmission of the two messages, one of the free bits of the \ac{can} protocol can be used to signal that a following message will be transmitted carrying the remaining \ac{mac} values.
In doing so, we let the protocol comply with the atomic authentication requirement.

\noindent\textbf{MaCAN} ensures atomic authentication by incorporating several values in the authentication request message, such as the desired message, the prescaler value, and the cryptographic \ac{mac} computed using the sender's key.
The recipient \ac{ecu}, upon receiving the request, processes it as a complete authentication request.
This ensures that the authentication process is atomic, with no intermediate or partial stages.

\noindent\textbf{VeCure} does not attach the \ac{mac} value to the authenticated message but transmits it in a subsequent frame.
However, the protocol defines that the authentication frame is transmitted immediately after the one carrying the data.
Hence, if a receiving \ac{ecu} sees a \ac{can} message with an $ECU_{ID}$ that is in a trusted group (thus will be authenticated), it will not attempt to access the bus due to the immediate transmission of the authentication message.
This guarantees that no \ac{ecu} can transmit in between the transmission of the two messages.
VeCure authentication procedure can thus be considered atomic.

\noindent\textbf{LCAP} uses a hash chain mechanism to authenticate the exchanged frames.
Chain values are attached to the same frame to authenticate, thus making the protocol compliant with the atomic authentication requirement.

\noindent\textbf{Woo-Auth} expects two possible operation modes for authentication.
If the protocol follows the \emph{basic method}, it complies with this requirement as the authentication tag is attached to the frame.
Otherwise, if using the \emph{enhanced method}, the tag is split into two messages, thus leading to possible interruptions and delays between their transmission, making it not compliant.

\noindent\textbf{CaCAN}, \textbf{LiBrA-CAN} and \textbf{TOUCAN} authenticate the exchanged frames by attaching the corresponding \ac{hmac} value to the same message they are authenticating, thus letting the receivers be able to verify the authenticity of the message immediately.
Therefore, the protocols comply with this requirement.

\noindent\textbf{LeiA}, \textbf{vatiCAN} and \textbf{VulCAN} do not comply with the atomic authentication requirement since they expect to transmit the authentication tag in a separate message with respect to the authenticated data.
Moreover, all protocols expect to transmit the authentication message with a different ID with respect to the corresponding message it wants to authenticate. As a drawback, if the number of data messages is greater than $2^{|ID\,bits|-1}$, where $|ID\,bits|$ is the number of bits used for the ID, then not all the messages can be authenticated.

\noindent\textbf{AuthentiCAN} does not require using any specific authentication tag as this is achieved thanks to the uniqueness of the private key used for encryption.
Therefore, if a message is decrypted, the transmitter uses the correct private key, proving its legitimacy.
The protocol complies with this requirement since a correct encryption grants authentication.

\noindent\textbf{S2-CAN} provides authentication through a correct encoding of the exchange payload, thus providing atomic authentication.

\subsection{Replay Attack Resistance}
\label{subsec:replayattack}
\textbf{CANAuth} and \textbf{LeiA} make use of a fresh element for each security object computed (i.e., session key, session key signature, payload signature, \ac{mac}).
Whenever a message with a nonce that has already been used is received, it is discarded a prior.

\noindent\textbf{Car2X} does not mention any usage of counter or nonces in the \ac{mac} computation.
Furthermore, while the protocol uses session keys, they last up to 48 hours, giving plenty of time for an attacker to flood the \ac{can} bus with previously sniffed authenticated messages.
Thus, Car2X is not resistant to replay attacks.

\noindent\textbf{LinAuth} uses an increment counter for \ac{mac} computation, thus protecting against replay attacks.
In the case of a replayed message, since the frame only carries the least significant bits of the counter, the verifier would increase the most significant part. This will result in a different \ac{mac} value than carried by the message, thus leading to message rejection.

\noindent\textbf{MaCAN} utilizes timestamps for message freshness in unidirectional, event-driven scenarios where replay attacks are a concern.
It introduces a Time Server (TS) to ensure synchronized time signals, transmitting timestamps regularly and offering on-demand authenticated timestamps.
However, replay attacks are still a threat when using group keys since all group members cannot verify the challenge sent by the \ac{ecu} to the TS in the authenticated time distribution protocol.

\noindent\textbf{LCAP} is resistant to reply attacks thanks to the hash values attached to each message.
Indeed, replaying any previously authenticated message leads the receiver to reject it, as the authentication check will fail.
The only non-authenticated message happens during the \emph{Channel Setup} phase with the transmission of the 32-bit nonce from the receiver.
It is worth noting that this message cannot be exploited for a reply attack as the sender will perform the hash computation over a non-valid nonce, forcing the legitimate receiver not to accept the transmitted parameters.

\noindent\textbf{VeCure} is resistant to replay attacks since, for a given $ECU_{ID}$, the \ac{mac} computation depends on the triple $\left(c_s, c_o, c_m\right)$, which is unique during the vehicle lifetime.
Thus, the receiver will reject any previously authenticated message.
It is worth mentioning that in our analysis, we only consider the authenticated messages.
Messages from \acp{ecu} in the low-trust group are not authenticated.
Nevertheless, we can provide replay attack resistance in this latter case by simply introducing authentication for the low-trust group.

\noindent\textbf{Woo-Auth} is resistant against replay attacks.
If during the \emph{Initial Session Key Distribution} phase, a malicious \ac{ecu} tries to replay an old $(S, MAC_1)$ message, the receiving \ac{ecu} would not accept it. Indeed, the $MAC_1$ value also depends on the previously exchanged $R$ value, which differs from the one in use.
Therefore, because of the security assumptions for the \ac{hmac} algorithms, the expected \ac{mac} value will differ from the received one, and the message will be discarded.
A similar reasoning applies to the \emph{Session Key Update} phase.
Lastly, messages are authenticated using a counter, which assures freshness and replay attack resistance.

\noindent\textbf{CaCAN} uses counter values for \ac{hmac} computation, avoiding that two equal messages return the same hash value.
However, it must be noticed that only the $8$ least significant bits of the \ac{hmac} are considered for authentication, and only the least significant part of the counter is exchanged.
Therefore, an attacker may have the chance to succeed in replaying a message.
Two equal messages may authenticate with two different counters but with the same least significant part, return different \ac{hmac} values but having the same $8$ least significant bits. In such a scenario, the central node would classify the message as legitimate.
We are aware that this scenario is very unlikely.
However, we believe that $8$ bits are not enough to guarantee a security level and resistance that can be considered absolute, as in previous protocols.
Therefore, we have to highlight such a difference in resistance strength.

\noindent\textbf{LiBrA-CAN} uses counters attached to each message to ensure freshness in messages and resistance against replay attacks.

\noindent\textbf{VulCAN} uses an increasing counter or nonce $c_i$ as a source of freshness in the \ac{mac} computation (where $i$ is the message identifier).
Nonces are always different when using the same key and associated data.
Thus, the authors provide a detailed description of the nonce initialization and resynchronization process in case a packet loss occurs.

\noindent\textbf{vatiCAN} uses a counter value to provide freshness for \ac{mac} computation.
However, the protocol may still be vulnerable to replay attacks.
Indeed, because no session key is used, a payload with the same $ID_i$ and $c_i$ will always return the same $MAC_i$.
Therefore, if for message $ID_i$, the counter used was $c_i = t$, but the Nonce Generator broadcasts a value $g < t$, an attacker will successfully replay all messages $ID_i$ such that $g < c_i < t$.

\noindent\textbf{TOUCAN} makes use of the Chaskey algorithm to compute the \acp{mac}.
Although the Chaskey algorithm does not use any fresh element for the computation, its design proves it is resistant to replay attacks~\cite{chaskey}.

\noindent\textbf{AuthentiCAN} attaches a fresh element to the exchange message's payload, thus avoiding possible replay attacks.

\noindent\textbf{S2-CAN} uses a counter value both in the Handshake and Operation phases, thus defending against possible replay attacks.

\subsection{Masquerade Attack Resistance}
\label{subsec:masquerateattack}
\textbf{CANAuth} may be vulnerable to both internal masquerade attacks (\textit{General} and \textit{Selective}) if the group keys approach is adopted and groups are poorly designed.
Instead, if the approach is adopted consistently to what we defined in Section~\ref{sec:criteria}, the protocol will be resilient to \textit{General-internal} masquerade attacks.
The attacker must first learn either the master key or session key to authenticate correctly a forged message. This is, however, not possible as $K_i$ is assumed to be stored in a secure memory.
At the same time, $K_{s,i}$ is computed using an \ac{hmac} algorithm that assures a very high level of security.
However, the protocol remains vulnerable to \textit{Selective-internal} masquerade attacks.
Lastly, the protocol is resilient against external masquerade attacks but vulnerable to the \textit{Selective-internal} masquerade attack.

\noindent\textbf{Car2X} is resistant against internal and external masquerade attacks thanks to its use of a Key Master (KM) entity, which handles the key distribution.
We assume the KM cannot be compromised since the authors define it as an external entity.
External masquerade attacks can be prevented since external devices do not own a key pair $k_{s,g}$ and $k_{s,v}$ needed to communicate with the KM.
Internal masquerade attacks, regardless of their type (i.e., \textit{General} or \textit{Selective}), are prevented since the KM authorizes communication between an \ac{ecu} $e_1$ and a group $g_x$ based on predefined policies.

\noindent\textbf{LinAuth} is secure against both internal and external masquerade attacks.
Since the protocol employs a pairwise key approach, a rogue \ac{ecu} would need to learn all the necessary keys to successfully compute the \acp{mac} for all the interested \acp{ecu}.
On the other hand, if a group keys approach were employed, groups must be carefully designed; otherwise, this may open to internal masquerade attacks vulnerability. 

\noindent\textbf{MaCAN} authors provide a detailed analysis of the risk of guessing the signature during the lifetime of a key.
Indeed, considering an attempt each 20 ms and a key lifetime of 48 hours, the total number of attempts becomes 8640000.
Therefore, with a 32-bit \ac{mac}, internal masquerade attacks have a 0.002 (1 in 500) chance of occurring with a brute force attack.
However, an intrusion detection system can easily mitigate attempts at a high frequency on the bus.
Instead, using a symmetric key to compute the \ac{mac}, \textit{Selective-internal} masquerade attacks can still occur.

\noindent\textbf{LCAP} may be vulnerable to a masquerade attack carried on by an internal attacker.
Consider transmitter A and receivers B, C, and D.
It should be noted that after the \emph{Channel Setup} phase, all the receivers will have the same session and \ac{hmac} keys and the hash chain exchanged values.
Hence, we can say that the \emph{Channel Setup} phase actually sets up all the needed parameters to receive messages from that transmitter.
As a result, a compromised \ac{ecu}-B could correctly encrypt a forged message and send it to \ac{ecu}-C on behalf of transmitter \ac{ecu}-A.
However, to have the message accepted, the attacker must guess the correct $M_j^i$, which is assumed impossible from the previous values thanks to the security properties granted by the \ac{hmac} algorithm.
In their proposal, authors assume hash values of $16$ bits.
Therefore, $2^{16}$ different values are possible.
The attacker can discard the already used values, making the attack success probability higher if performed close to the end of the hash chain.
We derive that the security of LCAP against a masquerade attack carried on by an internal attacker at step $k$ is: $(2^{16}-n)^{-1}$.
On the other hand, the protocol is secure against external attackers as they first have to learn the session or master key, which is assumed to be computationally difficult or not possible because it is stored in protected memory.

\noindent\textbf{VeCure} is resistant against external masquerade attacks since an external device would need to learn the secret key of the trust group of the target \ac{ecu}.
However, this is not possible thanks to the security assumption for the hash and BME functions.
On the contrary, VeCure is vulnerable to internal masquerade attacks: a compromised \ac{ecu} in trust-group $i$ could successfully authenticate messages on behalf of any \ac{ecu} belonging to group $j \leq i$.
Indeed, the rogue \ac{ecu} would have the proper keys and the session counter (which is the same for all the \acp{ecu}).
Moreover, it could monitor the bus to retrieve the message counter $c_m$ of the target \ac{ecu} and could align with the overflow counter $c_o$ by \textit{(i)} trials and errors from previous authenticated messages or \textit{(ii)} sniff on the bus since the beginning of the session.
Therefore, the rogue \ac{ecu} would have all the elements to authenticate messages on behalf of another node.

\noindent\textbf{Woo-Auth} is completely vulnerable to an internal masquerade attack.
It must be noticed that to verify the received messages, the $K_{EK}$ between sender and receiver must be the same.
The key is derived from the KDF function, which in turn depends on $ID_{GW}$ (that is the same for all \acp{ecu}), and the random seed $S$, provided by GW.
Therefore, seed $S$ must be the same for all \acp{ecu}.
We conclude that all \acp{ecu} in the network will have the same session keys.
As a result, any compromised \ac{ecu} can perform the $AES-128$ encryption correctly.
The compromised ECU can thus make trials to understand, given the legitimate exchanged message $M$, which is the
counter value returning the correct $C$ value, and consequently the \ac{mac}.
Once this operation is performed successfully, the attacker is aligned with the counter value and can correctly authenticate future messages on behalf of the target \ac{ecu}. On the other hand, external attackers must first learn the session keys to perform the operations described above.
Still, since the security property assures the secrecy of the $KDF$ process, we conclude the protocol resists external masquerade attacks.

\noindent\textbf{CaCAN} behaves similarly to WooAuth, and an analogous analysis can be made.
Indeed, even though the counter is not fully transmitted, the attacker must sniff an exchanged packet to be aligned with its least significant part.
Consequently, an internal attacker may forge a frame with the correct $LC_i$ value. At this point, they must guess the 8 bits of the \ac{hmac} value correctly.
It is worth noting that no advantage can be taken from previously exchanged authenticated frames, as the authentication value is part of the full \ac{hmac} value.
Thus, repetitions are not possible since the full hash value is guaranteed to be almost unique.
To conclude, the protocol offers a $2^{-8}$-security level against \textit{General-internal} masquerade attacks.
The protocol is instead secure against \textit{Selective-internal} attacks, as the only entity that verifies the authentication tags is the Monitor Node.
It is also secure against external attacks since the key should be learned first. However, this is assumed to be impossible.

\noindent\textbf{LeiA} is completely vulnerable to an internal masquerade attack.
In fact, it should be noticed that since the \emph{Setup} phase only depends on a common security parameter $\eta$, every \ac{ecu} will possess the long-term keys $K_i \forall ID_i$.
Therefore, an \ac{ecu} can compute $K^s_i$ for any $ID_i$ and authenticate any message successfully.
This clearly leads to a masquerade attack vulnerability.
On the other side, the protocol is secure against external attacks, as the attacker must first learn either the long-term key or session key, which is not possible because of assumptions for the \textit{(i)} secure storage and \textit{(ii)} security provided by the session key derivation function. 

\noindent\textbf{vatiCAN} masquerade resistance analysis is analogous to the one we did for CANAuth.
If the group keys are well-designed, the protocol resists masquerade attacks.
Otherwise, a vulnerability flow is left against the internal masquerade attacks. Resistance against external attacks is guaranteed.

\noindent\textbf{LiBrA-CAN} is resistant against both internal and external masquerade attacks.
Each \ac{ecu} is provided with the keys of the groups it belongs to.
Thus, if $ECU_j$ wants to impersonate $ECU_i$, it will need all the keys in $\Bar{K^i}$, but this is not possible by group construction because this would mean that $ECU_j$ has to belong to the same groups $ECU_i$ is part of.

\noindent\textbf{VulCAN} does not provide specific mechanisms for addressing masquerading attacks. Anyway, by using a symmetric key for the computation of the 64-bit \ac{mac}, it provides $2^{-6}$-security against \textit{General-internal} masquerade attacks.
\textit{Selective-internal} masquerade attacks can still occur since an attacker can send a malicious \ac{can} frame and subsequently compute its \ac{mac}, assuming the attacker has also gained knowledge on the counter.
Since the nonce increases in each sent message, the leakage of one counter, or the knowledge of the starting value, allows the attacker to infer it.
Furthermore, the authors explicitly state that nonce values are not confidential and that receivers can accept \acp{mac} generated with any nonce that is strictly higher than the previously authenticated nonce value.

\noindent\textbf{TOUCAN} protocol specifications do not directly define how the keys are managed and shared, whether all the \acp{ecu} have the same key, or if keys are \ac{ecu}-specific or message ID-specific.
Therefore, an analysis for each of these three scenario must be considered.
If just one single key is used, the protocol would be vulnerable to an internal attacker.
This also holds in the second case, in which we had \ac{ecu}-specific keys: it is reasonable to assume, indeed, that all the \acp{ecu} are equipped with the keys of (almost) all the others since they would need that for authentication verification.
This situation can be reduced to the case in which just one key is used, thus making the protocol vulnerable.
The last possible case is the only one in which the protocol is not vulnerable: to successfully authenticate a message, a malicious \ac{ecu} should first learn the corresponding key or guess the correct authentication tag.
The probability of this happening is $2^{24}$, which is very low.
Thus, we can consider the protocol resistant against \textit{General-internal} masquerade attacks.
For the rest of our analysis, we will assume TOUCAN will adopt this last solution as it is the most secure and reasonable.
This is also reflected by the security level reported in Table~\ref{tab:comparisonProtocol1}.
However, the protocol is instead fully vulnerable against a \textit{Selective-internal} attacker in any of the three presented scenarios.
TOUCAN is resistant to external attacks.

\noindent\textbf{AuthentiCAN} is masquerade attacks resistant since an attacker must learn the target \ac{ecu}'s private key for a correct authentication.
However, this is assumed not possible for asymmetric encryption's security properties.

\noindent\textbf{S2-CAN} is completely vulnerable against an internal masquerade attack for the same reasons presented during the analysis of LeiA protocol.
Indeed, after the handshake phase, all the \acp{ecu} will be provided with \textit{(i)} the same global encoding parameter $f$, and \textit{(ii)} all the parameters of the other \acp{ecu} in the network.
Therefore, a compromised \ac{ecu} can correctly encode (thus authenticate) a message on behalf of another node.
On the other hand, the protocol is resistant to external masquerade attacks.

\subsection{Hard Real-Time Compliance}
\label{subsec:hard-real}

\textbf{CANAuth} is compliant with the \emph{atomic authentication} requirement, making the only overhead introduced given by the generation and verification of the authentication message.
As $sigM_i$ is computed with an \ac{hmac} algorithm that is assumed to be fast, the protocol can be considered hard real-time compliant.

\noindent\textbf{LinAuth} only requires a counter comparison and \ac{mac} computation for authentication, which are considered fast operations.
However, it is worth noticing that an \ac{ecu} has to compute several \acp{mac}, one for each of the interested receiving \acp{ecu}.
This may lead to the case for which the computation of all the \acp{mac} introduces a significant time delay, not ensuring hard real-time properties.
Using a group keys approach (which the protocol allows for) can reduce the number of keys and \acp{mac} to compute, thus making the protocol hard real-time compliant.
On the other hand, as already discussed, when using a group keys solution, particular attention must be paid to group design as this could open to masquerade attacks.

\noindent\textbf{MaCAN} truncates the \ac{mac} signatures down to 32 bits to fit into the frame payload, allowing for \textit{atomic authentication} and protecting real-time capabilities. 
On the other hand, it is worth noting that Nowdehi et~al.~\cite{survey2} states that MaCAN does not have an acceptable overhead and thus would not comply with the hard real-time requirement.
In doing so, Nowdehi et~al.~\cite{survey2} cite a work of Vasile et~al., which carries out a performance analysis of broadcast authentication protocols in for \ac{can}~\cite{vasile2015performance}.
However, this work (Vasile et~al.) focuses on implementing protocols in  FlexRay and CAN-FD, while the authors of MaCAN explicitly state that their focus is on the standard \ac{can} standard.
To conclude, MaCAN satisfies the hard real-time requirements according to our definition.

\noindent\textbf{LCAP} complies with the \textit{atomic authentication} requirement attaching the hash value to the authenticated frame.
As the time needed for \textit{(i)} hash chain computation, \textit{(ii)} setup, and \textit{(iii)} synchronization can be considered amortized, the only significant time overhead is due to the hash verification. This is considered fast because of the \ac{hmac} algorithm.
Therefore, the protocol satisfies the hard real-time requirement.

\noindent\textbf{Woo-Auth} uses $AES-128$ encryption and hashing during the frame authentication process. Both algorithms are considered fast~\cite{survey3}.
Hence, the protocol offers hard real-time capabilities since the session key distribution and update phases can be considered amortized in time.

\noindent\textbf{CaCAN} uses an \ac{hmac} algorithm for authentication, which is considered fast by assumptions.
The protocol is thus hard real-time compliant.

\noindent\textbf{VeCure} satisfies the \textit{atomic authentication} requirement. Besides, thanks to the \ac{mac} design, the computation of the \emph{hash} value can be carried out offline, removing the most demanding operation from the online \ac{mac} computation and verification process. In conclusion, VeCure offers hard real-time capabilities.

\noindent\textbf{LeiA} is not hard real-time compliant as the authentication tag is transmitted in a separate message than the data frame.
This may raise issues for the processing of the messages.
Suppose an \ac{ecu} with a high priority accesses the bus after sending the data message.
In that case, the authentication message would be delayed, making the receiver wait for the \ac{mac}, unable to verify and process the data received timely.

\noindent\textbf{vatiCAN} hard real-time compliance analysis follows the same reasoning of LeiA: due to the non-atomic authentication, there can be possible delays in receiving the authentication tag.
Therefore, we classify the protocol as not hard real-time compliant.

\noindent\textbf{LiBrA-CAN} may raise some doubt about its hard real-time execution because of the computation of multiple \acp{mac} for a single message.
However, \ac{mmac} computation is still fast enough to ensure hard real-time capabilities, as shown by simulation results~\cite{Libracan}. 

\noindent\textbf{VulCAN} considers real-time compliance as a system-level guarantee that is provided by the protocol.
Indeed, while the authors do not consider the issue of denial-of-service attacks, they still acknowledge the safety-critical functionality provided by the data exchanged on the bus.
To provide this requirement, the \ac{mac} payload is truncated to adhere to AUTOSAR specifications.
However, due to the non-atomic authentication, several delays might occur, which can alter the normal functionality of an \ac{ecu}.

\noindent\textbf{TOUCAN} attaches the \ac{mac} to the authenticated message and encrypts it with AES-128.
The only overhead introduced is then due to the two cryptographic mechanisms.
As both AES-128 and Chaskey are considered to be fast algorithms, we conclude that the protocol is hard real-time compliant.

\noindent\textbf{AuthentiCAN} may raise concerns about the time overhead it may introduce due to its employment of asymmetric encryption.
However, simulations have shown that the encryption overhead is minimal~\cite{authentiCAN}.
Hence, because of this, the fact that \acp{ecu} are becoming increasingly powerful (reducing computational times), and that public key broadcasting and nonce list transmission phases can be considered amortized in time, the protocol can be considered hard real-time compliant.

\noindent\textbf{S2-CAN} is hard real-time compliant as the payload encoding is a fast operation (as also shown by experiment results~\cite{S2-CAN}), and the session update phase can be considered amortized in time.

\subsection{Resynchronization Capability}
\label{subsec:resync}
\textbf{CANauth} and \textbf{S2-CAN} provide no explicit mechanism for resynchronization.
However, since nodes will accept the authentication tag if the received nonce is greater than the stored one, even in case of losses, future legitimate messages can still be successfully authenticated.

\noindent\textbf{MaCAN} uses a Time Server (TS) to prevent issues that may occur when \acp{ecu} maintains the time signal using an internal counter.
The message the TS sends exclusively comprises a 32-bit timestamp transmitted regularly.
Each node can receive and compare the time signal to its internal clock.
If the signals differ significantly or resynchronization is needed, such as after an \ac{ecu} reboot, an authenticated time signal can be requested.

\noindent\textbf{LCAP}, \textbf{LeiA} and \textbf{AuthentiCAN} provide explicit procedures for resynchronization, being then robust to losses and letting \acp{ecu} be aligned to the last exchanged parameters and thus guaranteeing the correct functioning of the system.

\noindent\textbf{Woo-Auth} and \textbf{VeCure} do not provide any resynchronization mechanism.
Therefore, if a message is lost, the \ac{ecu} will not be aligned with the counter(s) value(s) and will not accept future messages because of verification failure until the next session update and counter(s) reset.

\noindent\textbf{LinAuth} does not provide any counter resynchronization mechanism in case of message losses. However, we may notice that such a mechanism is not needed, as the condition for message acceptance is not counter matching (i.e., received counter $=$ stored counter +1) but counter increasing (i.e., received counter $>$ stored counter). This allows \acp{ecu} to correctly verify the authentication even in case of message losses.

\noindent\textbf{CaCAN} does not offer any resynchronization mechanism as well.
Thus, if the monitor node misses one frame, all the following messages will be blocked by raising an error due to counter-value misalignment.
This situation will last until the $UC$ value of the counter is increased since the $OLC$ value will be set as the received $LC$.

\noindent\textbf{vatiCAN} provides a means to periodically align all the \acp{ecu} to the same counter values thanks to the Nonce Generator node.
Thus, if a node remains out of synchronization, it cannot accept authenticated messages for a limited time.

\noindent\textbf{LiBrA-CAN} assumes a resynchronization mechanism is in place, but possible implementations are not mentioned.

\noindent\textbf{VulCAN} provides nonce resynchronization by ``vulcanizing'' the process employed by vatiCAN.
In this implementation, 16-bit nonces are encoded in the extended \ac{can} identifier, allowing for the re-establishment of session keys or counters.

\noindent\textbf{TOUCAN} does not need any nonce or counter for authentication, which means that no resynchronization mechanism is needed: if a message loss occurs, it will not affect the possibility of authenticating the following messages.
\section{Evaluation of Operational Criteria}
\label{sec:comparison2}

In this section, we present an analysis based on the operational features previously defined, which are backward compatibility (Section~\ref{subsec:backcomp}), group keys (Section~\ref{subsec:groupkeys}), session keys (Section~\ref{subsec:sessionkeys}), and storage overhead (Section~\ref{subsec:memoryoverhead}).
Again, these comparison criteria do not change the security capabilities of the protocols.
Rather, they may increase their security level and give a more complete overview of their features.
Table~\ref{tab:comparisonProtocol2} summarizes what is discussed in this section.

\def\halfheaderheight{15pt}

\begin{table*}[!htpb]
    \renewcommand{\arraystretch}{1.25}
    \caption{Operational criteria comparison of the different protocols.}
    \label{tab:comparisonProtocol2}
    \centering
    \begin{tabular}{l|c|c|c|l}
    \hline
    \multicolumn{1}{c|}{\textbf{Protocol}} & \textbf{Backward Compatibility} & \textbf{Group Keys} & \textbf{Session Keys} & \multicolumn{1}{c}{\textbf{Storage Overhead}}\\
    \hline
    \rowcolor{gray!15}
    CANAuth~\cite{CANauth} & 
    \multicolumn{1}{c|}{\cmark*} & \multicolumn{1}{c|}{\cmark**} & \multicolumn{1}{c|}{\cmark} & $332 \cdot n\_groups + 29 \cdot x \cdot log_2\left(n\_groups\right)$\\
    \hline
    Car2X~\cite{car2x}& \multicolumn{1}{c|}{\xmark} & \multicolumn{1}{c|}{\cmark} & \multicolumn{1}{c|}{\cmark} & $\left(86 + 256 + 128\right) \cdot 2n$ \\
    \hline
    \rowcolor{gray!15}
    LCAP~\cite{LCAP} & \multicolumn{1}{c|}{\cmark} & \multicolumn{1}{c|}{\cmark} & \multicolumn{1}{c|}{\cmark} & \begin{tabular}[l]{@{}l@{}} $164 \cdot \left(n-1\right) + 96 \cdot n + \lambda \cdot \left(36 + 16 \cdot x\_send\right) + $ \\ $ + 16 \cdot \sum_{i=1}^{n-1} x\_send_i$ \end{tabular}\\
    \hline
    LinAuth~\cite{LinAuth} & \multicolumn{1}{c|}{\cmark} & \multicolumn{1}{c|}{\cmark**} & \multicolumn{1}{c|}{\xmark} & \begin{tabular}[l]{@{}l@{}} $L_K \cdot \left(n-1\right) + L_{ID} \cdot \left(\sum_j^{x_{send}} x_{receive_j} + 1\right) $ \\ $ + L_c \cdot\left(x + x_{send}\right)$ \end{tabular}\\
    \hline
    \rowcolor{gray!15}
    MaCAN~\cite{MaCAN} & \multicolumn{1}{c|}{\xmark} & \multicolumn{1}{c|}{\cmark**} & \multicolumn{1}{c|}{\cmark} & $L_K \cdot n$\\
    \hline
    CaCAN~\cite{CaCAN} & \multicolumn{1}{c|}{\xmark} & \multicolumn{1}{c|}{\xmark} & \multicolumn{1}{c|}{\xmark} & $L_{ss} + 512$\\
    \hline
    \rowcolor{gray!15}
    VeCure~\cite{VeCure} & \multicolumn{1}{c|}{\cmark} & \multicolumn{1}{c|}{\cmark} & \multicolumn{1}{c|}{\xmark} & \begin{tabular}[l]{@{}l@{}} $128 \cdot \left(i-1\right) + \left(16 + L_{oc}\right) \cdot \sum_{j=2}^i g_j + 24$ \\ where $i=2, ..., n\_groups$\end{tabular}\\
    \hline
    Woo-Auth~\cite{WooAuth} & \multicolumn{1}{c|}{\cmark} & \multicolumn{1}{c|}{\xmark} & \multicolumn{1}{c|}{\cmark****} & \begin{tabular}[l]{@{}l@{}} $2 \cdot L_{MK} + L_{SKs} + \left(L_c + L_{ID}\right) \cdot n$ \end{tabular}\\
    \hline
    \rowcolor{gray!15}
    LeiA~\cite{LeiA} & \multicolumn{1}{c|}{\cmark} & \multicolumn{1}{c|}{\cmark} & \multicolumn{1}{c|}{\cmark} & $328 \cdot \left|ID\right| + L_{\eta}$\\
    \hline
    vatiCAN~\cite{vatiCAN} & \multicolumn{1}{c|}{\cmark} & \multicolumn{1}{c|}{\cmark***} & \multicolumn{1}{c|}{\xmark}
    & \begin{tabular}[l]{@{}l@{}} $\left(128 + L_c\right) \cdot n\_groups + 29 \cdot$ \\ $\cdot \left(log_2(n\_groups) \cdot x_{groups} + p\right)$\end{tabular}\\
    \hline
    \rowcolor{gray!15}
    LiBrA-CAN~\cite{Libracan} & \multicolumn{1}{c|}{\cmark} & \multicolumn{1}{c|}{\cmark} & \multicolumn{1}{c|}{\cmark} & $\binom{n-1}{g-1} \cdot L_K + L_c$\\
    \hline
    VulCAN~\cite{VulCAN} & \multicolumn{1}{c|}{\cmark} & \multicolumn{1}{c|}{\xmark} & \multicolumn{1}{c|}{\cmark} & $\left(L_K + L_{EC} + 32\right)x$\\
    \hline
    \rowcolor{gray!15}
    TOUCAN~\cite{Toucan} & \multicolumn{1}{c|}{\cmark} & \multicolumn{1}{c|}{\xmark} & \multicolumn{1}{c|}{\xmark} & $128 \cdot \left(x + x\_send\right)$\\
    \hline
    AuthentiCAN~\cite{authentiCAN} & \multicolumn{1}{c|}{\cmark} & \multicolumn{1}{c|}{Not needed} & \multicolumn{1}{c|}{Not needed} & $\left(512 + L_K\right) \cdot n + L_K$\\
    \hline
    \rowcolor{gray!15}
    S2-CAN~\cite{S2-CAN} & \multicolumn{1}{c|}{\cmark} & \multicolumn{1}{c|}{\xmark} & \multicolumn{1}{c|}{\cmark} & $L_K + Q + 48$\\
    \hline
    \end{tabular}
        \footnotesize
        \vspace{.25em}
        \begin{itemize}
            \item[] \qquad \quad * Only for transceivers supporting up to 60 Mbit/s
            \item[] \qquad \quad ** Adoption may open to replay attacks.
            \item[] \qquad \quad *** Adoption may open to internal masquerade vulnerabilities without proper group design.
            \item[] \qquad \quad **** Effective only against external attackers.
        \end{itemize}
\end{table*}

\subsection{Backward Compatibility}
\label{subsec:backcomp}

\textbf{CANAuth} relies on CAN+, requiring transceivers to support up to 60 Mbit/s data rate. Therefore, the protocol is backward compatible for \acp{ecu} that supports such data rate.
Otherwise, \acp{ecu} not equipped with such transceivers will require a hardware modification or substituting.

\noindent\textbf{Car2X} is not backward compatible since it requires the implementation of the Key Master module to function properly.
The authors define this component as an external device and, as such, cannot be implemented on an existing device in the bus.

\noindent\textbf{LinAuth}, \textbf{LCAP}, \textbf{VeCure}, \textbf{Woo-Auth}, \textbf{LeiA}, \textbf{vatiCAN}, \textbf{LiBrA-CAN}, \textbf{VulCAN}, \textbf{TOUCAN}, and \textbf{S2-CAN} are fully backward compatible as they only require software update to the \acp{ecu} to implement the authentication protocol.

\noindent\textbf{MaCAN} is also backward compatible with the standard \ac{can} frame format but also needs the implementation of the Time Server.
As stated by the authors, this component can be added to an existing \ac{ecu}.
Therefore, since it requires introducing new specific hardware, the protocol is not backward compatible according to our definition.

\noindent\textbf{CaCAN} is not backward compatible as it requires adding a task-specific Monitor Node adopted with specific hardware (special \ac{can} controller)~\cite{CaCAN}.
Conversely, no hardware modification is needed for all the other standard \acp{ecu}.

\noindent\textbf{AuthentiCAN} is backward compatible as it relies on CAN-FD protocol, and no further hardware modification is needed.

\subsection{Group Keys}
\label{subsec:groupkeys}

\textbf{CANAuth} and \textbf{vatiCAN} use a group keys approach.
However, as already discussed, if this solution is not properly designed, the protocols will become fully vulnerable to an internal masquerade attack.

\noindent\textbf{Car2X} uses a group key approach thanks to the predefined policies enforced by the KM.
Indeed, the KM must first authorize \acp{ecu} that request to send messages to a group.

\noindent\textbf{LinAuth} allows a group keys approach in which \acp{ecu} are assigned to groups, and all share the same key.
However, as already discussed, the groups must be carefully designed to avoid open to internal masquerade attacks. 

\noindent\textbf{MaCAN} makes use of group keys.
However, using those keys allows the attacker to perform replay attacks during the authenticated time distribution protocol.

\noindent\textbf{LCAP} does not use any group keys approach.
Rather, it lets every receiver possess all the security parameters for each sender, which may lead to possible masquerade attacks, as previously discussed.

\noindent\textbf{Woo-Auth} makes use of no group key approach and, as well as LCAP protocol, it makes all the \acp{ecu} share the same session keys, opening them to the internal masquerade attack vulnerability previously discussed.

\noindent\textbf{VeCure} protocol is based on trust-level group \acp{ecu} partitioning the \acp{ecu}, and for each group a key is defined.
This represents a group keys approach.
However, as discussed in the \emph{masquerade attack resistance} section, VeCure's group design opens to masquerade attacks.

\noindent\textbf{CaCAN}, \textbf{VulCAN}, \textbf{TOUCAN}, and \textbf{S2-CAN} do not use any group keys approach.

\noindent\textbf{LiBrA-CAN} makes use of group keys securely.

\noindent\textbf{LeiA} uses one key for each message ID, thus being consistent with our definition of group keys.

\noindent\textbf{AuthentiCAN} does not use group keys because of the asymmetric encryption.

\subsection{Session Keys}
\label{subsec:sessionkeys}
\textbf{CANAuth}, \textbf{Car2X}, \textbf{MaCAN}, \textbf{LeiA}, and \textbf{LiBrA-CAN} make use of session keys, thus improving their security in the case in which a key is compromised.
Possible damages are then limited for the key lifetime, after which the new key must be learned.

\noindent\textbf{LCAP} and \textbf{Woo-Auth} make use of session keys as well.
However, we want to specify that because of their specifications, this solution only has beneficial effects against external attackers.
At the same time, no advantage is gained against internal attackers as all the receiver \acp{ecu} possess all the session keys.

\noindent\textbf{LinAuth}\textbf{CaCAN}, \textbf{VeCure} and \textbf{vatiCAN} do not make use of any session key approach.

\noindent\textbf{VulCAN} uses short-term session keys, which are refreshed at system boot time or whenever the session counter $c_i$ overflows.
These session keys are based on the long-term pre-shared secret and a larger epoch counter, safely allowing nonces to start from zero at each session.

\noindent\textbf{TOUCAN} does not make use of session keys.
Moreover, since Chaskey does not make use of nonces for the \ac{mac} computation, authors state that to avoid attacks with a practical complexity of off-line permutation evaluations, the total number of blocks to be authenticated with the same key shall be at most $2^{64}$~\cite{chaskey}.
Even though this is a very large number, we must consider that as the number of electronic components continuously increases, this number may not be sufficient anymore to cover a vehicle's lifetime.

\noindent\textbf{AuthentiCAN} does not make use of session keys.
However, considering the security behind asymmetric key generation, we recognize that using session keys in an end-to-end communication protocol does not offer any significant improvement in terms of security.
Rather, this could increase some non-necessary overhead and power consumption.

\noindent\textbf{S2-CAN} does not properly use session keys in the strict meaning of the term.
However, payload encoding is performed according to the global encoding parameter $f$, which can be considered a key.
As a result, since this is updated at each new session, we can consider that the protocol uses session keys.

\subsection{Storage Overhead}
\label{subsec:memoryoverhead}
Since protocols do not specify the number of bits reserved for all the parameters used for authentication, and given that part of the additional storage required to \acp{ecu} depends on the specific implementation scenario, our analysis will provide the more general analysis as possible.
Table~\ref{tab:ValuesMeaning} provides the meaning of the common symbols we use in our analysis between the protocols.
Values that are specific to a single protocol will be introduced in the specific analysis.

\def\aline{12.5pt}
\def\aaline{20pt}

\begin{table}[!htpb]
    \caption{Notation for the storage overhead analysis.}
    \centering
    \label{tab:ValuesMeaning}
    \begin{tabular}{l|p{6.75cm}}
    \hline
    \parbox[][\halfheaderheight][c]{1.25cm}{\textbf{Value}} & \multicolumn{1}{c}{\textbf{Meaning}} \\
    \hline
    \rowcolor{gray!15}
    \parbox[][\aline][c]{1.25cm}{\(n\)} & Number of ECUs in the network   \\ \hline
    \parbox[][\aline][c]{1.25cm}{\(n\_groups\)} & Number of groups when using the group keys approach \\\hline
    \rowcolor{gray!15}
    \parbox[][\aline][c]{1.25cm}{\(g_i\)}  & Size of the group $i$ when using the group keys approach  \\\hline
    \parbox[][\aaline][c]{1.25cm}{\(x\)} & \parbox{6.75cm}{Number of different messages (IDs) the ECU is interested in receiving} \\\hline
    \rowcolor{gray!15}
    \parbox[][\aline][c]{1.25cm}{\(x_{send_i}\)} & Number of IDs that \(ECU_i\) can transmit \\\hline
    \parbox[][\aline][c]{1.25cm}{\(x_{receive}\)} & Number of different receivers that the ECU considers \\\hline
    \rowcolor{gray!15}
    \parbox[][\aline][c]{1.25cm}{\(L_{MK}\)} & Number of bits for the master key \\\hline
    \parbox[][\aline][c]{1.25cm}{\(L_{SK_s}\)} & Sum of the number of bits of the session keys used \\\hline
    \rowcolor{gray!15}
    \parbox[][\aaline][c]{1.25cm}{\(L_K\)} & \parbox{6.75cm}{Number of bits used for the secret keys (no distinction between master or session keys)} \\\hline
    \parbox[][\aline][c]{1.25cm}{\(L_c\)} & Number of bits used for the counter \\\hline
    \rowcolor{gray!15}
    \parbox[][\aline][c]{1.25cm}{\(L_{ID}\)} & Number of bits used for the ECU-specific identifier \\\hline
    \parbox[][\aline][c]{1.25cm}{\(L_{ss}\)} & Number of bits used for the stored shared secret \\\hline
    \rowcolor{gray!15}
    \parbox[][\aline][c]{1.25cm}{\(L_{\eta}\)} & Number of bits used for the secret parameter \(\eta\) \\\hline
    \parbox[][\aline][c]{1.25cm}{\(\lambda\)} & Number of elements of the hash chain \\\hline
    \end{tabular}
\end{table}


\noindent\textbf{CANauth} requires an additional storage of 128 bits for $K_i$ and $K_{S,i}$ each, 24 bits for $ctrA_i$ and 32 bits for $ctrM_i$.
This holds for each defined group.
Moreover, \acp{ecu} must maintain a table that relates the IDs to the corresponding group.
The minimum size of the table depends on how many IDs the specific \ac{ecu} is interested in.
Therefore, we have that $29 \cdot \log_{2}\left(n\_groups\right) \cdot x$ additional bits are required for storing the table (analysis is done in the worst case where the extended identifier is used).
To conclude, CANAuth requires a total of $\left(128 + 128 + 24 + 32\right) \cdot n\_groups + 29 \cdot x \cdot \log_{2}\left(n\_groups\right) = 332 \cdot n\_groups + 29 \cdot x \cdot log_{2}\left(n\_groups\right)$ additional storage bits per \ac{ecu}.

\noindent\textbf{{Car2X}} stores in the \ac{hsm} of each \ac{ecu} two keys that carry a header field of 86 bits, key data of 256 bits, and an authentication code of 128 bits, as defined also in~\cite{weyl2010secure}.
These two keys are $K_{i,a}$ and $K_{i,t}$.
Each KM then stores the same two keys for each \ac{ecu} $i$ in the group the KM manages.
However, since the KM is an additional node in the bus, we do not count its storage in the overhead since we assume that external hardware is equipped with the necessary capabilities.
Thus, the additional storage is $\left(86 + 256 + 128\right) \cdot 2n$.

\noindent\textbf{LinAuth} requires that each \ac{ecu} stores a symmetric key with each of the other \acp{ecu} in the network, requiring $L_K \cdot \left(n-1\right)$ bits.
Besides, \acp{ecu} are required to store their own specific $ECU_{ID}$ and an ID-table listing for each of the possible message ID they can transmit, the $ECU_{ID}$ of those nodes interested in receiving the corresponding message.
For each ID $j$ that an \ac{ecu} can transmit, storing the table requires $L_{ID} \cdot \sum_j^{x_{send}} x_{receive_j}$ bits.
Furthermore, for every transmitted and received message ID, the \ac{ecu} must also store a counter for freshness in \ac{mac} computation/verification.
This requires $L_c \cdot\left(x + x_{send}\right)$ bits.
In conclusion, the protocol requires a total of $L_K \cdot \left(n-1\right) + L_{ID} \cdot \left(\sum_j^{x_{send}} x_{receive_j} + 1\right) + L_c \cdot\left(x + x_{send}\right)$ additional storage bits for each \ac{ecu}.

\noindent\textbf{MaCAN} requires all participating nodes to have secure memory to store the key material.
However, since session keys can be requested at any time to the key server, there is no need for the nodes to permanently store any information other than their long-term key (LTK).
However, the paper does not specify the length of the long-term key.
Thus, the storage overhead becomes $L_K \cdot n$
This is valid whether the \acp{ecu} that need to communicate belong to the same group, since in the former case, a common group key can be requested and stored by the group members.

\noindent\textbf{LCAP} protocol expects that for each couple of \acp{ecu}, a 128-bit master key is stored.
This will then require $128 \cdot \left(n-1\right)$ bits.
Further, \acp{ecu} will have to store their own and one for any other node 80-bit session key, requiring $80 \cdot n$ bits.
The same reasoning is for the \ac{hmac} key: $16 \cdot n$ bits.
Moreover, $2 \cdot 16 \cdot \lambda$ bits are required to store the handshake and channel hash chains, while just a single hash value is required for received messages verification, which leads to $2 \cdot 16 \cdot \left(n-1\right)$ bits.
As last, $16 \cdot \lambda \cdot x\_send$ bits are needed for the transmitted messages hash chain, and $16 \cdot \sum^{n-1}_{i=1} x\_send_i$ are needed for the received messages.
Therefore, the protocol requires a total of $128 \cdot \left(n-1\right) + 80 \cdot n + 16 \cdot n + 2 \cdot 16 \cdot \lambda + 2 \cdot 16 \cdot \left(n-1\right) + 16 \cdot \lambda \cdot x\_send + 16 \cdot \sum_{i=1}^{n-1} x\_send_i = 164 \cdot \left(n-1\right) + 96 \cdot n + \lambda \cdot \left(36 + 16 \cdot x\_send\right) + 16 \cdot \sum_{i=1}^{n-1} x\_send_i$ bits storage overhead.

\noindent\textbf{VeCure} protocol introduces the \ac{ecu} identifier of 8-bits length and a session counter $c_s$ of 2 bytes. Then, assume having \emph{n\_groups} trust level groups that are enumerated starting from 1, corresponding to the lowest trust level group.
An \ac{ecu} in group $i$ has to store a 128-bit cryptographic key for each group $j=2, ..., i$ (remember the lowest group does not provide authentication; hence it has no secret key), thus $128 \cdot \left(i-1\right)$ are needed.
Besides, each \ac{ecu} has to store a 2-byte message counter $c_m$ and an overflow counter $c_o$ for itself and each of the \ac{ecu} belonging to security groups $j=2, ..., i$.
This will then require $16 \cdot \sum_{j=2}^i g_j + L_{oc} \cdot \sum_{j=2}^i g_j$ bits, where $L_{oc}$ is the length of the overflow counter since it is not specified in VeCure protocol.
To conclude, VeCure requires each \ac{ecu} belonging to group $i=2, ..., n\_groups$ to handle $128 \cdot \left(i-1\right) + 16 \cdot \sum_{j=2}^i g_j + L_{oc} \cdot \sum_{j=2}^i g_j + 16 + 8 = 128 \cdot \left(i-1\right) + \left(16 + L_{oc}\right) \cdot \sum_{j=2}^i g_j + 24$ additional bits for authentication. 

\noindent\textbf{WooAuth} assumes that the gateway \ac{ecu} has higher computing power than standard \acp{ecu}, so our analysis will focus on the standard \acp{ecu}.
Moreover, protocol specifications do not give any information about the length of the master and session keys or the counter values.
Therefore, we will perform a very general analysis.
Each $ECU_i$ has to store its own $ECU_{ID}$, its long-term key $K_i$, the gateway key $K_{GW}$, and all the other session keys $\left(K_E, K_A, K_{EK}, K_G, K_U\right)$.
Besides, \acp{ecu} also has to store its counter and one for each other \ac{ecu}.
Since the counter is \ac{ecu} specific, there must be maintained a table associating 
it
to the corresponding \ac{ecu}: $\left(L_c + L_{ID}\right) \cdot n$.
This results in a total of $2 \cdot L_{MK} + L_{SKs} + \left(L_c + L_{ID}\right) \cdot n$ additional storage bits.

\noindent\textbf{CaCAN} assumes the participation of a special Control Node, for which it is reasonable to assume it is provided with all the needed memory storage and computing power.
Hence, our analysis will be based solely on the standard \acp{ecu}.
A standard \ac{ecu} must store the shared secret of length $L_{ss}$, the 512-bit encryption key, and a 32-bit counter value.
This requires a $L_{ss} + 512 + 32 $ additional storage bits.

\noindent\textbf{LeiA} protocol expects that each \ac{ecu} stores two 128-bit keys (master and session), one 56-bit epoch value, and one 16-bit counter for each message ID.
Besides, each \ac{ecu} has to store its security parameter $\eta$.
Assuming that $\left|ID\right|$ different message IDs are exchanged in the network, each \ac{ecu} will have to store a total of $2 \cdot 128 \cdot \left|ID\right| + 56 \cdot \left|ID\right| + 16 \cdot \left|ID\right| + L_{\eta} = 328 \cdot \left|ID\right| + L_{\eta}$ additional bits.

\noindent\textbf{VatiCAN} expects that each \ac{ecu} stores a 128-bit key for each message it is interested in or for each group it is part of, and a $L_c$ bits for the counter.
Moreover, since the possibility of the group key approach, a table that relates the IDs with the corresponding group must be maintained within each network node.
Hence, following the same analysis we did for CANAuth, $29 \times log_2\left(n\_groups\right)$ additional bits are required for this purpose.
Lastly, since not every ID is authenticated, a table of critical messages must be maintained.
To conclude, given $x_{groups}$ the number of groups the \ac{ecu} is interested in, and $p$ the number of protected IDs, the protocol requires an additional amount of $128 \cdot n\_groups + L_c \cdot n\_groups + 29 \times log_2\left(n\_groups\right) \cdot x_{groups} + 29 \cdot p = \left(128 + L_c\right) \cdot n\_groups + 29 \cdot \left(log_2\left(n\_groups\right) \cdot x_{groups} + p\right)$ storage bits.

\noindent\textbf{LiBrA-CAN} expects that each \ac{ecu} is provided with an encryption key for each group it belongs to and a personal counter value.
Hence, knowing that each \ac{ecu} will belong to $\binom{n-1}{g-1}$ groups, each \ac{ecu} is required of $\binom{n-1}{g-1} \cdot L_K + L_c$ additional bits storage capacity.

\noindent\textbf{VulCAN} uses 128-bit session keys for each connection.
The session keys are generated by both parties based on the long-term pre-shared secret and a larger epoch counter.
The session keys are not stored on every \ac{ecu} but rather generated by the Attestation Server (AS) and distributed to all Protected Modules (PMs) during load-time attestation.
However, PMs are hosted in each \ac{ecu} as a protected software module.
It also uses a 32-bit nonce for each authenticated connection.
By defining $L_K$ as the length of the long-term pre-shared secret and $L_{EC}$ as the length of the epoch counter, the additional storage required is $\left(L_K + L_{EC} + 32\right)x$.

\noindent\textbf{TOUCAN} storage overhead analysis will consider the message ID-based keys scenario since it is the only situation for which the protocol offers \textit{General-internal} masquerade attack resistance properties.
Hence, given $x$ as the number of IDs an \ac{ecu} is interested in, and $x\_send$ the number of IDs the \ac{ecu} can transmit, recalling that the protocol uses 128-bit keys, the total amount of additional storage an \ac{ecu} must support is $128 \cdot \left(x + x\_send\right)$.

\noindent\textbf{AuthentiCAN} protocol specifications suggest defining a nonce length of 8 bits with a total list length such that it fulfills the payload size.
CAN-FD protocol offers up to 64 bytes of payload size.
Thus, we derive that a nonce list can contain up to 64 nonces.
A list is kept for communication with each node in the network.
Additionally, an \ac{ecu} must store its private and public keys and all the other nodes' public keys.
Hence, an \ac{ecu} is required to maintain a total additional memory storage of $8 \cdot 64 \cdot n + L_K + L_K \cdot n = \left(512 + L_K\right) \cdot n + L_K$ bits.

\noindent\textbf{S2-CAN} protocol specifications require each \ac{ecu} to store a shared symmetric key (whose length is not given), the 3-byte global encoding parameter $f$, the 1-byte $int\_ID_j$ parameter, the $Q$-byte $pos_{int, j}$ parameter, and a 2-byte counter $cnt_j$.
This results in a total amount of $L_K + 24 + 8 + Q + 16 = L_K + Q + 48$ additional storage bit requirement.
\section{Security Classification}
\label{sec:result}

From the analysis above and relative summary tables, we notice that among the security requirements, the most discriminating one is the masquerade attack resistance.
Table~\ref{tab:ranking} presents our ranking of the protocols resulting from our security analysis.
The ranking divides protocols into four categories, which are \textit{secure protocols} (Section~\ref{subsec:secure}), \textit{partially secure protocols} (Section~\ref{subsec:almost}), \textit{non-secure protocols} (Section~\ref{subsec:nonsecure}), and \textit{non-suitable protocols} (Section~\ref{subsec:nonsuitable}).

\begin{table}[!htpb]
    \renewcommand{\arraystretch}{1.25}
    \caption{Protocols' ranking.}
    \label{tab:ranking}
    \centering
    \begin{tabular}{p{3cm}|p{3cm}}
        \hline
        \parbox[][\halfheaderheight][c]{3cm}{\textbf{Security Level}} & \multicolumn{1}{c}{\textbf{Protocols}}\\
        \hline
        \rowcolor[HTML]{D9EAD3} 
        \parbox[][\fiveline][c]{3cm}{Secure Protocols} & 
            \parbox{3cm}{\begin{enumerate}
                \item LiBrA-CAN~\cite{Libracan}
                \item AuthentiCAN~\cite{authentiCAN}
                \item LCAP~\cite{LCAP}
                \item CaCAN~\cite{CaCAN}
                \item LinAuth~\cite{LinAuth}
            \end{enumerate}}\\
        \hline
        \rowcolor[HTML]{FFF2CC} 
        \parbox[][\threeline][c]{3cm}{Partially Secure Protocols} &
            \parbox{3cm}{\begin{enumerate}
                \setcounter{enumi}{5}
                \item TOUCAN~\cite{Toucan}
                \item CANAuth~\cite{CANauth}
                \item MaCAN~\cite{MaCAN}
            \end{enumerate}}\\
        \hline
        \rowcolor[HTML]{F4CCCC} 
        \parbox[][\threeline][c]{3cm}{Non-Secure Protocols} &
            \parbox{3cm}{\begin{enumerate}
                \setcounter{enumi}{8}
                \item S2-CAN~\cite{S2-CAN}
                \item Woo-Auth~\cite{WooAuth}
                \item VeCure~\cite{VeCure}
            \end{enumerate}}\\
        \hline
        \rowcolor{gray!15}
        \parbox[][\fourline][c]{3cm}{Non-Suitable Protocols} &
            \parbox{3cm}{\begin{enumerate}
                \setcounter{enumi}{11}
                \item VulCAN~\cite{VulCAN}     
                \item LeiA~\cite{LeiA}
                \item Car2X~\cite{car2x}
                \item vatiCAN~\cite{vatiCAN}
            \end{enumerate}}\\
        \hline
    \end{tabular}
\end{table}

\subsection{Secure Protocols}
\label{subsec:secure}
In this class belong protocols that satisfy all the security requirements.
However, one may notice that CaCAN does not satisfy the \emph{resynchronization capability} requirement.
Since we believe designing a proper one is not very demanding, we decide to include CaCAN in this class anyway.
In particular, what discriminates these protocols from the \emph{almost secure protocols} class is the resistance against \textit{Selective-internal} internal masquerade attacks.
Therefore, we conclude that approaches based on H-MAC, hash chain, or asymmetric encryption are the most secure in this particular framework of in-vehicle communications.
Besides, LinAuth also shows how using pair-wise shared secret keys allows one to successfully defend against internal masquerade attacks; however, this may raise issues with possible hard real-time compliance.

We classify LCAP and CACAN protocols one and two steps lower than LiBrA-CAN and AuthentiCAN because they offer weaker resistance against internal masquerade attacks.
However, we recall that this lower resistance is given by the fact that the length of the authentication tags has been constrained due to the limited size of standard \ac{can} payloads.
Therefore, if protocols were adapted to work on top of CAN-FD, they would benefit from the increased payload size, having the chance to increase the length of the authentication tag, which results in an increased level of security.
Finally, LinAuth is last classified within this group due to the possible issues for hard real-time deriving from the multiple MACs computation due to the pairwise secret keys approach.

\subsection{Partially Secure Protocols}
\label{subsec:almost}

In this class belong protocols that satisfy all the security requirements apart from the resistance against the \textit{Selective-internal} masquerade attacks.
This class of protocols offers a high-security level with the only exception of that particular situation.

We classify TOUCAN with a higher ranking than CANAuth because \textit{(i)} it is fully backward compatible and \textit{(ii)} tends to introduce a lower storage overhead.
MaCAN instead scores even lower due to the problematic use of group keys that might allow for replay attacks in specific scenarios.

\subsection{Non-Secure Protocols}
\label{subsec:nonsecure}
In this class, some protocols do not offer protection against internal masquerade attacks.
This clearly is a significant security lack that makes the protocol insecure for our scenario.
This is because the security elements of the protocols all depend on one or more common security parameters. Thus, these approaches are not a good solution and are not applicable for secure authentication over \ac{can}.

We classify S2-CAN better than WooAuth and VeCure because these two protocols lack of a resynchronization mechanism, and WooAuth is better than VeCure as it tends to require less storage overhead.

\subsection{Non-Suitable Protocols}
\label{subsec:nonsuitable}

These protocols are unsuitable for hard real-time in-vehicle communication applications as they are not secure and do not satisfy the hard real-time and atomic authentication requirements.
However, we have to specify that if CAN-FD was considered, all protocols could become hard real-time compliant with respect to our definition since they would take advantage of the much larger payload size.
This would, in fact, allow the transmission of the authentication tag together with the authenticated message without the need to split them into two frames.
As a result, if this were the case, LeiA and VulCAN would jump into the \emph{Non-Secure Protocols} class, while vatiCAN and Car2X would still be in this class due to their non-compliance with the replay attack resistance requirement.
\section{Lessons Learned and Future Directions}
\label{sec:Takeaways}

In this section, we point out the key takeaways and possible future research directions that result from our security analysis concerning the adoption of cryptographic authentication solutions for the CAN-bus in the automotive application.

\emph{Lesson 1}. By looking at the results of our security requirements comparison in Table~\ref{tab:comparisonProtocol1} and the protocols classification, we can identify the resistance to masquerade attacks as being the most discriminating requirement.
In particular, thanks to our analysis, we highlight that the major focus in the literature for \ac{can} bus cryptographic authentication protocols is protecting against external attackers, as defined in Section~\ref{sec:adversarial}.
While all the considered protocols are resistant to external masquerade attacks, only half of them are secure against \textit{General-internal} internal masquerade attacks.
Further, just one-third of the considered protocols are fully secure, thus also offering protection against the specific scenario identified by the \emph{Selective-internal} masquerade attack (\ref{subsec:secreq}).

\begin{formal}
\textbf{Lesson 1} -- \textit{Masquerade attack resistance is one of the most important security requirements to ensure protocol security.}
\end{formal}

\emph{Lesson 2}. One of the major criticalities in designing an authentication protocol for \ac{can} is making the authentication tag fit into the frame payload.
This is due to the limited space in a \ac{can} frame.
The most common solutions to face this issue are either a tag truncation or splitting the authentication phase into two messages and embedding the tag in a frame after the data frame.
However, as we have extensively discussed throughout our analysis, these approaches lead to a lower security level in the case of tag truncation or to non-atomic authentication and hard real-time issues, in the case of message splitting.
On the other hand, these problems can be solved by adopting the CAN-FD protocol as the new standard for in-vehicle communications.
Indeed, CAN-FD not only allows for an extended data payload while maintaining an acceptable transmission time and bus load but also offers full backward compatibility.

\begin{formal}
\textbf{Lesson 2} -- \textit{Limited space in a \ac{can} frame compromises the security level, and thus CAN-FD should be preferred.}
\end{formal}

\emph{Lesson 3}. The maturity of cryptographic authentication schemes for \ac{can} seems promising, as demonstrated by the fact that as a result of our analysis, 5 out of 15 protocols (LiBrA-CAN, AuthentiCAN, LCAP, CaCAN, LinAuth) are perfectly suitable for a real-life implementation offering a very high level of security and reliability.
Besides, there are other three protocols (TOUCAN, CANAuth, MaCAN) that offer a high level of security but leave open vulnerabilities only against the very specific scenario of a \emph{Selective-internal} internal masquerade attack.

\begin{formal}
\textbf{Lesson 3} -- \textit{To meet security standards, novel authentication schemes should align with the maturity level of the highest-ranking protocols in our assessment.}
\end{formal}

\emph{Lesson 4}. The best approaches for secure authentication in \ac{can} are based on \textit{(i)} hash chains, \textit{(ii)} asymmetric encryption, and \textit{(iii)} M-MACs, as they offer full security against the threat model considered in this work.
By adopting M-MACs in a broadcast scenario, we gain the advantage that an internal attacker must first learn the (symmetric) secret keys for all the receiving nodes in order to successfully carry out a masquerade attack and correctly generate the ensemble of \ac{mac} values.
By adopting a hash-chain solution, we rather gain advantage of the pre-image resistance property guaranteed by the hash function.
Therefore, given the hash value $H_i$, an attacker cannot guess the following value $H_{i+1}$ (which computation comes before $H_i$ during chain development) for correct authentication.

\begin{formal}
\textbf{Lesson 4} -- \textit{Most secure approaches rely on hash chains, asymmetric encryption, or M-MACs.}
\end{formal}

\emph{Future Research Directions}. Many of the considered protocols designed the authentication procedure without considering the key distribution phase and maintenance, rather than assuming them as a service.
However, in real-life scenarios, these are still open issues that harden the adoption of these protocols, although being mature enough, from the authentication security point of view, to be deployed.
Future research should focus on implementing new methods and optimizing existing ones for key distribution and management tasks.
\section{Conclusions}
\label{sec:conclusions}

In addressing the critical challenge of securing \ac{can} communications in modern vehicles, this study presents and compares the most promising authentication protocols currently under discussion.
Identifying essential security requirements as foundational, our analysis evaluates the strengths and weaknesses of these protocols.
Additionally, we extend our comparison to include operational criteria, providing a comprehensive understanding of the protocols' specifications and features.
In our findings, we observe a prevalent emphasis in existing solutions on fortifying defenses against external attackers, leaving potential vulnerabilities open to internal threats.
However, a notable subset of protocols, namely LiBrA-CAN, AuthentiCAN, LCAP, CaCAN, and LinAuth, emerges as capable of providing comprehensive security for message authenticity and system reliability.
These protocols, leveraging approaches such as hash chains, asymmetric encryption, and \acp{mmac}, showcase robustness in safeguarding \ac{can} communications.
In particular, the efficacy of the hash-chain approach, coupled with asymmetric encryption and \ac{mmac} techniques, stands out as the optimal solution.
This comparative analysis underscores the availability of secure and viable authentication solutions within the literature, dispelling any reservations about adoption.
Such protocols not only ensure secure communication exchanges but also substantially mitigate potential harm from malicious attacks.

In conclusion, this work contributes valuable insights to the realm of \ac{can} security, advocating for the widespread adoption of authentication mechanisms in in-vehicle communication.
By fortifying \ac{can} bus security, our findings aim to motivate and persuade manufacturing industries to embrace these proven and secure authentication solutions, thereby enhancing the overall resilience of automotive systems against cybersecurity threats.

\balance
\bibliographystyle{IEEEtran}
\bibliography{bibliography}

\begin{thebibliography}{100}
\providecommand{\url}[1]{#1}
\csname url@samestyle\endcsname
\providecommand{\newblock}{\relax}
\providecommand{\bibinfo}[2]{#2}
\providecommand{\BIBentrySTDinterwordspacing}{\spaceskip=0pt\relax}
\providecommand{\BIBentryALTinterwordstretchfactor}{4}
\providecommand{\BIBentryALTinterwordspacing}{\spaceskip=\fontdimen2\font plus
\BIBentryALTinterwordstretchfactor\fontdimen3\font minus \fontdimen4\font\relax}
\providecommand{\BIBforeignlanguage}[2]{{%
\expandafter\ifx\csname l@#1\endcsname\relax
\typeout{** WARNING: IEEEtran.bst: No hyphenation pattern has been}%
\typeout{** loaded for the language `#1'. Using the pattern for}%
\typeout{** the default language instead.}%
\else
\language=\csname l@#1\endcsname
\fi
#2}}
\providecommand{\BIBdecl}{\relax}
\BIBdecl

\bibitem{HighEfficiency}
J.~Liu, J.~Wan, D.~Jia, B.~Zeng, D.~Li, C.-H. Hsu, and H.~Chen, ``High-efficiency urban traffic management in context-aware computing and 5g communication,'' vol.~55, no.~1.\hskip 1em plus 0.5em minus 0.4em\relax IEEE, 2017, pp. 34--40.

\bibitem{IoV}
J.~Contreras-Castillo, S.~Zeadally, and J.~A. Guerrero-Iba{\~n}ez, ``Internet of vehicles: architecture, protocols, and security,'' vol.~5, no.~5.\hskip 1em plus 0.5em minus 0.4em\relax IEEE, 2017, pp. 3701--3709.

\bibitem{toor2008vehicle}
Y.~Toor, P.~Muhlethaler, A.~Laouiti, and A.~De~La~Fortelle, ``Vehicle ad hoc networks: Applications and related technical issues,'' \emph{IEEE communications surveys \& tutorials}, vol.~10, no.~3, pp. 74--88, 2008.

\bibitem{hartenstein2008tutorial}
H.~Hartenstein and L.~Laberteaux, ``A tutorial survey on vehicular ad hoc networks,'' \emph{IEEE Communications magazine}, vol.~46, no.~6, pp. 164--171, 2008.

\bibitem{djahel2014communications}
S.~Djahel, R.~Doolan, G.-M. Muntean, and J.~Murphy, ``A communications-oriented perspective on traffic management systems for smart cities: Challenges and innovative approaches,'' \emph{IEEE Communications Surveys \& Tutorials}, vol.~17, no.~1, pp. 125--151, 2014.

\bibitem{checkoway2011comprehensive}
S.~Checkoway, D.~McCoy, B.~Kantor, D.~Anderson, H.~Shacham, S.~Savage, K.~Koscher, A.~Czeskis, F.~Roesner, and T.~Kohno, ``Comprehensive experimental analyses of automotive attack surfaces,'' in \emph{20th USENIX security symposium (USENIX Security 11)}, 2011.

\bibitem{LeiA}
A.-I. Radu and F.~D. Garcia, ``Leia: A lightweight authentication protocol for can,'' in \emph{European Symposium on Research in Computer Security}.\hskip 1em plus 0.5em minus 0.4em\relax Springer, 2016, pp. 283--300.

\bibitem{CAN}
J.~Cook and J.~Freudenberg, ``Controller area network (can),'' in \emph{EECS}, vol. 461, 2007, pp. 1--5.

\bibitem{CAN2}
S.~C. HPL, ``Introduction to the controller area network (can).''\hskip 1em plus 0.5em minus 0.4em\relax Texas instruments, 2002, pp. 1--17.

\bibitem{otherCAN1}
I.~Foster and K.~Koscher, ``Exploring controller area networks,'' \emph{login. USENIX Association}, vol.~40, no.~6, 2015.

\bibitem{CANissues}
R.~Buttigieg, M.~Farrugia, and C.~Meli, ``Security issues in controller area networks in automobiles,'' in \emph{2017 18th International Conference on Sciences and Techniques of Automatic Control and Computer Engineering (STA)}.\hskip 1em plus 0.5em minus 0.4em\relax IEEE, 2017, pp. 93--98.

\bibitem{WooAuth}
S.~Woo, H.~J. Jo, and D.~H. Lee, ``A practical wireless attack on the connected car and security protocol for in-vehicle can,'' vol.~16, no.~2, 2015, pp. 993--1006.

\bibitem{CANauth}
A.~Van~Herrewege, D.~Singelee, and I.~Verbauwhede, ``Canauth-a simple, backward compatible broadcast authentication protocol for can bus,'' in \emph{ECRYPT Workshop on Lightweight Cryptography}, vol. 2011, 2011, p.~20.

\bibitem{LCAP}
A.~Hazem and H.~Fahmy, ``Lcap-a lightweight can authentication protocol for securing in-vehicle networks,'' in \emph{10th escar Embedded Security in Cars Conference, Berlin, Germany}, vol.~6, 2012, p. 172.

\bibitem{kang2018automated}
T.~U. Kang, H.~M. Song, S.~Jeong, and H.~K. Kim, ``Automated reverse engineering and attack for can using obd-ii,'' in \emph{2018 IEEE 88th Vehicular Technology Conference (VTC-Fall)}.\hskip 1em plus 0.5em minus 0.4em\relax IEEE, 2018, pp. 1--7.

\bibitem{zhang2016controlling}
Y.~Zhang, B.~Ge, X.~Li, B.~Shi, and B.~Li, ``Controlling a car through obd injection,'' in \emph{2016 IEEE 3rd International Conference on Cyber Security and Cloud Computing (CSCloud)}.\hskip 1em plus 0.5em minus 0.4em\relax IEEE, 2016, pp. 26--29.

\bibitem{palanca2017stealth}
A.~Palanca, E.~Evenchick, F.~Maggi, and S.~Zanero, ``A stealth, selective, link-layer denial-of-service attack against automotive networks,'' in \emph{Detection of Intrusions and Malware, and Vulnerability Assessment: 14th International Conference, DIMVA 2017, Bonn, Germany, July 6-7, 2017, Proceedings 14}.\hskip 1em plus 0.5em minus 0.4em\relax Springer, 2017, pp. 185--206.

\bibitem{bloom2021weepingcan}
G.~Bloom, ``Weepingcan: A stealthy can bus-off attack,'' in \emph{Workshop on Automotive and Autonomous Vehicle Security}, 2021.

\bibitem{iehira2018spoofing}
K.~Iehira, H.~Inoue, and K.~Ishida, ``Spoofing attack using bus-off attacks against a specific ecu of the can bus,'' in \emph{2018 15th IEEE Annual Consumer Communications \& Networking Conference (CCNC)}.\hskip 1em plus 0.5em minus 0.4em\relax IEEE, 2018, pp. 1--4.

\bibitem{bozdal2020evaluation}
M.~Bozdal, M.~Samie, S.~Aslam, and I.~Jennions, ``Evaluation of can bus security challenges,'' \emph{Sensors}, vol.~20, no.~8, p. 2364, 2020.

\bibitem{remoteExploit}
C.~Miller and C.~Valasek, ``Remote exploitation of an unaltered passenger vehicle,'' vol. 2015, no. S 91, 2015.

\bibitem{lokman2019intrusion}
S.-F. Lokman, A.~T. Othman, and M.-H. Abu-Bakar, ``Intrusion detection system for automotive controller area network (can) bus system: a review,'' \emph{EURASIP Journal on Wireless Communications and Networking}, vol. 2019, pp. 1--17, 2019.

\bibitem{hossain2020lstm}
M.~D. Hossain, H.~Inoue, H.~Ochiai, D.~Fall, and Y.~Kadobayashi, ``Lstm-based intrusion detection system for in-vehicle can bus communications,'' \emph{IEEE Access}, vol.~8, pp. 185\,489--185\,502, 2020.

\bibitem{hanselmann2020canet}
M.~Hanselmann, T.~Strauss, K.~Dormann, and H.~Ulmer, ``Canet: An unsupervised intrusion detection system for high dimensional can bus data,'' \emph{Ieee Access}, vol.~8, pp. 58\,194--58\,205, 2020.

\bibitem{headlights}
K.~Tindell, ``Can injection: keyless car theft,'' \url{https://kentindell.github.io/2023/04/03/can-injection/}, April 2023.

\bibitem{VehicleNet}
P.~Carsten, T.~R. Andel, M.~Yampolskiy, and J.~T. McDonald, ``In-vehicle networks: Attacks, vulnerabilities, and proposed solutions,'' in \emph{Proceedings of the 10th Annual Cyber and Information Security Research Conference}, 2015, pp. 1--8.

\bibitem{survey4}
M.~Wolf, A.~Weimerskirch, and C.~Paar, ``Security in automotive bus systems,'' in \emph{Workshop on Embedded Security in Cars}.\hskip 1em plus 0.5em minus 0.4em\relax Citeseer, 2004, pp. 1--13.

\bibitem{survey13}
I.~Studnia, V.~Nicomette, E.~Alata, Y.~Deswarte, M.~Kaâniche, and Y.~Laarouchi, ``Survey on security threats and protection mechanisms in embedded automotive networks,'' in \emph{2013 43rd Annual IEEE/IFIP Conference on Dependable Systems and Networks Workshop (DSN-W)}, 2013, pp. 1--12.

\bibitem{survey11}
J.~Liu, S.~Zhang, W.~Sun, and Y.~Shi, ``In-vehicle network attacks and countermeasures: Challenges and future directions,'' \emph{IEEE Network}, vol.~31, no.~5, pp. 50--58, 2017.

\bibitem{survey2}
N.~Nowdehi, A.~Lautenbach, and T.~Olovsson, ``In-vehicle can message authentication: An evaluation based on industrial criteria,'' in \emph{2017 IEEE 86th Vehicular Technology Conference (VTC-Fall)}.\hskip 1em plus 0.5em minus 0.4em\relax IEEE, 2017, pp. 1--7.

\bibitem{survey1}
O.~Avatefipour and H.~Malik, ``State-of-the-art survey on in-vehicle network communication "can-bus" security and vulnerabilities,'' 2018.

\bibitem{survey3}
B.~Groza and P.-S. Murvay, ``Security solutions for the controller area network: Bringing authentication to in-vehicle networks,'' vol.~13, no.~1, 2018, pp. 40--47.

\bibitem{survey7}
M.~Gmiden, M.~H. Gmiden, and H.~Trabelsi, ``Cryptographic and intrusion detection system for automotive can bus: Survey and contributions,'' in \emph{2019 16th International Multi-Conference on Systems, Signals \& Devices (SSD)}, 2019, pp. 158--163.

\bibitem{survey10}
M.~Bozdal, M.~Samie, S.~Aslam, and I.~Jennions, ``Evaluation of can bus security challenges,'' \emph{Sensors}, vol.~20, no.~8, 2020.

\bibitem{survey9}
S.~Hartzell, C.~Stubel, and T.~Bonaci, ``Security analysis of an automobile controller area network bus,'' \emph{IEEE Potentials}, vol.~39, no.~3, pp. 19--24, 2020.

\bibitem{survey6}
E.~Aliwa, O.~Rana, C.~Perera, and P.~Burnap, ``Cyberattacks and countermeasures for in-vehicle networks,'' \emph{ACM Comput. Surv.}, vol.~54, no.~1, mar 2021.

\bibitem{survey5}
H.~J. Jo and W.~Choi, ``A survey of attacks on controller area networks and corresponding countermeasures.''\hskip 1em plus 0.5em minus 0.4em\relax IEEE, 2021.

\bibitem{survey8}
T.~M. Fakhfakh~F. and M.~M., ``Cybersecurity attacks on can bus based vehicles: a review and open challenges,'' \emph{Library Hi Tech,}, vol.~40, no.~5, pp. 1179--1203, 2022.

\bibitem{S2-CAN}
M.~D. Pes{\'e}, J.~W. Schauer, J.~Li, and K.~G. Shin, ``S2-can: Sufficiently secure controller area network,'' in \emph{Annual Computer Security Applications Conference}, 2021, pp. 425--438.

\bibitem{CAN+}
T.~Ziermann, S.~Wildermann, and J.~Teich, ``Can+: A new backward-compatible controller area network (can) protocol with up to 16× higher data rates.'' in \emph{2009 Design, Automation Test in Europe Conference Exhibition}, 2009, pp. 1088--1093.

\bibitem{CAN-FD}
F.~Hartwich \emph{et~al.}, ``Can with flexible data-rate,'' in \emph{Proc. iCC}.\hskip 1em plus 0.5em minus 0.4em\relax Citeseer, 2012, pp. 1--9.

\bibitem{wey2013enhancement}
C.-L. Wey, C.-H. Hsu, K.-C. Chang, and P.-C. Jui, ``Enhancement of controller area network (can) bus arbitration mechanism,'' in \emph{2013 International Conference on Connected Vehicles and Expo (ICCVE)}.\hskip 1em plus 0.5em minus 0.4em\relax IEEE, 2013, pp. 898--902.

\bibitem{murvay2017attacks}
P.-S. Murvay and B.~Groza, ``Dos attacks on controller area networks by fault injections from the software layer,'' in \emph{Proceedings of the 12th International Conference on Availability, Reliability and Security}, 2017, pp. 1--10.

\bibitem{comparingCAN-FD}
\BIBentryALTinterwordspacing
``Comparing can fd with classical can.'' [Online]. Available: \url{https://www.kvaser.com/wp-content/uploads/2016/10/comparing-can-fd-with-classical-can.pdf}
\BIBentrySTDinterwordspacing

\bibitem{mavrovouniotis2013hardware}
S.~Mavrovouniotis and M.~Ganley, ``Hardware security modules,'' in \emph{Secure Smart Embedded Devices, Platforms and Applications}.\hskip 1em plus 0.5em minus 0.4em\relax Springer, 2013, pp. 383--405.

\bibitem{Evita}
\BIBentryALTinterwordspacing
``E-safety vehicle intrusion protected applications (evita).'' [Online]. Available: \url{https://evita-project.org/}
\BIBentrySTDinterwordspacing

\bibitem{Evita1}
L.~Apvrille, R.~El~Khayari, O.~Henniger, Y.~Roudier, H.~Schweppe, H.~Seudi{\'e}, B.~Weyl, and M.~Wolf, ``Secure automotive on-board electronics network architecture,'' in \emph{FISITA 2010 world automotive congress, Budapest, Hungary}, vol.~8, 2010.

\bibitem{Evita2}
C.~Labrado and H.~Thapliyal, ``Hardware security primitives for vehicles,'' vol.~8, no.~6.\hskip 1em plus 0.5em minus 0.4em\relax IEEE, 2019, pp. 99--103.

\bibitem{Evita3}
O.~Henniger, A.~Ruddle, H.~Seudi{\'e}, B.~Weyl, M.~Wolf, and T.~Wollinger, ``Securing vehicular on-board it systems: The evita project,'' in \emph{VDI/VW Automotive Security Conference}, 2009, p.~41.

\bibitem{autosar}
S.~F{\"u}rst, J.~M{\"o}ssinger, S.~Bunzel, T.~Weber, F.~Kirschke-Biller, P.~Heitk{\"a}mper, G.~Kinkelin, K.~Nishikawa, and K.~Lange, ``Autosar--a worldwide standard is on the road,'' in \emph{14th International VDI Congress Electronic Systems for Vehicles, Baden-Baden}, vol.~62, 2009, p.~5.

\bibitem{mcivor2006hardware}
C.~J. McIvor, M.~McLoone, and J.~V. McCanny, ``Hardware elliptic curve cryptographic processor over $ rm gf (p) $,'' \emph{IEEE Transactions on Circuits and Systems I: Regular Papers}, vol.~53, no.~9, pp. 1946--1957, 2006.

\bibitem{stallings2006whirlpool}
W.~Stallings, ``The whirlpool secure hash function,'' \emph{Cryptologia}, vol.~30, no.~1, pp. 55--67, 2006.

\bibitem{dworkin2001advanced}
M.~J. Dworkin, E.~B. Barker, J.~R. Nechvatal, J.~Foti, L.~E. Bassham, E.~Roback, and J.~F. Dray~Jr, ``Advanced encryption standard (aes),'' 2001.

\bibitem{SAEprotocol}
\BIBentryALTinterwordspacing
``Introduction to j1939.'' [Online]. Available: \url{https://cdn.vector.com/cms/content/know-how/_application-notes/AN-ION-1-3100_Introduction_to_J1939.pdf}
\BIBentrySTDinterwordspacing

\bibitem{burakova2016truck}
Y.~Burakova, B.~Hass, L.~Millar, and A.~Weimerskirch, ``Truck hacking: An experimental analysis of the $\{$SAE$\}$ j1939 standard,'' in \emph{10th USENIX Workshop on Offensive Technologies (WOOT 16)}, 2016.

\bibitem{murvay2018security}
P.-S. Murvay and B.~Groza, ``Security shortcomings and countermeasures for the sae j1939 commercial vehicle bus protocol,'' \emph{IEEE Transactions on Vehicular Technology}, vol.~67, no.~5, pp. 4325--4339, 2018.

\bibitem{wolf2012design}
M.~Wolf and T.~Gendrullis, ``Design, implementation, and evaluation of a vehicular hardware security module,'' in \emph{Information Security and Cryptology-ICISC 2011: 14th International Conference, Seoul, Korea, November 30-December 2, 2011. Revised Selected Papers 14}.\hskip 1em plus 0.5em minus 0.4em\relax Springer, 2012, pp. 302--318.

\bibitem{car2x}
H.~Schweppe, Y.~Roudier, B.~Weyl, L.~Apvrille, and D.~Scheuermann, ``Car2x communication: Securing the last meter - a cost-effective approach for ensuring trust in car2x applications using in-vehicle symmetric cryptography,'' in \emph{2011 IEEE Vehicular Technology Conference (VTC Fall)}, 2011, pp. 1--5.

\bibitem{LinAuth}
C.-W. Lin and A.~Sangiovanni-Vincentelli, ``Cyber-security for the controller area network (can) communication protocol,'' in \emph{2012 International Conference on Cyber Security}, 2012, pp. 1--7.

\bibitem{MaCAN}
O.~Hartkopp, C.~Reuber, and R.~Schilling, ``Macan - message authenticated can,'' in \emph{Escar Conference}, 2012.

\bibitem{CaCAN}
R.~Kurachi, Y.~Matsubara, H.~Takada, N.~Adachi, Y.~Miyashita, and S.~Horihata, ``Cacan-centralized authentication system in can (controller area network),'' in \emph{14th Int. Conf. on Embedded Security in Cars (ESCAR 2014)}, 2014.

\bibitem{VeCure}
Q.~Wang and S.~Sawhney, ``Vecure: A practical security framework to protect the can bus of vehicles,'' in \emph{2014 International Conference on the Internet of Things (IOT)}, 2014, pp. 13--18.

\bibitem{vatiCAN}
S.~N{\"u}rnberger and C.~Rossow, ``-- vatican-- vetted, authenticated can bus,'' in \emph{Cryptographic Hardware and Embedded Systems -- CHES 2016}, B.~Gierlichs and A.~Y. Poschmann, Eds.\hskip 1em plus 0.5em minus 0.4em\relax Springer Berlin Heidelberg, 2016, pp. 106--124.

\bibitem{Libracan}
B.~Groza, S.~Murvay, A.~V. Herrewege, and I.~Verbauwhede, ``Libra-can: Lightweight broadcast authentication for controller area networks,'' vol.~16, no.~3.\hskip 1em plus 0.5em minus 0.4em\relax Association for Computing Machinery, apr 2017.

\bibitem{VulCAN}
J.~Van~Bulck, J.~T. M\"{u}hlberg, and F.~Piessens, ``Vulcan: Efficient component authentication and software isolation for automotive control networks,'' in \emph{Proceedings of the 33rd Annual Computer Security Applications Conference}, ser. ACSAC '17.\hskip 1em plus 0.5em minus 0.4em\relax Association for Computing Machinery, 2017, p. 225–237.

\bibitem{Toucan}
G.~Bella, P.~Biondi, G.~Costantino, and I.~Matteucci, ``Toucan: A protocol to secure controller area network,'' in \emph{Proceedings of the ACM Workshop on Automotive Cybersecurity}, ser. AutoSec '19.\hskip 1em plus 0.5em minus 0.4em\relax Association for Computing Machinery, 2019, p. 3–8.

\bibitem{authentiCAN}
E.~O. Marasco and F.~Quaglia, ``Authentican: a protocol for improved security over can,'' in \emph{2020 Fourth World Conference on Smart Trends in Systems, Security and Sustainability (WorldS4)}.\hskip 1em plus 0.5em minus 0.4em\relax IEEE, 2020, pp. 533--538.

\bibitem{bogdanov2007present}
A.~Bogdanov, L.~R. Knudsen, G.~Leander, C.~Paar, A.~Poschmann, M.~J. Robshaw, Y.~Seurin, and C.~Vikkelsoe, ``Present: An ultra-lightweight block cipher,'' in \emph{Cryptographic Hardware and Embedded Systems-CHES 2007: 9th International Workshop, Vienna, Austria, September 10-13, 2007. Proceedings 9}.\hskip 1em plus 0.5em minus 0.4em\relax Springer, 2007, pp. 450--466.

\bibitem{chaskey}
N.~Mouha, B.~Mennink, A.~Van~Herrewege, D.~Watanabe, B.~Preneel, and I.~Verbauwhede, ``Chaskey: an efficient mac algorithm for 32-bit microcontrollers,'' in \emph{International Conference on Selected Areas in Cryptography}.\hskip 1em plus 0.5em minus 0.4em\relax Springer, 2014, pp. 306--323.

\bibitem{groza2019tricks}
B.~Groza, L.~Popa, and P.-S. Murvay, ``Tricks—time triggered covert key sharing for controller area networks,'' \emph{IEEE Access}, vol.~7, pp. 104\,294--104\,307, 2019.

\bibitem{wang2017probing}
H.~Wang, D.~Forte, M.~M. Tehranipoor, and Q.~Shi, ``Probing attacks on integrated circuits: Challenges and research opportunities,'' \emph{IEEE Design \& Test}, vol.~34, no.~5, pp. 63--71, 2017.

\bibitem{pinto2019demystifying}
S.~Pinto and N.~Santos, ``Demystifying arm trustzone: A comprehensive survey,'' \emph{ACM computing surveys (CSUR)}, vol.~51, no.~6, pp. 1--36, 2019.

\bibitem{AG}
\BIBentryALTinterwordspacing
I.~T. AG, ``Aurixtm family – tc29xtx.'' [Online]. Available: \url{https://www.infineon.com/cms/en/product/microcontroller/32-bit-tricore-microcontroller/32-bit-tricore-aurix-tc2xx/aurix-family-tc29xtx/}
\BIBentrySTDinterwordspacing

\bibitem{xie2017hardware}
G.~Xie, Y.~Chen, R.~Li, and K.~Li, ``Hardware cost design optimization for functional safety-critical parallel applications on heterogeneous distributed embedded systems,'' \emph{IEEE Transactions on Industrial Informatics}, vol.~14, no.~6, pp. 2418--2431, 2017.

\bibitem{zou2018hardware}
W.~Zou, R.~Li, W.~Wu, and L.~Zeng, ``Hardware cost and energy consumption optimization for safety-critical applications on heterogeneous distributed embedded systems,'' in \emph{2018 IEEE 24th International Conference on Parallel and Distributed Systems (ICPADS)}.\hskip 1em plus 0.5em minus 0.4em\relax IEEE, 2018, pp. 1--10.

\bibitem{xie2019price}
G.~Xie, W.~Ma, H.~Peng, R.~Li, and K.~Li, ``Price performance-driven hardware cost optimization under functional safety requirement in large-scale heterogeneous distributed embedded systems,'' \emph{IEEE Transactions on Industrial Electronics}, vol.~68, no.~5, pp. 4485--4497, 2019.

\bibitem{xie2021security}
Y.~Xie, Y.~Guo, S.~Yang, J.~Zhou, and X.~Chen, ``Security-related hardware cost optimization for can fd-based automotive cyber-physical systems,'' \emph{Sensors}, vol.~21, no.~20, p. 6807, 2021.

\bibitem{NadhirMansour}
N.~M. Ben~Lakhal, O.~Nasri, L.~Adouane, and J.~B. Hadj~Slama, ``Controller area network reliability: overview of design challenges and safety related perspectives of future transportation systems,'' \emph{IET Intelligent Transport Systems}, vol.~14, no.~13, pp. 1727--1739, 2020.

\bibitem{davis2011controller}
R.~I. Davis, S.~Kollmann, V.~Pollex, and F.~Slomka, ``Controller area network (can) schedulability analysis with fifo queues,'' in \emph{2011 23rd Euromicro Conference on Real-Time Systems}.\hskip 1em plus 0.5em minus 0.4em\relax IEEE, 2011, pp. 45--56.

\bibitem{tindell1994guaranteeing}
K.~Tindell and A.~Burns, ``Guaranteeing message latencies on control area network (can),'' in \emph{Proceedings of the 1st International CAN Conference}.\hskip 1em plus 0.5em minus 0.4em\relax Citeseer, 1994.

\bibitem{tindell1995calculating}
K.~Tindell, A.~Burns, and A.~J. Wellings, ``Calculating controller area network (can) message response times,'' \emph{Control engineering practice}, vol.~3, no.~8, pp. 1163--1169, 1995.

\bibitem{tindell1994analysing}
Tindell, Hansson, and Wellings, ``Analysing real-time communications: controller area network (can),'' in \emph{1994 Proceedings Real-Time Systems Symposium}.\hskip 1em plus 0.5em minus 0.4em\relax IEEE, 1994, pp. 259--263.

\bibitem{thirumavalavasethurayar2021implementation}
P.~Thirumavalavasethurayar and T.~Ravi, ``Implementation of replay attack in controller area network bus using universal verification methodology,'' in \emph{2021 International Conference on Artificial Intelligence and Smart Systems (ICAIS)}.\hskip 1em plus 0.5em minus 0.4em\relax IEEE, 2021, pp. 1142--1146.

\bibitem{jo2021survey}
H.~J. Jo and W.~Choi, ``A survey of attacks on controller area networks and corresponding countermeasures,'' \emph{IEEE Transactions on Intelligent Transportation Systems}, vol.~23, no.~7, pp. 6123--6141, 2021.

\bibitem{lin2012cyber}
C.-W. Lin and A.~Sangiovanni-Vincentelli, ``Cyber-security for the controller area network (can) communication protocol,'' in \emph{2012 International Conference on Cyber Security}.\hskip 1em plus 0.5em minus 0.4em\relax IEEE, 2012, pp. 1--7.

\bibitem{sagong2018cloaking}
S.~U. Sagong, X.~Ying, A.~Clark, L.~Bushnell, and R.~Poovendran, ``Cloaking the clock: Emulating clock skew in controller area networks,'' in \emph{2018 ACM/IEEE 9th International Conference on Cyber-Physical Systems (ICCPS)}.\hskip 1em plus 0.5em minus 0.4em\relax IEEE, 2018, pp. 32--42.

\bibitem{choi2018identifying}
W.~Choi, H.~J. Jo, S.~Woo, J.~Y. Chun, J.~Park, and D.~H. Lee, ``Identifying ecus using inimitable characteristics of signals in controller area networks,'' \emph{IEEE Transactions on Vehicular Technology}, vol.~67, no.~6, pp. 4757--4770, 2018.

\bibitem{salmani2005contribution}
H.~Salmani and S.~G. Miremadi, ``Contribution of controller area networks controllers to masquerade failures,'' in \emph{11th Pacific Rim International Symposium on Dependable Computing (PRDC'05)}.\hskip 1em plus 0.5em minus 0.4em\relax IEEE, 2005, pp. 5--pp.

\bibitem{zhang2018new}
H.~Zhang, Y.~Shi, J.~Wang, and H.~Chen, ``A new delay-compensation scheme for networked control systems in controller area networks,'' \emph{IEEE Transactions on Industrial Electronics}, vol.~65, no.~9, pp. 7239--7247, 2018.

\bibitem{groza2013efficient}
B.~Groza and S.~Murvay, ``Efficient protocols for secure broadcast in controller area networks,'' \emph{IEEE Transactions on Industrial Informatics}, vol.~9, no.~4, pp. 2034--2042, 2013.

\bibitem{groza2012broadcast}
B.~Groza and P.-S. Murvay, ``Broadcast authentication in a low speed controller area network,'' in \emph{E-Business and Telecommunications: International Joint Conference, ICETE 2011, Seville, Spain, July 18-21, 2011, Revised Selected Papers}.\hskip 1em plus 0.5em minus 0.4em\relax Springer, 2012, pp. 330--344.

\bibitem{cena2015improving}
G.~Cena, I.~C. Bertolotti, T.~Hu, and A.~Valenzano, ``Improving compatibility between can fd and legacy can devices,'' in \emph{2015 IEEE 1st International Forum on Research and Technologies for Society and Industry Leveraging a better tomorrow (RTSI)}.\hskip 1em plus 0.5em minus 0.4em\relax IEEE, 2015, pp. 419--426.

\bibitem{backcomp}
\BIBentryALTinterwordspacing
Lenovo, ``What is backward compatible?'' [Online]. Available: \url{https://www.lenovo.com/us/en/glossary/backward-compatible}
\BIBentrySTDinterwordspacing

\bibitem{opendbc}
\BIBentryALTinterwordspacing
commaai, ``opendbc,'' 2017. [Online]. Available: \url{https://github.com/commaai/opendbc}
\BIBentrySTDinterwordspacing

\bibitem{musuroi2021fast}
A.~Musuroi, B.~Groza, L.~Popa, and P.-S. Murvay, ``Fast and efficient group key exchange in controller area networks (can),'' \emph{IEEE Transactions on Vehicular Technology}, vol.~70, no.~9, pp. 9385--9399, 2021.

\bibitem{jain2016physical}
S.~Jain and J.~Guajardo, ``Physical layer group key agreement for automotive controller area networks,'' in \emph{Cryptographic Hardware and Embedded Systems--CHES 2016: 18th International Conference, Santa Barbara, CA, USA, August 17-19, 2016, Proceedings 18}.\hskip 1em plus 0.5em minus 0.4em\relax Springer, 2016, pp. 85--105.

\bibitem{wang2018delay}
Q.~Wang, Y.~Qian, Z.~Lu, Y.~Shoukry, and G.~Qu, ``A delay based plug-in-monitor for intrusion detection in controller area network,'' in \emph{2018 Asian Hardware Oriented Security and Trust Symposium (AsianHOST)}.\hskip 1em plus 0.5em minus 0.4em\relax IEEE, 2018, pp. 86--91.

\bibitem{sagong2019mitigating}
S.~U. Sagong, R.~Poovendran, and L.~Bushnell, ``Mitigating vulnerabilities of voltage-based intrusion detection systems in controller area networks,'' \emph{arXiv preprint arXiv:1907.10783}, 2019.

\bibitem{hoppe2009applying}
T.~Hoppe, S.~Kiltz, and J.~Dittmann, ``Applying intrusion detection to automotive it-early insights and remaining challenges,'' \emph{Journal of Information Assurance and Security (JIAS)}, vol.~4, no.~6, pp. 226--235, 2009.

\bibitem{vasile2015performance}
P.~Vasile, B.~Groza, and S.~Murvay, ``Performance analysis of broadcast authentication protocols on can-fd and flexray,'' in \emph{Proceedings of the WESS'15: Workshop on Embedded Systems Security}, 2015, pp. 1--8.

\bibitem{weyl2010secure}
B.~Weyl, M.~Wolf, F.~Zweers, T.~Gendrullis, M.~S. Idrees, Y.~Roudier, H.~Schweppe, H.~Platzdasch, R.~El~Khayari, O.~Henniger \emph{et~al.}, ``Secure on-board architecture specification,'' \emph{Evita Deliverable D}, vol.~3, p.~2, 2010.

\end{thebibliography}

\clearpage
\nobalance
\appendices

\section{Cryptography}
\label{sec:cryptosec}

This section presents an overview of the cryptographic primitives that are introduced in this paper.
We assume we have two entities, Alice and Bob, that want to communicate with each other and an eavesdropper, Eve.

\subsection{Symmetric Encryption Cryptosystem}
Alice and Bob want to communicate secretly, meaning that a possible eavesdropper, Eve, listening to the communication, cannot grasp the meaning of the conversation. 
Assume that Alice and Bob share a secret \emph{K}.
We call it \emph{secret key}.
In a symmetric encryption cryptosystem, Alice encrypts a secret message \emph{u} through a predefined encryption algorithm $E\left(u, K\right)$.
The algorithm takes as input the secret message \emph{u}, the secret key \emph{K}, and returns in output the encrypted message \emph{x}: $x = E\left(u, K\right)$.
Then, Alice transmits \emph{x} to Bob, which in turn execute the decryption algorithm $D\left(x, K\right)$, giving as input the received encrypted message \emph{x} and the key \emph{K}: $u = D\left(x, K\right).$
We can then derive that the following relation must hold: $D_K = E_K^{-1}.$
It is worth noticing that the secrecy of the communication between Alice and Bob comes from the fact that Eve does not know the secret key \emph{K}, which is rather shared between Alice and Bob.

\subsection{Symmetric Authentication}
We want to authenticate the communication, ensuring that Bob distinguishes whether a received message comes from Alice or Eve masquerading as Alice.
Assume that Alice and Bob share a symmetric key $k$.
Alice generates a signature $x$ of the transmitted message in a symmetric authentication mechanism with a signing function $S\left(u, k\right)$. The function receives as input the message $u$ and the key $K$: $x = S\left(u, K\right)$.
Eve wants Bob to accept a forged message $u'$ rather than $u$.
The authentication problem can thus be modeled as a binary problem with a switch $b = 0$ if the message comes from Alice, $b = 1$ otherwise.
Bob verifies the authenticity of the message thanks to the verification function $V\left(\Tilde{x}, K\right)$, taking as input the received signature $\Tilde{x}$ and the key $K$. Hence: $\hat{b} = V\left(\Tilde{x}, K\right)$.
If $\hat{b} = 0$, then Bob accepts the message, as this is coming from the legitimate transmitter; if instead $\hat{b} = 1$, the message is rejected.

\subsection{Asymmetric Encryption Cryptosystem}
In an asymmetric encryption cryptosystem, there is no shared key between Alice and Bob.
Rather, each entity possesses a couple of keys: a \emph{private key} $K$ and a \emph{public key} $K' = f\left(K\right)$. While the former is private and only known by the owner, the latter is known to anybody.
There is a special relationship between the two keys, so if Alice encrypts a message using Bob's public key, only Bob can decrypt it as he is the only one who possesses the corresponding private key.
This system ensures that anyone can encrypt a message, but only the legitimate receiver can remove the encryption.
The security of an asymmetric encryption cryptosystem is based on the following assumptions: \textit{(i)} it is computationally hard to derive $K$ from $K'$; \textit{(ii)} it is computationally hard to derive the message $u$ given $K'$ and the encrypted message $x$; \textit{(iii)} it is computationally hard to derive $K$ from $u$ and $x$.

\subsection{Cryptographic Hash Functions \& Hash Chains}
In general, a \textit{hash function} $h: \mathcal{X} \xrightarrow{} \mathcal{Y}$ maps an arbitrarily long input $x \in \mathcal{X}$ into a fixed-length output $y \in \mathcal{Y}$, called as \emph{hash value}.
The hash value is unique to the input data and should be uniform in $\mathcal{Y}$.
A \textit{cryptographic hash function} is a special class of hash functions for which the following properties must hold:
\begin{enumerate}
    \item \underline{\emph{Pre-Image Resistance}} -- Given $y \in \mathcal{Y}$ it must be computationally difficult retrieve $x | h\left(x\right) = y$.
    \item \underline{\emph{Second Pre-image Resistance}} -- Given $x_1 \in \mathcal{X}$, it must be computationally difficult to find $x_2 \in \mathcal{X}, x_2 \neq x_1 | h\left(x_2\right) = h\left(x_2\right)$.
    \item \underline{\emph{Strong Collision Resistance}} -- It must be computationally hard to find any pair $\left(x_1, x_2\right) \in \mathcal{X}^2, x1 \neq x_2 | h\left(x_2\right) = h\left(x_2\right)$.
\end{enumerate}
A \textit{hash chain} is a collection of hash values $\left\{y_0, \dots, y_n\right\}$ where the following relation holds: $y_0 = h\left(x\right), y_i = h\left(y_{i-1}\right), i = 1, \dots, n$, and $h\left(\cdot\right)$ being a cryptographic hash function.

\end{document}